\documentclass[reprint,amsmath,amssymb,aps,floatfix,superscriptaddress]{revtex4-2}

\usepackage{graphicx}
\usepackage[caption=false]{subfig}
\usepackage{dcolumn}
\usepackage{bm}
\usepackage{braket}
\usepackage{appendix}
\usepackage{todonotes}
\usepackage{soul}
\usepackage[colorlinks = true, linkcolor = blue, citecolor = blue, urlcolor = blue]{hyperref}
\usepackage{academicons}

\newcommand{\orcid}[1]{\href{https://orcid.org/#1}{\textcolor[HTML]{A6CE39}{\aiOrcid}}}

\newcommand{\manny}[1]{%
  \bgroup
  \hskip0pt\color{red!80!black}%
  #1%
  \egroup
}

\newcommand{\pete}[1]{%
  \bgroup
  \hskip0pt\color{green!80!black}%
  #1%
  \egroup
}

\begin{document}

\preprint{APS/123-QED}

\title{Digital control of a superconducting qubit using a Josephson pulse generator at 3~K}

\author{L. Howe}
    \email{logan.howe@nist.gov}
    \email{logan.alfred.howe@gmail.com}
    \affiliation{National Institute of Standards and Technology, Boulder CO 80305, USA}
\author{M. Castellanos-Beltran}
    \email{manuel.castellanosbeltran@nist.gov}
    \affiliation{National Institute of Standards and Technology, Boulder CO 80305, USA}
\author{A. J. Sirois}
    \affiliation{National Institute of Standards and Technology, Boulder CO 80305, USA}
\author{D. Olaya}
    \affiliation{National Institute of Standards and Technology, Boulder CO 80305, USA}
    \affiliation{University of Colorado, Boulder CO 80309, USA}
\author{J. Biesecker}
    \affiliation{National Institute of Standards and Technology, Boulder CO 80305, USA}
\author{P. D. Dresselhaus}
    \affiliation{National Institute of Standards and Technology, Boulder CO 80305, USA}
\author{S. P. Benz}
    \affiliation{National Institute of Standards and Technology, Boulder CO 80305, USA}
\author{P. F. Hopkins}
    \affiliation{National Institute of Standards and Technology, Boulder CO 80305, USA}

\date{\today}


\begin{abstract}
Scaling of quantum computers to fault-tolerant levels relies critically on the integration of energy-efficient, stable, and reproducible qubit control and readout electronics. In comparison to traditional semiconductor control electronics (TSCE) located at room temperature, the signals generated by Josephson junction (JJ) based rf sources benefit from small device sizes, low power dissipation, intrinsic calibration, superior reproducibility, and insensitivity to ambient fluctuations. Previous experiments to co-locate qubits and JJ-based control electronics resulted in quasiparticle poisoning of the qubit; degrading the qubit's coherence and lifetime. In this paper, we digitally control a 0.01~K transmon qubit with pulses from a Josephson pulse generator (JPG) located at the 3~K stage of a dilution refrigerator. We directly compare the qubit lifetime $T_1$, coherence time $T_2^*$, and thermal occupation $P_{th}$ when the qubit is controlled by the JPG circuit versus the TSCE setup. We find agreement to within the daily fluctuations on $\pm 0.5~\mu$s and $\pm 2~\mu$s for $T_1$ and $T_2^*$, respectively, and agreement to within the 1\% error for $P_{th}$. Additionally, we perform randomized benchmarking to measure an average JPG gate error of $2.1 \times 10^{-2}$. In combination with a small device size ($<25$~mm$^2$) and low on-chip power dissipation ($\ll 100~\mu$W), these results are an important step towards demonstrating the viability of using JJ-based control electronics located at temperature stages higher than the mixing chamber stage in highly-scaled superconducting quantum information systems.
\end{abstract}
\keywords{Quantum computing, scalable, cryogenic control, Josephson junction (JJ), Single Flux Quantum (SFQ), transmon, superconductor, voltage metrology}
                              

\maketitle



\section{\label{sec:intro}Introduction}
Error-corrected quantum computers are projected to require large numbers, $\mathcal{O}(10^6)$, of qubits \cite{fowler2012surface, kelly2015state, andersen2020repeated, ai2021exponential}; placing stringent requirements on the per-qubit hardware overhead. Superconducting quantum circuits are a leading technology for scaling existing systems into the noisy intermediate-scale quantum (NISQ) era of $\gtrsim 1000$ qubits. In present systems, qubit gates and entangling operations are performed using shaped microwave pulses synthesized using instrumentation at room temperature \cite{krantz2019quantum} -- here referred to as traditional semiconductor control electronics (TSCE). Signals are routed into a dilution refrigerator (DR) to the $\sim 0.01$~K qubits and typically attenuated by 40--60~dB to suppress thermal noise on drive lines \cite{krinner2019engineering}.

Limitations in cryogenic cooling power, TSCE power instability \cite{vanDijk2019impact}, and system complexity mandate a shift to miniaturize and enhance the stability/precision of waveform generation in superconducting quantum information systems. Recently, the pulse-shaping DAC/mixers used to generate qubit control signals have been successfully integrated at 3~K \cite{vanDijk2020scalable, bardin2019design} and 100~mK \cite{pauka2021cryogenic} using cryogenic CMOS (cryoCMOS) technology. While an impressive step towards miniaturization and large-scale integration, a significant gap exists between these devices and scalable qubit control. Specifically, gate fidelity; power dissipation; and the accuracy, stability, and repeatability of the signals need improvement \cite{sirois2020josephson, vanDijk2019impact, ball2016role}.

The scalability constraints of physical size and power consumption per channel may be satisfied by superconducting Josephson junction (JJ) signal generator circuits or aforementioned cryoCMOS controllers. Comparable to cryoCMOS devices, JJ circuits have small device sizes ($<1 \times 1$~cm$^2$) and very low on-chip power dissipation ($\ll 100~\mu$W); while also leveraging the intrinsically-calibrated nature of single flux quantum (SFQ) pulses. This feature provides avenues for improving waveform quality and repeatability beyond what is achievable using semiconductor-based generators. Capitalizing on pulse area quantization enables use of JJ arrays to construct exceptionally stable and repeatable voltage sources from dc to a few gigahertz \cite{rufenacht2015cryocooled, burroughs2011nist, brevik2018radio, hopkins2019rfwaveform}. Similar devices are used to realize intrinsically accurate voltages for the international system of units, and are disseminated worldwide as primary dc and ac voltage standards \cite{rufenacht2018impact}. Furthermore, the use of SFQ pulses has been proposed as a scalable paradigm for digitally controlling qubits \cite{mcdermott2014accurate, liebermann2016optimal, mcdermott2018quantum}, and was recently demonstrated with a SFQ driver and qubit circuit co-fabricated on the same chip \cite{leonard2019digital}.

\begin{figure*}[ht]
    \centering
    \includegraphics[width = .98 \textwidth]{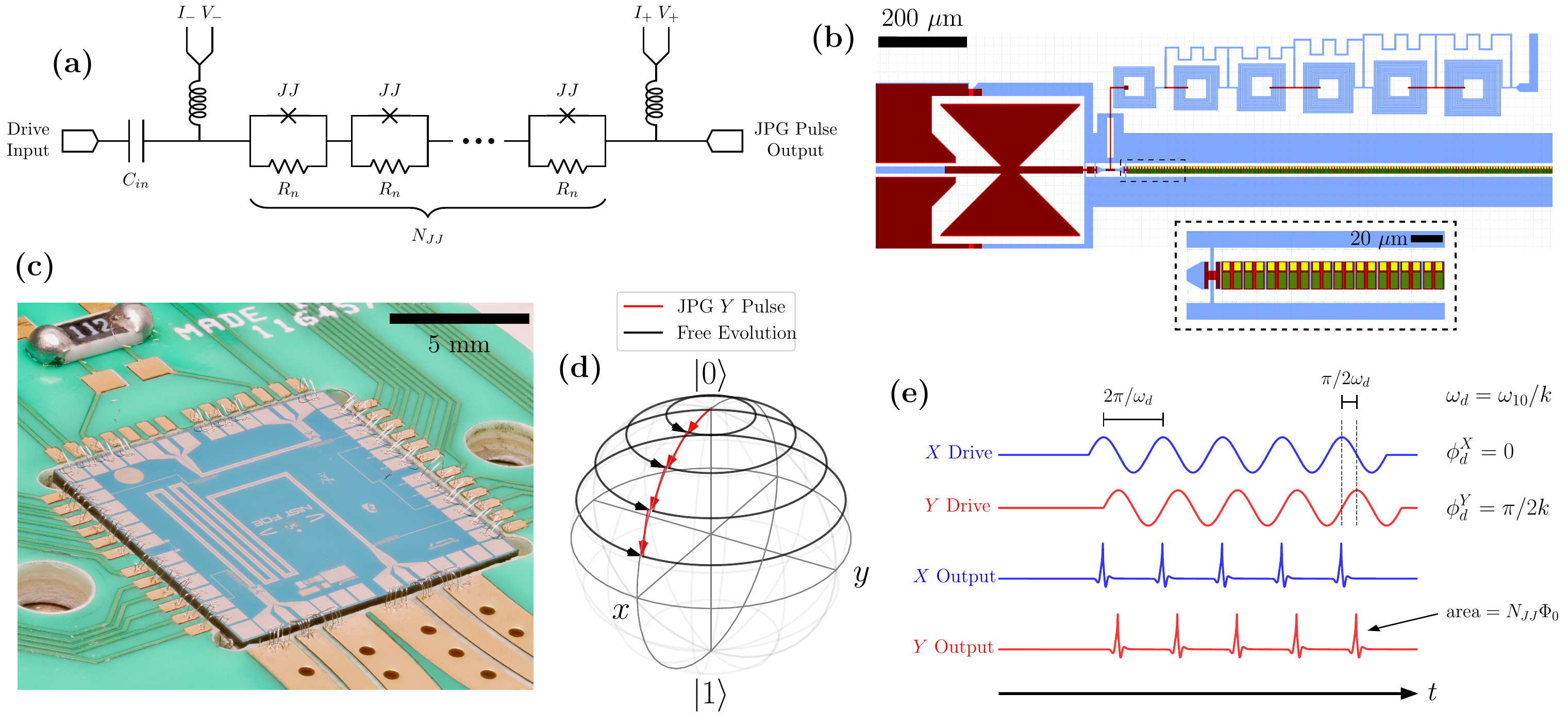}
    \caption{(Color) Digital qubit control with a 3~K JPG constructed using JJs with amorphous silicon ($\alpha$Si) barriers and palladium-gold (PdAu) shunt resistors. \textbf{(a)} Circuit schematic of the JPG containing $N_{JJ} = 4650$ series JJ cells and their shunt resistors $R_n$. Low-frequency inductive taps are used to provide a dc bias current $I_{b}$ and permit measurement of \mbox{$I$-$V$} curves. A dc block ($C_{in}$) is used on the drive input. \textbf{(b)} Rendering of the input portion of the JPG circuit layout showing approximately 250 JJs. Light blue (red) is the base (top) niobium metal layer, green is the $\alpha$Si JJ barrier, and yellow the PdAu shunts, $R_n$. The bowtie structure (left) forms $C_{in}$ and the square washer inductor network (top) forms the low-frequency current and voltage taps. The dashed region shows a magnified view to display the first 20 JJs and shunt resistors in the array. \textbf{(c)} Photo of the packaged JPG chip prior to installation in the DR. The drive input (JPG output) is the left (right) coplanar waveguide microwave launch. \textbf{(d)} Cartoon qubit state evolution on the Bloch sphere under an artificially large coupling ($\delta \theta = \pi / 10$) with $\omega_d = \omega_{10}$. \textbf{(e)} Schematic example of sinusoidal JPG drive and corresponding output. Under proper conditions, one JPG pulse is output per drive period, which may be done at the qubit frequency ($\omega_d = \omega_{10}$) or at any subharmonic $\omega_{10} / m$. The orthogonal $\hat{y}$ control axis is realized by phase-shifting, by $\phi_d^Y$, the drive tone with respect to the timing reference established via the $\hat{x}$ drive. The JPG emits no pulses when the drive is inactive.}
    \label{fig:digital_jpg_qubit_control}
\end{figure*}

The primary limitation of SFQ operation proximal to quantum arrays is degradation of qubit lifetimes from quasiparticles created during pulse generation \cite{patel2017phonon, martinis2009energy}. A solution that mitigates these quasiparticles must be implemented -- such as physically separating the SFQ elements and qubits. In this work we locate the JJ control circuitry on the 3~K DR stage to interrupt quasiparticle-qubit propagation. Similar to \cite{leonard2019digital}, we deliver sparse trains of pulses subresonantly to enact control -- giving our device its name of the Josephson pulse generator (JPG). A 0.01~K bump-bonded multi-chip configuration \cite{ballard2021single, liu2021single} and/or introduction of normal-metal quasiparticle traps can also be effective for quasiparticle mitigation \cite{hosseinkhani2017optimal, martinis2021saving}.

Location of the cryogenic control electronics at a higher temperature stage liberates physical volume at 0.01~K -- commonly monopolized by the quantum array and readout hardware -- and leverages higher cooling powers. This approach may also benefit from integration with \mbox{cryoCMOS} circuits by exploiting the advantages of cryoCMOS-implemented logic/memory elements \cite{bardin2019design, vanDijk2020scalable, pauka2021cryogenic}. Location of the control electronics at 3~K does increase wiring complexity and the parasitic heat loads to the $< 3$~K stages; however, solutions are under development which demonstrate low thermal loading and crosstalk \cite{smith2020flexible, walter2018laminated, tuckerman2016flexible}.

While the aforementioned merits of JJ-based sources \cite{rufenacht2018impact} are expected to apply for qubit control, this work is the first validation of using JJ-based pulse generation at 3~K to control a  0.01~K qubit. Here we show that the JPG does not adversely affect the qubit by separately measuring the qubit energy relaxation time $T_1$, coherence time $T_2^*$, and thermal occupancy $P_{th}$ with both a TSCE setup and the JPG. Our findings show good agreement in all three metrics with each control setup. Additionally, we measure the JPG gate fidelity to be within an order of magnitude of the qubit coherence limit, and provide discussion of future devices expected to yield coherence-limited gates.

\section{\label{sec:sfq_qubit_control}JPG-based Qubit Control}

An input current evolving the JJ superconducting phase difference by $2\pi$ generates a voltage pulse whose time-integrated area equals the magnetic flux quantum $\Phi_0 \equiv h / 2e$:
\begin{equation}
    \int V dt = \Phi_0.
    \label{eq:quantized_voltage_pulse}
\end{equation}
The duration of this SFQ pulse is approximately $\tau = \Phi_0 / I_c R_n$, where $\tau$ is the JJ characteristic time, $I_c$ is its critical current, and $R_n$ is its normal resistance \cite{tinkham2004introduction}. SFQ signal amplification can be achieved by connecting a series of $N_{JJ}$ junctions whose pulses add coherently. This yields a larger pulse of area $N_{JJ} \Phi_0$, which we call a \textit{JPG pulse}. Depending on qubit coupling to the control line, arrays with $N_{JJ} \sim 10^2-10^4$ are required if located at 3~K. In this work our JPG has $N_{JJ} = 4650$, $I_c = 3.05$~mA, and $R_n = 6.93$~m$\Omega$ -- resulting in a characteristic frequency of $f_c = 1 / \tau = 10.2$~GHz. \mbox{Fig.~\ref{fig:digital_jpg_qubit_control}(a)--(c)} show a schematic, portion of the JPG layout, and image of the packaged device.

If $f_c$ is much larger than the qubit transition frequency, $\omega_{10} / 2 \pi$, then during pulse arrival the qubit undergoes a discrete rotation
\begin{equation}\label{eq:delta_theta_sfq}
    \delta \theta = N_{JJ} A C_c \Phi_0 \sqrt{\frac{2 \omega_{10}}{\hbar C_T}},
\end{equation}
where $A$ is the JPG-qubit amplitude attenuation, $C_c$ is the control line-qubit coupling capacitance, and $C_T$ is the qubit capacitance \cite{mcdermott2014accurate}. $N_{JJ}$, $A$, and $C_c$ may be treated as free design parameters to realize a combination of adequate control line thermalization and tip-angle per pulse, $\delta \theta$. For 3D readout cavity configurations $C_c$ also encapsulates attenuation from pulse transit of the cavity resonance. A train of sharp pulses arriving resonantly to the qubit ($\omega_d = \omega_{10}$), or at a subharmonic ($\omega_d = \omega_{10} / k$, where $k \geq 2$ is an integer), discretely rotate the qubit around the Bloch sphere during pulse arrival, while between pulses the qubit precesses for $k$ periods at fixed $\theta$ (see Fig.~\ref{fig:digital_jpg_qubit_control}(d)).

In our implementation the JPG must be driven using a sinusoidal signal at $k \geq 2$ because there is no isolation between the drive input and device output. Otherwise, the large drive signal dominates and induces spurious qubit rotations. Generation of an integer number of JPG pulses $\ell$ is performed by sending an integer number of sinusoidal drive periods, $\nu$. Under the correct bias parameters there is a one-to-one correspondence between the number of JPG pulses generated and the number of drive periods ($\nu = \ell$). Orthogonal axis control, realized by phasing the drive signal relative to a timing reference, is depicted in Fig.~\ref{fig:digital_jpg_qubit_control}(e). More details are found in Supplementary Material~\ref{supp:jpg_orthogonal_axis_control}.

\begin{figure}[t!]
    \centering
    \includegraphics[width = .48 \textwidth]{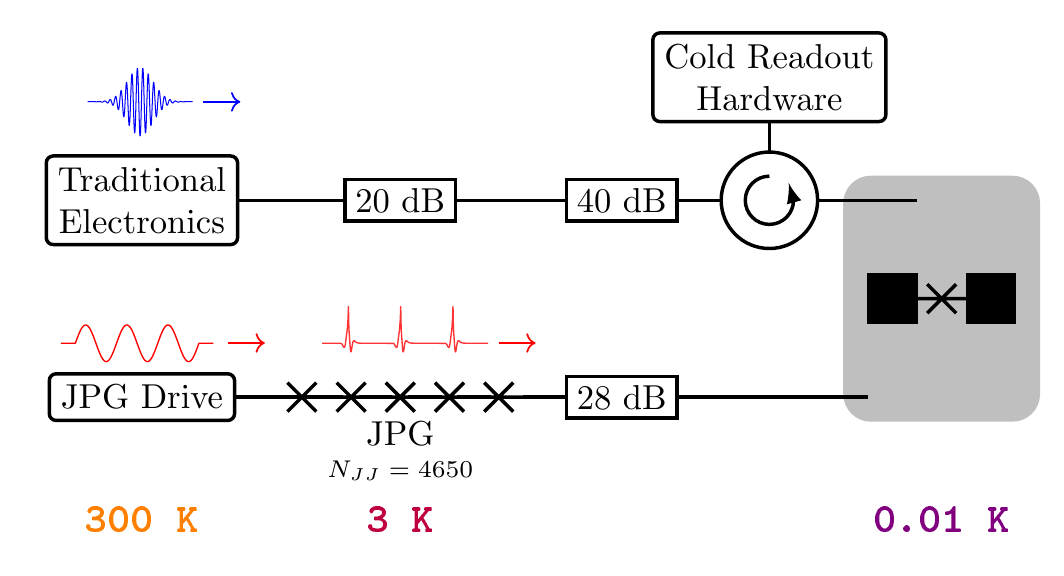}
    \caption{(Color) Simplified schematic of the experimental setup. JPG pulses are routed to the qubit through the weakly coupled cavity port. The conventional TSCE qubit control/readout synthesizers, and cold readout components (JPA, isolators, filters, etc.) are attached to the strongly coupled port. JPG pulse generation is driven by a commercial 65~GSa/s arbitrary waveform generator, which is also semiconductor-based, located at ambient temperature. Control line thermalization is achieved with attenuators or lossy microwave components at the \{3~K, 1~K, 0.05~K, 0.01~K\} DR temperature stages. The attenuation stacks (in dB units) are \{20, 10, 10, 20\} and \{9, 3, 6, 10\} for the TSCE and JPG control lines respectively. See Fig.~\ref{fig:full_schematic} for more details.}
    \label{fig:simplified_schematic}
\end{figure}

\section{\label{sec:experiment_desc}Experimental Details}

In this work we use a transmon qubit dispersively coupled to a 3D aluminum readout cavity possessing two control lines with different coupling strengths. A simplified experimental schematic is shown in Fig~\ref{fig:simplified_schematic}. The JPG is connected to the cavity's weakly-coupled port (0.175~MHz coupling rate) and the TSCE control and readout line to its strongly-coupled port (2.01~MHz coupling rate).
With this setup, a direct comparison of qubit performance with both control schemes is possible during the same cooldown. This qubit was measured for a previous publication; for other parameters see \cite{lecocq2021control}.

Qubit state readout is performed by probing the qubit-state-dependent frequency shift of the cavity. The dressed cavity frequency is $\omega_{\ket{0}, \ket{1}} = \omega_r \pm \chi$, where $\omega_r$ is the bare frequency and $\chi$ is the shift due to cavity-qubit coupling \cite{bianchetti2009dynamics}. A Josephson parametric amplifier \cite{castellanos2007widely} is operated with a gain of 20~dB (phase-insensitive) to enable single-shot measurements. To minimize measurement-induced transitions the cavity probe tone amplitude is typically $n_r \approx 6 \lesssim n_{crit} / 20$, where $n_r$ and $n_{crit}$ are the readout and critical photon numbers \cite{sank2016measurement}. We perform passive qubit state reset via relaxation over a period $ \geq 15 T_1$. The same readout procedure and instrumentation is used for both TSCE and JPG measurements.

\begin{figure*}
    \centering
    \includegraphics[width = .99 \textwidth]{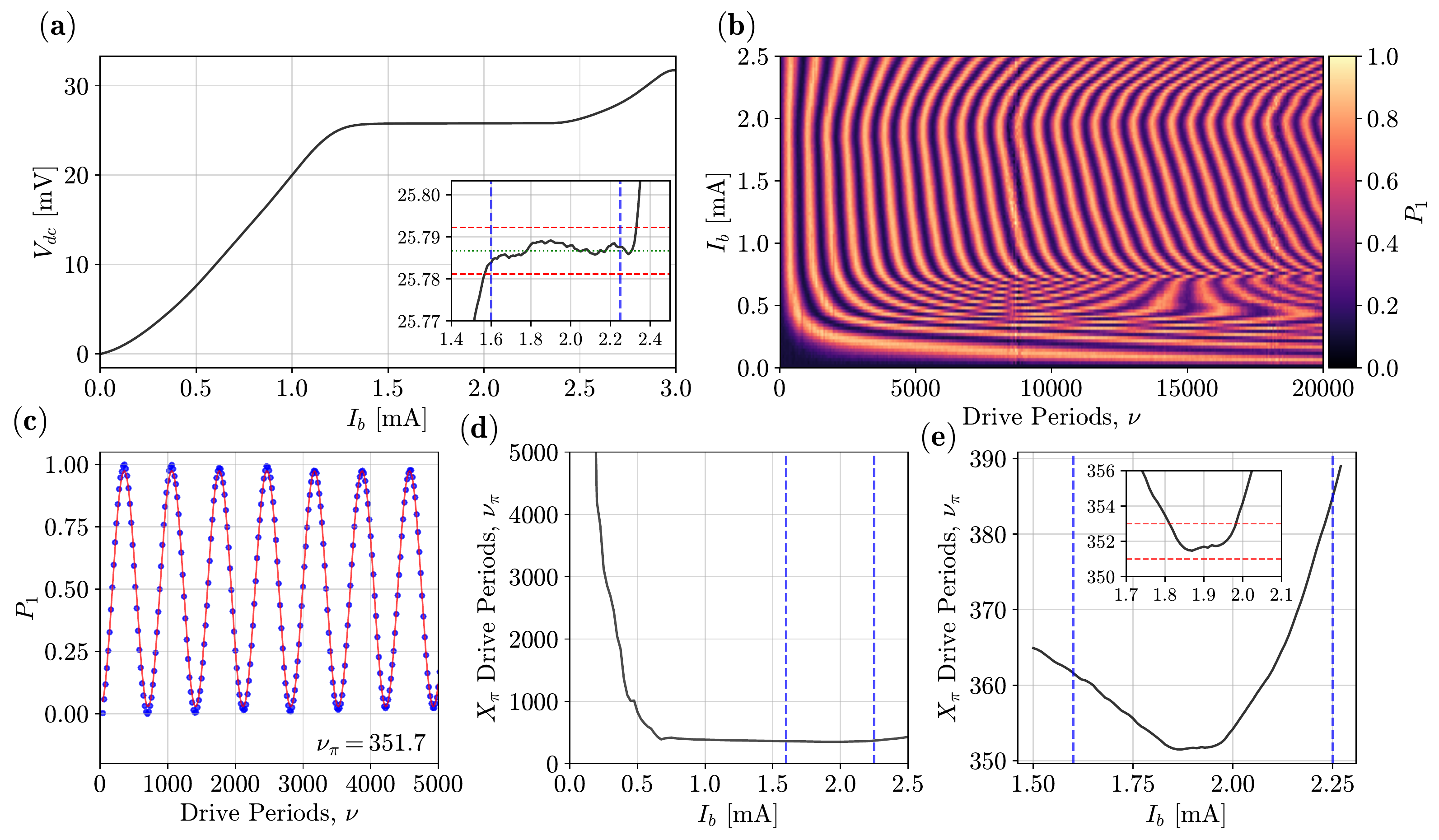}
    \caption{(Color) JPG $X_\pi$ calibration procedure. \textbf{(a)} JPG \mbox{$I$-$V$} curves -- shown here with $\omega_d / 2 \pi = 2.679$~GHz -- are first used to establish rough bounds on the range of $I_b$ giving a constant Rabi oscillation period with respect to the number of drive periods $\nu$. The inset shows behavior on smaller scales near the calculated Shapiro voltage (green dotted line, given by Eq.~\ref{eq:shapiro_voltage}). Red dashed lines correspond to the Shapiro voltage if there were one missing or one additional JJ. From this dc measurement we extract a locking range of 1.6--2.25~mA, which is indicated in all plots with blue dashed lines. \textbf{(b)} Next, we perform a JPG Rabi oscillation scan in which both the JPG current bias $I_b$ and the number of drive periods, $\nu$, are swept. \textbf{(c)} Single $I_b = 1.9$~mA JPG Rabi oscillation measurement. A fit to the functional form $\sin(2\pi \nu / 2 \nu_\pi)$ of the oscillation period yields the corresponding value for $\nu_\pi$ of 351.7 for $I_b = 1.9$~mA -- which is rounded to 352 as we can only send an integer number of pulses. For these data the JPG is driven at 2.685~GHz, resulting in an $X_\pi$ gate time of $t_{gate} = 131$~ns. \textbf{(d)} Fitted $\nu_\pi$ at each $I_b$ from data in (b) emphasizing that at low $I_b$ the JPG does not pulse efficiently (or at all for $I_b = 0$). \textbf{(e)} Analyzed data of a finer resolution Rabi scan around the locking range as determined in (a). The inset again shows smaller scale features with red dashed lines bounding $\nu_\pi$ by $\pm 1$ pulse. Here the Rabi oscillation period of the qubit is insensitive to variations in $I_b$ over the range of \mbox{1.82--1.97}~mA and $\nu_\pi = 352$ is fixed.}
    \label{fig:jpg_bringup}
\end{figure*}

\subsection{\label{sec:jpg_specs}JPG Operation and $X_\pi$ Calibration}
After characterization with the TSCE setup, we establish operating parameters, specifically the rf drive power and dc current bias $I_b$, for the JPG in which the number of output JPG pulses is equal to the number of input drive periods -- called the \textit{locking range}. Under sinusoidal rf drive at frequency $f_d$, a constant voltage Shapiro step manifests at
\begin{equation}
    V = N_{JJ} \Phi_0 f_d
    \label{eq:shapiro_voltage}
\end{equation}
When the measured voltage is constant and equal to Eq.~\ref{eq:shapiro_voltage} over a range of $I_b$, then for any $I_b$ on the Shapiro step the device is locked. Thus, we first maximize the locking range by determining the drive power giving the largest Shapiro steps. Fig.~\ref{fig:jpg_bringup}(a) shows the JPG \mbox{$I$-$V$} curve with maximized locking range.

For all measurements in this work we drive at subharmonic $k = 2$ ($\omega_d = \omega_{10} / 2$) and use the second harmonic power of the JPG pulse train to control the qubit. As we are restricted to subharmonic drive, $k = 2$ maximizes the locking range by making $f_d$ as close as possible to $f_c$ \cite{benz1996pulse} and provides the highest fidelity (fastest) gates. Next, we measure JPG-induced Rabi oscillations to characterize the JPG-qubit interaction. At the optimal drive power we measure the number of drive periods $\nu$ required for a $\pi$ rotation, $\nu_\pi$, versus $I_b$. Results of this procedure are shown in Fig.~\ref{fig:jpg_bringup}(b)--(e). Fitting these Rabi oscillations at constant $I_b$ yields $\nu_\pi(I_b)$ and we look for regions where $\nu_\pi$ is insensitive to the number of Rabi periods (i.e. drive time). This demonstrates locking of the JPG, where $\nu = \ell$, and a stable JPG-qubit interaction as the drive pattern is lengthened.

One may expect the entire locking range in Fig.~\ref{fig:jpg_bringup}(a) to give a constant $\nu_\pi$, however, this is not observed in Fig~\ref{fig:jpg_bringup}(e). This is because the pulse width for Josephson devices operated at $f_d \sim f_c$ varies as $I_b$ traverses the Shapiro step. Widening of the pulses results in a reduction in $\delta \theta$ and is discussed further in the following section, and in Supplementary Material~\ref{supp:digital_pulse_fidelity}. Our simulations for pulse width variation across the Shapiro step agree with previous work \cite{donnelly2020oneghz, babenko2020characterization} and a variation of $< 10\%$ is expected. This restricts the region of constant $\nu_\pi$ in the Rabi measurements relative to the dc locking range measurement. Despite these effects, Fig.~\ref{fig:jpg_bringup}(e) nevertheless demonstrates a range of 150~$\mu$A where $\nu_\pi$ is constant to within one pulse.

\begin{figure}[ht]
    \centering
    \includegraphics[width = .48 \textwidth]{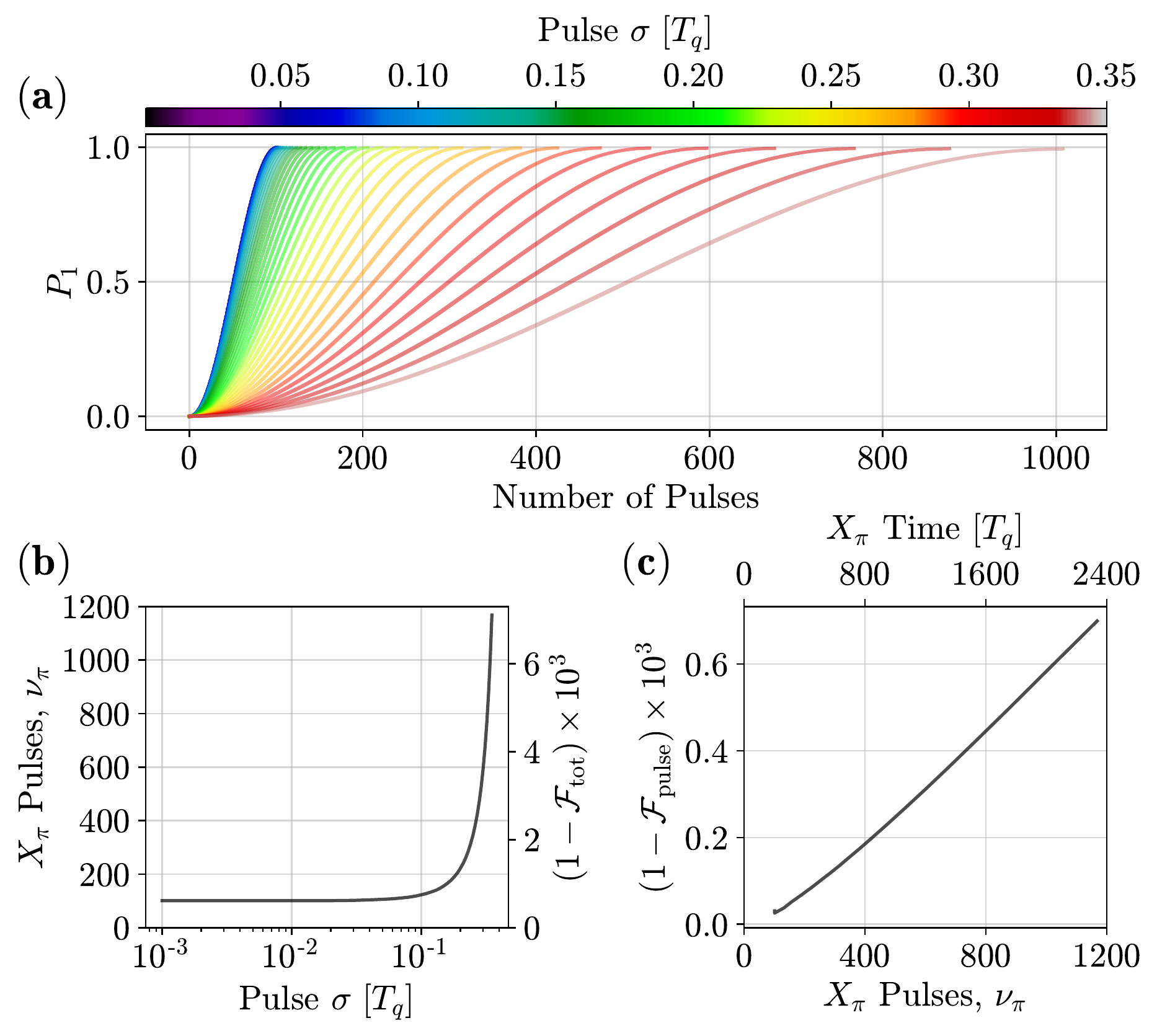}
    \caption{(Color) Simulations of a qubit digitally driven by a Gaussian pulse train to extract the expected $X_\pi$ fidelity as a function of the pulse standard deviation $\sigma$. Here $\sigma$ is in units of qubit periods $T_q$. Pulses are delivered at $\omega_d = \omega_{10} / 2$ so a pulse arrives at every other qubit period. We use energy relaxation and dephasing rates consistent with our qubit (see Sec.~\ref{sec:randomized_benchmarking}). Fits to single JPG pulses measured at room temperature yield an upper bound of $\sigma_{\mathrm{JPG}} = 0.19~T_q$~(35~ps) for the width of pulses delivered to the qubit when parameterized as a Gaussian. \textbf{(a)} Qubit excitation probability as a function of number of pulses for increasing $\sigma$ and for the first half of the first Rabi oscillation. \textbf{(b)} Extracted value of $\nu_\pi$ versus $\sigma$ and the corresponding total $X_\pi$ total infidelity, $1 - \mathcal{F}_{\mathrm{tot}}$. The pulse-qubit coupling is normalized such that for the Dirac delta function limit $\nu_\pi = 100$. With the coupling re-adjusted to match our experimental values of $\nu_\pi = 352$ and $\sigma_{\mathrm{JPG}} = 0.19~T_q$, we obtain $1 - \mathcal{F}_{tot} = 2 \times 10^{-3}$ and $1 - \mathcal{F}_{pulse} = 1 \times 10^{-4}$. \textbf{(c)} Digital-pulse-only $X_\pi$ infidelity, $1 - \mathcal{F}_{\mathrm{pulse}}$, as function of $\nu_\pi$ -- which is parameterized by $\sigma$ in (b). $1 - \mathcal{F}_{\mathrm{pulse}}$ is calculated by subtracting the coherence limit contribution to $1 - \mathcal{F}_{\mathrm{tot}}$.}
    \label{fig:gaussian_pulse_sims}
\end{figure}

\subsection{\label{sec:gauss_pulse_sims}Finite-Width Pulses}

Production of perfectly sharp pulses is not possible, so $\delta \theta$ also depends on JPG pulse width. Indeed, we have demonstrated this by broadening the JPG pulses (we heat the JPG to reduce $I_c$) and observe $\nu_\pi$ to increase by approximately the same factor $I_c$ decreases. To explore the pulse width dependence of $\delta \theta$ we perform simulations \cite{johansson2012qutip} of a qubit driven at $\omega_d = \omega_{10} / 2$ by Gaussian pulses whose width (standard deviation $\sigma$ in units of the qubit period, $T_q$) we control. Fig.~\ref{fig:gaussian_pulse_sims} shows our simulation results and illustrates a strongly nonlinear dependence on the qubit response for $\sigma > 0.25~T_q$. For short pulses in the Dirac delta function limit $\sigma < 0.01~T_q$, the qubit response is independent of $\sigma$ ($\nu_\pi$ changes by less than one pulse). Supplementary Material~\ref{supp:jpg_simulations} discusses the relationship between JJ $\tau$ and $\sigma$ of a Gaussian fitted to the pulses. For our JPG with $\tau = 98$~ps the (on-chip) Gaussian-parameterized pulse width is $\sigma = 17$~ps.

To measure the pulse width, the JPG output is split (see Fig.~\ref{fig:full_schematic}) and recorded with an oscilloscope at room temperature. We find $\sigma = 35$~ps which, for our 5.37~GHz qubit, gives $\sigma = 0.19~T_q$. This is an upper bound for the widths of pulses delivered to the qubit due to added dispersion in the additional 2~m of JPG-oscilloscope cabling compared to the JPG-qubit cable length. From the simulations (after adjusting coupling so $\nu_\pi = 352$) and pulse width measurements we obtain a lower bound of our expected JPG $X_\pi$ infidelity of $2 \times 10^{-3}$. While the total infidelity is coherence-limit-dominated, subtraction of this contribution gives the infidelity due only to the nature of control via digital pulses, $1 - \mathcal{F}_{\mathrm{pulse}}$. Fig.~\ref{fig:gaussian_pulse_sims}(c) demonstrates that, for our current gate times, this digital-pulse-only infidelity is $\sim 10\%$ of the total infidelity. Furthermore, this is competitive with state-of-the-art TSCE techniques, reaching $\sim 10^{-4}$ infidelity, and shows there is no fundamental limitation imposed by digital control with sharp pulses \footnote{These simulations were repeated using simulated SFQ pulses with the JPG $\tau = 98$~ps (Supplementary Material~\ref{supp:jpg_simulations}) and show no significant change in fidelity.}. See Supplementary Material~\ref{supp:digital_pulse_fidelity} for more discussion.

When driven at a frequency below $f_c$, our prototype JPG's locking range decreases by a factor $\propto f_d / f_c$ but a lower $f_c$ also compromises ideal digital qubit control dynamics. More ideal control may be realized at the expense of locking range (or vice versa) by tuning $f_c = I_c R_n / \Phi_0$ -- with the caveat that $\sigma \gtrsim 0.3~T_q$ pulses are too wide for efficient digital control. We keep $\sigma < 0.2~T_q$ to balance locking range and optimal qubit dynamics. Future devices (see Sec.~\ref{sec:scalability_and_discussion}) will not possess this limitation.

\begin{figure*}
    \centering
    \includegraphics[width = .98 \textwidth]{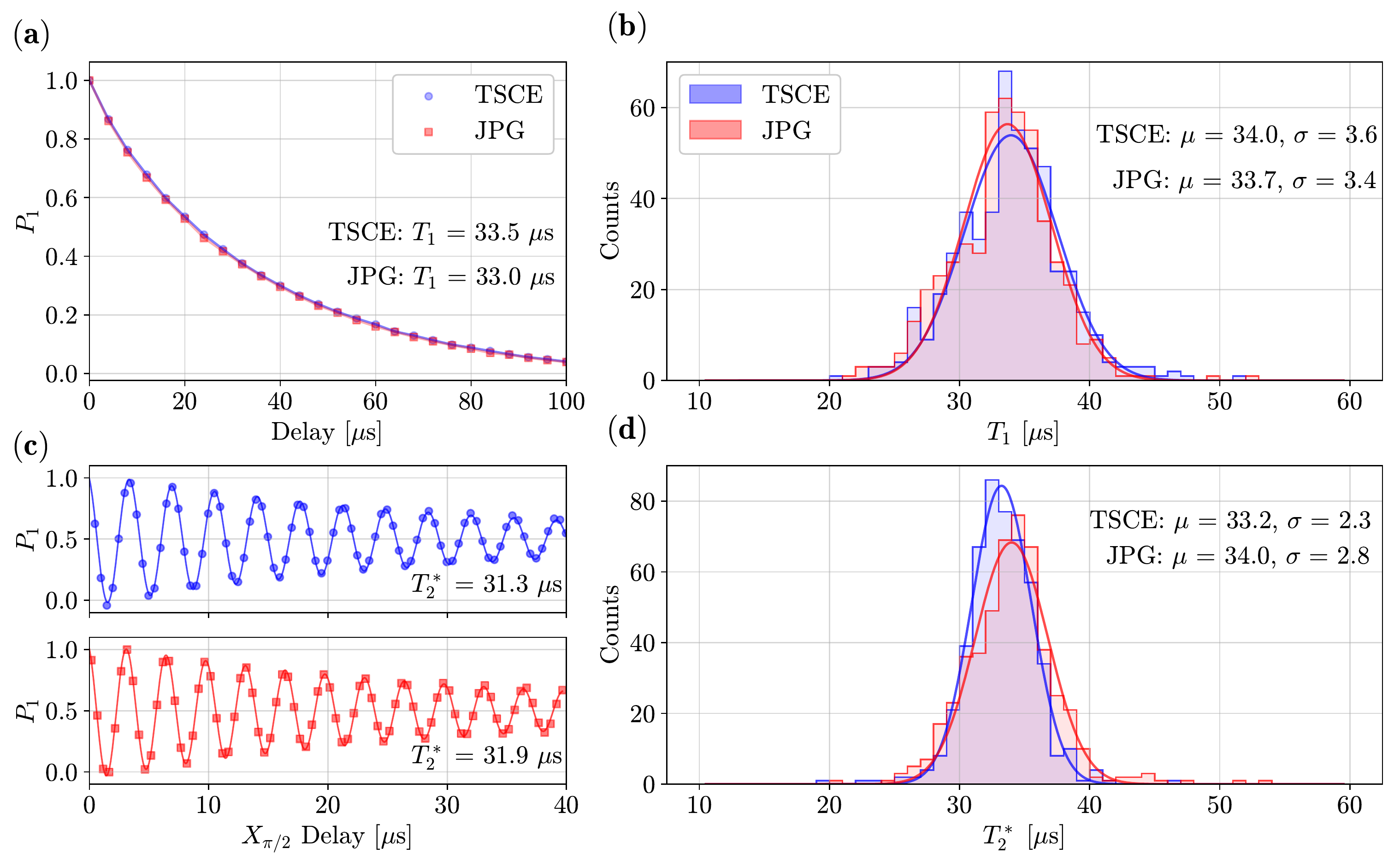}
    \caption{(Color) Comparison of the measured qubit lifetime $T_1$ and Ramsey coherence time $T_2^*$ using a TSCE setup and a JPG at 3~K. For each each setup 500 individual measurements of $T_1$ and $T_2^*$ are taken. Data in blue (circles) are data taken with the TSCE setup, while data in red (squares) are with the JPG. Observed daily fluctuations in the $T_1$ and $T_2^*$ distribution means -- i.e. the uncertainty in the means -- are $\pm 0.5~\mu$s and $\pm 2~\mu$s, respectively. Solid lines are fits to the $T_1$ decay curves and Ramsey fringes. \textbf{(a)} $T_1$ decay curve averaged across all measurements and fitted $T_1$ value for each setup. \textbf{(b)} Histograms and Gaussian fits of the $T_1$ distributions showing excellent agreement in the mean ($\mu$) and standard deviation ($\sigma$) of the distributions. \textbf{(c)} Ramsey fringe averaged across all measurements and fitted $T_2^*$ value for each setup. \textbf{(d)} Histograms and Gaussian fits for $T_2^*$, again showing good agreement with each setup.}
    \label{fig:T1_T2_comparison}
\end{figure*}

\section{\label{sec:qubit_performance_comparison}Qubit Performance Comparison}

Here we describe the side-by-side comparison of the TSCE and JPG setups through measurements of $T_1$, $T_2^*$, and $P_{th}$. For the $T_1$ comparison, a JPG $X_\pi$ rotation is constructed of $\nu_\pi = 352$ drive periods (131~ns drive time) obtained with the calibration shown in Fig.~\ref{fig:jpg_bringup}. For the $T_2^*$ comparison a JPG $X_{\pi/2}$ rotation is created with a $\nu_\pi /2 = 176$ period drive waveform. We gather statistics on 500 measurements of $T_1$ and $T_2^*$ with each setup. Data are compiled in Fig.~\ref{fig:T1_T2_comparison} and show energy decay curves and Ramsey fringes averaged over all measurements, as well as histograms of the extracted $T_1$ and $T_2^*$. The small discrepancy in the distribution means are well within the expected variation in $T_1$ and $T_2^*$ for superconducting qubits \cite{burnett2019decoherence, vepsalainen2020impact, martinis2021saving, mcrae2020materials, ai2021exponential, mcewen2021resolving} and within the observed daily fluctuations for this device of $\pm 0.5~\mu$s and $\pm 2~\mu$s for the $T_1$ and $T_2^*$ mean values, respectively. Indeed, excellent agreement is found in $T_1$ and $T_2^*$ as measured with each setup -- showing that JPG operation does not enhance relaxation or dephasing from quasiparticle poisoning or larger cavity photon number fluctuations.

An important validation of JPG-qubit compatibility is to demonstrate adequate thermalization when controlled with the JPG. State inversions from elevated qubit thermal occupancy can be $\gtrsim 10$\% in transmon qubits with 3D aluminum readout cavities \cite{wenner2013excitation, jin2015thermal, jeffrey2014fast}. We define the qubit thermal occupancy $P_{th}$ as the probability of incorrect state identification based on the desired preparation.

Measurement of $P_{th}$ is performed in a two-part experiment \cite{walter2017rapid}. First, no qubit rotation is applied and the state is simply measured. Second, we apply an $X_\pi$ rotation to invert the qubit population, and then measure. Measurements are single-shot and we do not perform heralding. The total state preparation and measurement (SPAM) fidelity is
\begin{equation}
    \mathcal{F} = 1 - P(1 | 0) - P(0 | 1),
    \label{eq:readout_fidelity}
\end{equation}
where $P(i | j)$ is the probability of measuring state $\ket{i}$ when the qubit was intended to be prepared in $\ket{j}$. Eq.~(\ref{eq:readout_fidelity}) describes the combined preparation fidelity and ability for the single-shot measurement to correctly distinguish between $\ket{0}$ or $\ket{1}$. Ideally, each $P(i | j)$ contains contributions only from thermal occupancy. In reality, both $P(i | j)$ include decays from correctly prepared $\ket{1}$ and spuriously excited $\ket{0}$ initial states, and the $\ket{0}$ and $\ket{1}$ distribution overlap. The SPAM fidelity thus gives an upper bound on $P_{th}$ and we minimize the effects of overlap infidelity and decays during measurement to improve our estimate of $P_{th}$.

We limit these decays to $< 1\%$, which become the dominant uncertainty in the $P_{th}$ measurement, by shortening the readout pulse to 400~ns and combat the corresponding reduction of the single-shot SNR using an optimal mode-matching integration weight function \cite{walter2017rapid}. Overlap infidelity is minimized by increasing the cavity drive strength to separate the primary $\ket{0}$ and $\ket{1}$ distribution lobes. For the 400~ns readout pulse this occurs at $n_r = 50 \approx n_{crit} / 2.3$, which still avoids measurement-induced transitions \cite{sank2016measurement}. 

\begin{figure}[t]
    \centering
    \includegraphics[width = .45 \textwidth]{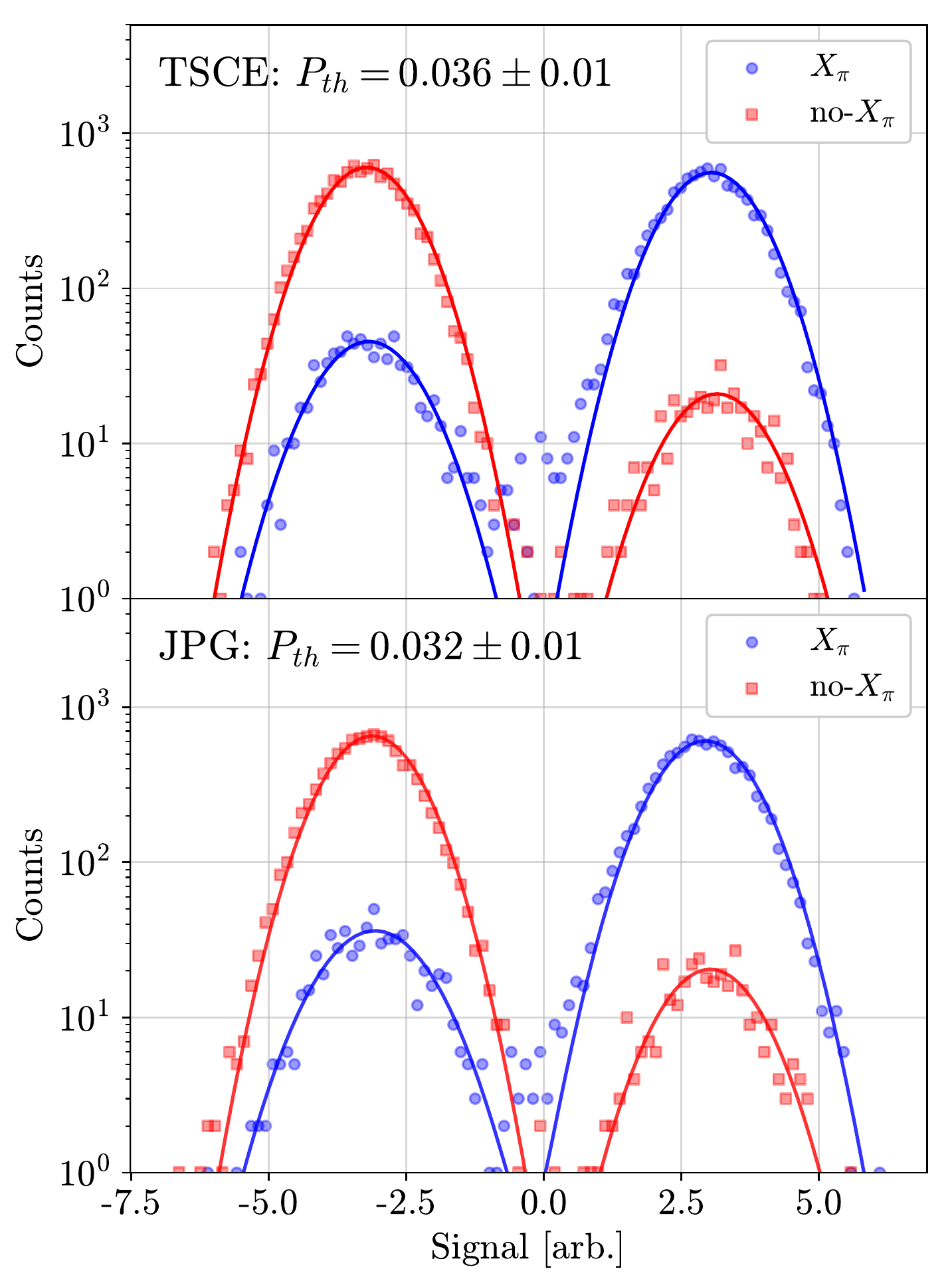}
    \caption{(Color) Measurement of the qubit thermal population, $P_{th}$, to compare control with the TSCE (top) and JPG (bottom). Solid lines are a bimodal Gaussian fit to the data. The $P_{th}$ estimate is obtained by integrating the fit to the no-$X_\pi$ population with voltage levels greater than zero. These data points correspond to instances in which a spurious state inversion occurred; spoiling the desired $\ket{0}$ preparation. The primary uncertainty in $P_{th}$ using this method is due to decays during measurement, i.e. a $\sim 1\%$ error.}
    \label{fig:thermal_pop_comparison}
\end{figure}

Fig.~\ref{fig:thermal_pop_comparison} shows data from $10^4$ measurements with each setup and bimodal Gaussian fits to the data. As the no-$X_\pi$ case (desired preparation in $\ket{0}$) additionally removes decays during preparation, we choose this population to bound $P_{th}$ with the most accuracy. Integration of all spurious $\ket{1}$ outcomes for this case yields excellent agreement in $P_{th}$ of $0.036 \pm 0.01$ and $0.032 \pm 0.01$ for the TSCE and JPG setups, respectively. These results demonstrate that qubit thermalization is not affected when the JPG is used -- our final compatibility metric of digital control using JJ-based pulses from 3~K.

\section{\label{sec:randomized_benchmarking}JPG Randomized Benchmarking}

We now characterize JPG gate fidelities through a randomized benchmarking (RB) routine \cite{knill2008randomized, magesan2012efficient, magesan2011scalable, chen2018metrology, rol2017restless} where we apply a sequence of $m$ random gates followed by a single sequence-inverting gate. The sequence fidelity is an exponential decay
\begin{equation}
    \mathcal{F}(m) = a p^m + b,
    \label{eq:rb_fidelity}
\end{equation}
where the constants $a$ and $b$ encapsulate SPAM errors and errors on the final gate. For single qubit gates, the depolarizing parameter $p$ is related to the per-gate error $r$ by
\begin{equation}
    r = \frac{1}{2}(1 - p).
    \label{eq:per_gate_error_rb}
\end{equation}
Results of the same routine using the TSCE setup are provided as a reference gate error.

We choose the set of primitive and Pauli gates: \mbox{\{$I$, $\pm X_{\pi/2}$, $\pm Y_{\pi/2}$, $\pm X_\pi$, $\pm Y_\pi$\}}, where the idle $I$ and $X_\pi$ gates lengths are equal \cite{petit2020universal, mckay2019three}. Given that $\pi$ gates are twice as long as $\pi/2$ gates, and the fact we use a Clifford group subset, rescaling $m$ by a factor of 1.125 permits comparison to full-Clifford-group RB \cite{chen2018metrology,leonard2019digital}. JPG gates are constructed as described above, while TSCE gates use a $\sigma = 35$~ns Gaussian pulse truncated at $\pm 2 \sigma$ to closely match the JPG gate time.

\begin{figure}
    \centering
    \includegraphics[width = .47 \textwidth]{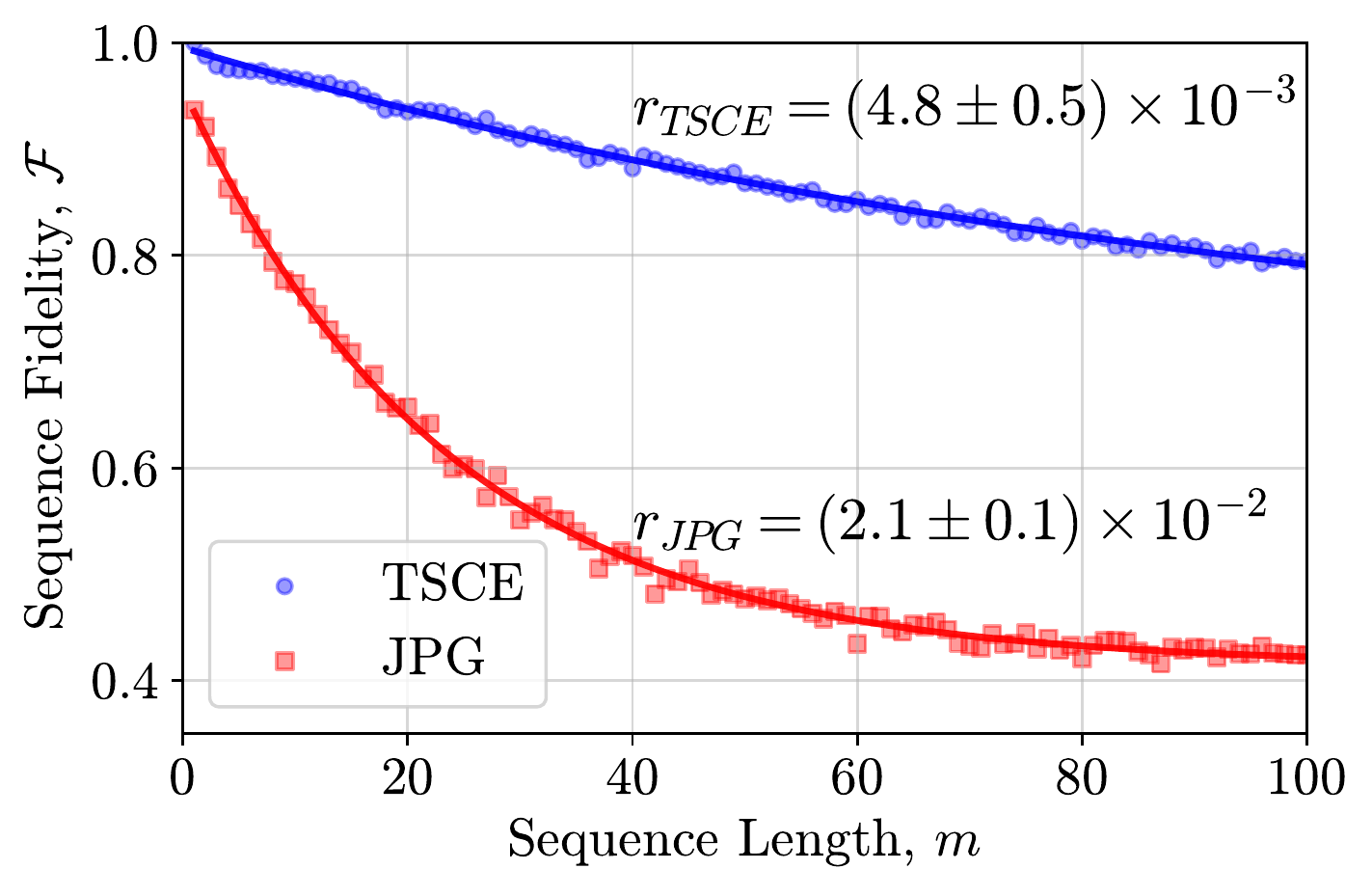}
    \caption{(Color) Depolarizing curve for single qubit RB using the TSCE and 3~K JPG qubit control setups. Solid lines are a fit to Eq.~\ref{eq:rb_fidelity}. TSCE rotations are performed with Gaussian pulses with $\sigma = 35$~ns and truncated at $\pm 2 \sigma$ to match the JPG $X_{\pi}$ gate length of 131~ns. We extract an average error per gate of $r_{TSCE} = (4.8 \pm 0.5) \times 10^{-3}$ and $r_{JPG} = (2.1 \pm 0.1) \times 10^{-2}$. The former is just over twice the coherence limit of the qubit, while the latter is a factor of 10 higher than the $X_\pi$ fidelity of $2 \times 10^{-3}$ as determined in simulations -- which is dominated by qubit coherence. Uncertainties in $r$ are determined as the standard error of the fits to Eq.~\ref{eq:rb_fidelity}.}
    \label{fig:jpg_rb}
\end{figure}

In Fig.~\ref{fig:jpg_rb} we show the results of the RB routine with both setups giving $r_{TSCE} = (4.8 \pm 0.5) \times 10^{-3}$ and $r_{JPG} = (2.1 \pm 0.1) \times 10^{-2}$, where the uncertainties are from the Eq.~\ref{eq:rb_fidelity} fit standard error. The JPG $r$ is approximately a factor of 10 higher than the simulated single $X_\pi$ gate error of $2.1 \times 10^{-3}$ \footnote{Note that a small measurement error is present in all the JPG RB data, which were taken first and on a separate cooldown than the TSCE RB. This error was remedied before the TSCE measurement was performed and is responsible for the offset in the JPG curve (Fig.~\ref{fig:jpg_rb}). We emphasize this has no influence on the extracted $r$ for either setup.}. Performing a detailed accounting of known possible errors (Supplementary Material~\ref{supp:digital_pulse_fidelity}) from digitization, finite pulse widths, higher state leakage, and pulse timing jitter gives an estimated infidelity of $6 \times 10 ^{-3}$. This is a factor of three below the RB result. We attribute the remaining error to possible systematic or coherent errors which are presently under investigation. Regardless, these measurements serve as an excellent proof-of-concept demonstration of qubit control using 3~K JJ-based digital pulses.

\section{\label{sec:scalability_and_discussion}Scalability and Discussion of Future Devices}
The scalability of digital qubit control using JJ devices at 3~K is promising even with the current device and configuration -- which were not optimized for size or power dissipation. The JPG circuit is $<15$~mm$^2$ and the power dissipated at duty cycle $\eta_d$ is approximately
\begin{equation}
    P = I_c V = \Phi_0 N_{JJ} I_c f_d \eta_d.
    \label{eq:jpg_power}
\end{equation}
For the maximum $\eta_d$ in our experiment of 0.02, this yields $P = 1.6~\mu$W. Commercial cryocoolers delivering $\sim 1$~W of cooling power at 3~K \cite{tsan2021effects} permit use of over $\sim$500k similar devices; occupying $\sim 10^{-2}$~m$^3$. The JPG pulse output power is $\sim -56$~dBm (at $\eta_d = 0.02$), giving off-chip dissipation of the qubit drive signal of \{2~nW, 0.3~nW, 0.02~nW, 0.008~nW\} at the \{3~K, 1~K, 0.05~K, 0.01~K\} stages (see Supplementary Material~\ref{supp:full_schematic} for details of the JPG control line attenuation stack). Of larger consideration is dissipation of the large drive signal, which is $\sim -6$~dBm at the JPG input at 3~K, however the new devices discussed below offer techniques to circumvent this limitation.

Plans for NISQ systems require \mbox{$\sim 10$~m$^3$} of cryostat volume and numerous cryocoolers. We conclude that neither power dissipation, nor device size (volume of silicon) present significant obstacles in scaling the number of 3~K JJ devices to control NISQ-era quantum arrays. Furthermore, our experimental and JPG architectures can both be adjusted in a straightforward manner to reduce device size and on-chip power dissipation each by more than a factor of 10. Thus, the primary obstacle in scaling a 3~K JJ-based qubit control architecture, at least in the near-term, is one shared by many competing qubit control technologies: wiring and signal routing logistics. Multi-chip modules \cite{holman2021three, gold2021entanglement, hidakajapanese}, high-density/bandwidth interconnects \cite{smith2020flexible, tuckerman2016flexible}, and out-of-plane coupling \cite{rahamim2017double} make overcoming these challenges feasible.

The present experiment architecture, where JPG pulses are heavily attenuated by the cavity resonance before reaching the qubit, necessitates use of a large $N_{JJ}$ to yield appropriate signal levels. This increases device size, power dissipation, and can reduce qubit coherence by allowing the higher pulse train harmonics to populate the cavity (not observed here). These limitations can be eliminated using 2D readout cavity qubit devices and an independent control line. With such a device we expect to reduce the array factor to $N_{JJ} \approx 500$ without sacrificing thermalization or gate time.

Next-generation devices will implement an SFQ logic shift register and voltage multiplier pulse amplification with no increase in JJ count \cite{hopkins2019rfwaveform, castellanos2021single}. Use of a high speed clock far above the qubit spectrum eliminates the qubit-drive interaction and permits pulse delivery at $\omega_d \geq \omega_{10}$ or at variable timing. The latter has been theoretically shown to reach 99.99\% fidelity with under 10~ns gates \cite{liebermann2016optimal}. Voltage multiplier amplification minimizes on-chip dispersion and permits narrower output pulses; enabling more ideal digital qubit dynamics. Finally, these devices permit signal routing which eliminates dissipation of the drive signal in cryogenic attenuators, which is a major consideration for the present JPG device and would strongly limit scaling of these prototype devices.

\section{\label{sec:conclusion}Conclusion}

In this article we have demonstrated, for the first time, successful digital control of a transmon qubit at 0.01~K using a superconducting Josephson pulse generator located at 3~K. Through dual characterization of the system, using both traditional semiconductor control electronics \cite{krantz2019quantum} and the 3~K JPG, we see no reduction in intrinsic qubit performance. Specifically we measured no negative impact to $T_1$, $T_2^*$, or $P_{th}$ -- indicative that quasiparticle propagation is effectively broken by locating the JJ elements and quantum circuits on separate temperature stages. Additionally, we measure an average JPG gate error of $r = 2.1 \times 10^{-2}$ which, considering the improvements of the future JJ devices discussed in the previous section, are expected to reach the simulated coherence-dominated infidelity of $r_{min} = 2 \times 10^{-3}$.

These results enable scaled quantum information systems which leverage the merits of Josephson-based sources for qubit control: signal stability, reproducibility, SFQ pulse self-calibration, small device size, and low power dissipation. Straightforward alterations in the qubit and JPG architectures enable factors of $\geq 10$ reduction in dissipation and size, and future devices are expected to bring JJ-based digital gates into competition with contemporary TSCE gates. Such improvements further increase the potential value of 3~K JJ-based qubit control, as current device sizes and dissipation are commensurate with operation of over $\sim$500k devices. Integration with cryoCMOS devices \cite{bardin2019design, vanDijk2020scalable, pauka2021cryogenic} is also possible; potentially yielding a hybrid cryogenic controller exploiting the advantages of both technologies.

\begin{acknowledgements}
    We acknowledge the NIST Advanced Microwave Photonics Group for the qubit, JPA, and readout microwave components. We thank N. Flowers-Jacobs for readout microwave components. We thank F. Lecocq, R. McDermott, and B Plourde for fruitful discussions, and A. Babenko and A. Fox for design and fabrication of initial devices. L. Howe was funded through the NRC Postdoctoral Research Associateship Program.
\end{acknowledgements}


\newpage
\bibliography{references}

\begin{thebibliography}{64}%
\makeatletter
\providecommand \@ifxundefined [1]{%
 \@ifx{#1\undefined}
}%
\providecommand \@ifnum [1]{%
 \ifnum #1\expandafter \@firstoftwo
 \else \expandafter \@secondoftwo
 \fi
}%
\providecommand \@ifx [1]{%
 \ifx #1\expandafter \@firstoftwo
 \else \expandafter \@secondoftwo
 \fi
}%
\providecommand \natexlab [1]{#1}%
\providecommand \enquote  [1]{``#1''}%
\providecommand \bibnamefont  [1]{#1}%
\providecommand \bibfnamefont [1]{#1}%
\providecommand \citenamefont [1]{#1}%
\providecommand \href@noop [0]{\@secondoftwo}%
\providecommand \href [0]{\begingroup \@sanitize@url \@href}%
\providecommand \@href[1]{\@@startlink{#1}\@@href}%
\providecommand \@@href[1]{\endgroup#1\@@endlink}%
\providecommand \@sanitize@url [0]{\catcode `\\12\catcode `\$12\catcode
  `\&12\catcode `\#12\catcode `\^12\catcode `\_12\catcode `\%12\relax}%
\providecommand \@@startlink[1]{}%
\providecommand \@@endlink[0]{}%
\providecommand \url  [0]{\begingroup\@sanitize@url \@url }%
\providecommand \@url [1]{\endgroup\@href {#1}{\urlprefix }}%
\providecommand \urlprefix  [0]{URL }%
\providecommand \Eprint [0]{\href }%
\providecommand \doibase [0]{https://doi.org/}%
\providecommand \selectlanguage [0]{\@gobble}%
\providecommand \bibinfo  [0]{\@secondoftwo}%
\providecommand \bibfield  [0]{\@secondoftwo}%
\providecommand \translation [1]{[#1]}%
\providecommand \BibitemOpen [0]{}%
\providecommand \bibitemStop [0]{}%
\providecommand \bibitemNoStop [0]{.\EOS\space}%
\providecommand \EOS [0]{\spacefactor3000\relax}%
\providecommand \BibitemShut  [1]{\csname bibitem#1\endcsname}%
\let\auto@bib@innerbib\@empty
\bibitem [{\citenamefont {Fowler}\ \emph {et~al.}(2012)\citenamefont {Fowler},
  \citenamefont {Mariantoni}, \citenamefont {Martinis},\ and\ \citenamefont
  {Cleland}}]{fowler2012surface}%
  \BibitemOpen
  \bibfield  {author} {\bibinfo {author} {\bibfnamefont {A.~G.}\ \bibnamefont
  {Fowler}}, \bibinfo {author} {\bibfnamefont {M.}~\bibnamefont {Mariantoni}},
  \bibinfo {author} {\bibfnamefont {J.~M.}\ \bibnamefont {Martinis}},\ and\
  \bibinfo {author} {\bibfnamefont {A.~N.}\ \bibnamefont {Cleland}},\
  }\bibfield  {title} {\bibinfo {title} {\textit{Surface codes: Towards
  practical large-scale quantum computation}},\ }\href
  {https://doi.org/10.1103/PhysRevA.86.032324} {\bibfield  {journal} {\bibinfo
  {journal} {Phys. Rev. A}\ }\textbf {\bibinfo {volume} {86}},\ \bibinfo
  {pages} {032324} (\bibinfo {year} {2012})}\BibitemShut {NoStop}%
\bibitem [{\citenamefont {Kelly}\ \emph {et~al.}(2015)\citenamefont {Kelly},
  \citenamefont {Barends}, \citenamefont {Fowler}, \citenamefont {Megrant},
  \citenamefont {Jeffrey}, \citenamefont {White}, \citenamefont {Sank},
  \citenamefont {Mutus}, \citenamefont {Campbell}, \citenamefont {Chen} \emph
  {et~al.}}]{kelly2015state}%
  \BibitemOpen
  \bibfield  {author} {\bibinfo {author} {\bibfnamefont {J.}~\bibnamefont
  {Kelly}}, \bibinfo {author} {\bibfnamefont {R.}~\bibnamefont {Barends}},
  \bibinfo {author} {\bibfnamefont {A.~G.}\ \bibnamefont {Fowler}}, \bibinfo
  {author} {\bibfnamefont {A.}~\bibnamefont {Megrant}}, \bibinfo {author}
  {\bibfnamefont {E.}~\bibnamefont {Jeffrey}}, \bibinfo {author} {\bibfnamefont
  {T.~C.}\ \bibnamefont {White}}, \bibinfo {author} {\bibfnamefont
  {D.}~\bibnamefont {Sank}}, \bibinfo {author} {\bibfnamefont {J.~Y.}\
  \bibnamefont {Mutus}}, \bibinfo {author} {\bibfnamefont {B.}~\bibnamefont
  {Campbell}}, \bibinfo {author} {\bibfnamefont {Y.}~\bibnamefont {Chen}},
  \emph {et~al.},\ }\bibfield  {title} {\bibinfo {title} {\textit{State
  preservation by repetitive error detection in a superconducting quantum
  circuit}},\ }\href {https://www.nature.com/articles/nature14270} {\bibfield
  {journal} {\bibinfo  {journal} {Nature}\ }\textbf {\bibinfo {volume} {519}},\
  \bibinfo {pages} {66} (\bibinfo {year} {2015})}\BibitemShut {NoStop}%
\bibitem [{\citenamefont {Andersen}\ \emph {et~al.}(2020)\citenamefont
  {Andersen}, \citenamefont {Remm}, \citenamefont {Lazar}, \citenamefont
  {Krinner}, \citenamefont {Lacroix}, \citenamefont {Norris}, \citenamefont
  {Gabureac}, \citenamefont {Eichler},\ and\ \citenamefont
  {Wallraff}}]{andersen2020repeated}%
  \BibitemOpen
  \bibfield  {author} {\bibinfo {author} {\bibfnamefont {C.~K.}\ \bibnamefont
  {Andersen}}, \bibinfo {author} {\bibfnamefont {A.}~\bibnamefont {Remm}},
  \bibinfo {author} {\bibfnamefont {S.}~\bibnamefont {Lazar}}, \bibinfo
  {author} {\bibfnamefont {S.}~\bibnamefont {Krinner}}, \bibinfo {author}
  {\bibfnamefont {N.}~\bibnamefont {Lacroix}}, \bibinfo {author} {\bibfnamefont
  {G.~J.}\ \bibnamefont {Norris}}, \bibinfo {author} {\bibfnamefont
  {M.}~\bibnamefont {Gabureac}}, \bibinfo {author} {\bibfnamefont
  {C.}~\bibnamefont {Eichler}},\ and\ \bibinfo {author} {\bibfnamefont
  {A.}~\bibnamefont {Wallraff}},\ }\bibfield  {title} {\bibinfo {title}
  {\textit{Repeated quantum error detection in a surface code}},\ }\href
  {https://www.nature.com/articles/s41567-020-0920-y} {\bibfield  {journal}
  {\bibinfo  {journal} {Nature Physics}\ }\textbf {\bibinfo {volume} {16}},\
  \bibinfo {pages} {875} (\bibinfo {year} {2020})}\BibitemShut {NoStop}%
\bibitem [{\citenamefont {{Google Quantum AI}}(2021)}]{ai2021exponential}%
  \BibitemOpen
  \bibfield  {author} {\bibinfo {author} {\bibnamefont {{Google Quantum AI}}},\
  }\bibfield  {title} {\bibinfo {title} {\textit{Exponential suppression of bit
  or phase errors with cyclic error correction}},\ }\href
  {https://www.nature.com/articles/s41586-021-03588-y} {\bibfield  {journal}
  {\bibinfo  {journal} {Nature}\ }\textbf {\bibinfo {volume} {595}},\ \bibinfo
  {pages} {383} (\bibinfo {year} {2021})}\BibitemShut {NoStop}%
\bibitem [{\citenamefont {Krantz}\ \emph {et~al.}(2019)\citenamefont {Krantz},
  \citenamefont {Kjaergaard}, \citenamefont {Yan}, \citenamefont {Orlando},
  \citenamefont {Gustavsson},\ and\ \citenamefont
  {Oliver}}]{krantz2019quantum}%
  \BibitemOpen
  \bibfield  {author} {\bibinfo {author} {\bibfnamefont {P.}~\bibnamefont
  {Krantz}}, \bibinfo {author} {\bibfnamefont {M.}~\bibnamefont {Kjaergaard}},
  \bibinfo {author} {\bibfnamefont {F.}~\bibnamefont {Yan}}, \bibinfo {author}
  {\bibfnamefont {T.~P.}\ \bibnamefont {Orlando}}, \bibinfo {author}
  {\bibfnamefont {S.}~\bibnamefont {Gustavsson}},\ and\ \bibinfo {author}
  {\bibfnamefont {W.~D.}\ \bibnamefont {Oliver}},\ }\bibfield  {title}
  {\bibinfo {title} {\textit{A quantum engineer's guide to superconducting
  qubits}},\ }\href {https://aip.scitation.org/doi/10.1063/1.5089550}
  {\bibfield  {journal} {\bibinfo  {journal} {Applied Physics Reviews}\
  }\textbf {\bibinfo {volume} {6}},\ \bibinfo {pages} {021318} (\bibinfo {year}
  {2019})}\BibitemShut {NoStop}%
\bibitem [{\citenamefont {Krinner}\ \emph {et~al.}(2019)\citenamefont
  {Krinner}, \citenamefont {Storz}, \citenamefont {Kurpiers}, \citenamefont
  {Magnard}, \citenamefont {Heinsoo}, \citenamefont {Keller}, \citenamefont
  {Luetolf}, \citenamefont {Eichler},\ and\ \citenamefont
  {Wallraff}}]{krinner2019engineering}%
  \BibitemOpen
  \bibfield  {author} {\bibinfo {author} {\bibfnamefont {S.}~\bibnamefont
  {Krinner}}, \bibinfo {author} {\bibfnamefont {S.}~\bibnamefont {Storz}},
  \bibinfo {author} {\bibfnamefont {P.}~\bibnamefont {Kurpiers}}, \bibinfo
  {author} {\bibfnamefont {P.}~\bibnamefont {Magnard}}, \bibinfo {author}
  {\bibfnamefont {J.}~\bibnamefont {Heinsoo}}, \bibinfo {author} {\bibfnamefont
  {R.}~\bibnamefont {Keller}}, \bibinfo {author} {\bibfnamefont
  {J.}~\bibnamefont {Luetolf}}, \bibinfo {author} {\bibfnamefont
  {C.}~\bibnamefont {Eichler}},\ and\ \bibinfo {author} {\bibfnamefont
  {A.}~\bibnamefont {Wallraff}},\ }\bibfield  {title} {\bibinfo {title}
  {\textit{Engineering cryogenic setups for 100-qubit scale superconducting
  circuit systems}},\ }\href
  {https://epjquantumtechnology.springeropen.com/articles/10.1140/epjqt/s40507-019-0072-0}
  {\bibfield  {journal} {\bibinfo  {journal} {EPJ Quantum Technology}\ }\textbf
  {\bibinfo {volume} {6}},\ \bibinfo {pages} {2} (\bibinfo {year}
  {2019})}\BibitemShut {NoStop}%
\bibitem [{\citenamefont {van Dijk}\ \emph {et~al.}(2019)\citenamefont {van
  Dijk}, \citenamefont {Kawakami}, \citenamefont {Schouten}, \citenamefont
  {Veldhorst}, \citenamefont {Vandersypen}, \citenamefont {Babaie},
  \citenamefont {Charbon},\ and\ \citenamefont
  {Sebastiano}}]{vanDijk2019impact}%
  \BibitemOpen
  \bibfield  {author} {\bibinfo {author} {\bibfnamefont {J.}~\bibnamefont {van
  Dijk}}, \bibinfo {author} {\bibfnamefont {E.}~\bibnamefont {Kawakami}},
  \bibinfo {author} {\bibfnamefont {R.}~\bibnamefont {Schouten}}, \bibinfo
  {author} {\bibfnamefont {M.}~\bibnamefont {Veldhorst}}, \bibinfo {author}
  {\bibfnamefont {L.}~\bibnamefont {Vandersypen}}, \bibinfo {author}
  {\bibfnamefont {M.}~\bibnamefont {Babaie}}, \bibinfo {author} {\bibfnamefont
  {E.}~\bibnamefont {Charbon}},\ and\ \bibinfo {author} {\bibfnamefont
  {F.}~\bibnamefont {Sebastiano}},\ }\bibfield  {title} {\bibinfo {title}
  {\textit{Impact of Classical Control Electronics on Qubit Fidelity}},\ }\href
  {https://doi.org/10.1103/PhysRevApplied.12.044054} {\bibfield  {journal}
  {\bibinfo  {journal} {Phys. Rev. Applied}\ }\textbf {\bibinfo {volume}
  {12}},\ \bibinfo {pages} {044054} (\bibinfo {year} {2019})}\BibitemShut
  {NoStop}%
\bibitem [{\citenamefont {{Van Dijk}}\ \emph {et~al.}(2020)\citenamefont {{Van
  Dijk}}, \citenamefont {{Patra}}, \citenamefont {{Subramanian}}, \citenamefont
  {{Xue}}, \citenamefont {{Samkharadze}}, \citenamefont {{Corna}},
  \citenamefont {{Jeon}}, \citenamefont {{Sheikh}}, \citenamefont
  {{Juarez-Hernandez}}, \citenamefont {{Esparza}}, \citenamefont
  {{Rampurawala}}, \citenamefont {{Carlton}}, \citenamefont {{Ravikumar}},
  \citenamefont {{Nieva}}, \citenamefont {{Kim}} \emph
  {et~al.}}]{vanDijk2020scalable}%
  \BibitemOpen
  \bibfield  {author} {\bibinfo {author} {\bibfnamefont {J.~P.~G.}\
  \bibnamefont {{Van Dijk}}}, \bibinfo {author} {\bibfnamefont
  {B.}~\bibnamefont {{Patra}}}, \bibinfo {author} {\bibfnamefont
  {S.}~\bibnamefont {{Subramanian}}}, \bibinfo {author} {\bibfnamefont
  {X.}~\bibnamefont {{Xue}}}, \bibinfo {author} {\bibfnamefont
  {N.}~\bibnamefont {{Samkharadze}}}, \bibinfo {author} {\bibfnamefont
  {A.}~\bibnamefont {{Corna}}}, \bibinfo {author} {\bibfnamefont
  {C.}~\bibnamefont {{Jeon}}}, \bibinfo {author} {\bibfnamefont
  {F.}~\bibnamefont {{Sheikh}}}, \bibinfo {author} {\bibfnamefont
  {E.}~\bibnamefont {{Juarez-Hernandez}}}, \bibinfo {author} {\bibfnamefont
  {B.~P.}\ \bibnamefont {{Esparza}}}, \bibinfo {author} {\bibfnamefont
  {H.}~\bibnamefont {{Rampurawala}}}, \bibinfo {author} {\bibfnamefont {B.~R.}\
  \bibnamefont {{Carlton}}}, \bibinfo {author} {\bibfnamefont {S.}~\bibnamefont
  {{Ravikumar}}}, \bibinfo {author} {\bibfnamefont {C.}~\bibnamefont
  {{Nieva}}}, \bibinfo {author} {\bibfnamefont {S.}~\bibnamefont {{Kim}}},
  \emph {et~al.},\ }\bibfield  {title} {\bibinfo {title} {\textit{A Scalable
  Cryo-CMOS Controller for the Wideband Frequency-Multiplexed Control of Spin
  Qubits and Transmons}},\ }\href {https://doi.org/10.1109/JSSC.2020.3024678}
  {\bibfield  {journal} {\bibinfo  {journal} {IEEE Journal of Solid-State
  Circuits}\ }\textbf {\bibinfo {volume} {55}},\ \bibinfo {pages} {2930}
  (\bibinfo {year} {2020})}\BibitemShut {NoStop}%
\bibitem [{\citenamefont {{Bardin}}\ \emph {et~al.}(2019)\citenamefont
  {{Bardin}}, \citenamefont {{Jeffrey}}, \citenamefont {{Lucero}},
  \citenamefont {{Huang}}, \citenamefont {{Das}}, \citenamefont {{Sank}},
  \citenamefont {{Naaman}}, \citenamefont {{Megrant}}, \citenamefont
  {{Barends}}, \citenamefont {{White}}, \citenamefont {{Giustina}},
  \citenamefont {{Satzinger}}, \citenamefont {{Arya}}, \citenamefont
  {{Roushan}}, \citenamefont {{Chiaro}} \emph {et~al.}}]{bardin2019design}%
  \BibitemOpen
  \bibfield  {author} {\bibinfo {author} {\bibfnamefont {J.~C.}\ \bibnamefont
  {{Bardin}}}, \bibinfo {author} {\bibfnamefont {E.}~\bibnamefont {{Jeffrey}}},
  \bibinfo {author} {\bibfnamefont {E.}~\bibnamefont {{Lucero}}}, \bibinfo
  {author} {\bibfnamefont {T.}~\bibnamefont {{Huang}}}, \bibinfo {author}
  {\bibfnamefont {S.}~\bibnamefont {{Das}}}, \bibinfo {author} {\bibfnamefont
  {D.~T.}\ \bibnamefont {{Sank}}}, \bibinfo {author} {\bibfnamefont
  {O.}~\bibnamefont {{Naaman}}}, \bibinfo {author} {\bibfnamefont {A.~E.}\
  \bibnamefont {{Megrant}}}, \bibinfo {author} {\bibfnamefont {R.}~\bibnamefont
  {{Barends}}}, \bibinfo {author} {\bibfnamefont {T.}~\bibnamefont {{White}}},
  \bibinfo {author} {\bibfnamefont {M.}~\bibnamefont {{Giustina}}}, \bibinfo
  {author} {\bibfnamefont {K.~J.}\ \bibnamefont {{Satzinger}}}, \bibinfo
  {author} {\bibfnamefont {K.}~\bibnamefont {{Arya}}}, \bibinfo {author}
  {\bibfnamefont {P.}~\bibnamefont {{Roushan}}}, \bibinfo {author}
  {\bibfnamefont {B.}~\bibnamefont {{Chiaro}}}, \emph {et~al.},\ }\bibfield
  {title} {\bibinfo {title} {\textit{Design and Characterization of a 28-nm
  Bulk-CMOS Cryogenic Quantum Controller Dissipating Less Than 2 mW at 3 K}},\
  }\href {https://doi.org/10.1109/JSSC.2019.2937234} {\bibfield  {journal}
  {\bibinfo  {journal} {IEEE Journal of Solid-State Circuits}\ }\textbf
  {\bibinfo {volume} {54}},\ \bibinfo {pages} {3043} (\bibinfo {year}
  {2019})}\BibitemShut {NoStop}%
\bibitem [{\citenamefont {Pauka}\ \emph {et~al.}(2021)\citenamefont {Pauka},
  \citenamefont {Das}, \citenamefont {Kalra}, \citenamefont {Moini},
  \citenamefont {Yang}, \citenamefont {Trainer}, \citenamefont {Bousquet},
  \citenamefont {Cantaloube}, \citenamefont {Dick}, \citenamefont {Gardner}
  \emph {et~al.}}]{pauka2021cryogenic}%
  \BibitemOpen
  \bibfield  {author} {\bibinfo {author} {\bibfnamefont {S.}~\bibnamefont
  {Pauka}}, \bibinfo {author} {\bibfnamefont {K.}~\bibnamefont {Das}}, \bibinfo
  {author} {\bibfnamefont {R.}~\bibnamefont {Kalra}}, \bibinfo {author}
  {\bibfnamefont {A.}~\bibnamefont {Moini}}, \bibinfo {author} {\bibfnamefont
  {Y.}~\bibnamefont {Yang}}, \bibinfo {author} {\bibfnamefont {M.}~\bibnamefont
  {Trainer}}, \bibinfo {author} {\bibfnamefont {A.}~\bibnamefont {Bousquet}},
  \bibinfo {author} {\bibfnamefont {C.}~\bibnamefont {Cantaloube}}, \bibinfo
  {author} {\bibfnamefont {N.}~\bibnamefont {Dick}}, \bibinfo {author}
  {\bibfnamefont {G.}~\bibnamefont {Gardner}}, \emph {et~al.},\ }\bibfield
  {title} {\bibinfo {title} {\textit{A cryogenic CMOS chip for generating
  control signals for multiple qubits}},\ }\href
  {https://www.nature.com/articles/s41928-020-00528-y} {\bibfield  {journal}
  {\bibinfo  {journal} {Nature Electronics}\ }\textbf {\bibinfo {volume} {4}},\
  \bibinfo {pages} {64} (\bibinfo {year} {2021})}\BibitemShut {NoStop}%
\bibitem [{\citenamefont {Sirois}\ \emph {et~al.}(2020)\citenamefont {Sirois},
  \citenamefont {Castellanos-Beltran}, \citenamefont {Fox}, \citenamefont
  {Benz},\ and\ \citenamefont {Hopkins}}]{sirois2020josephson}%
  \BibitemOpen
  \bibfield  {author} {\bibinfo {author} {\bibfnamefont {A.~J.}\ \bibnamefont
  {Sirois}}, \bibinfo {author} {\bibfnamefont {M.}~\bibnamefont
  {Castellanos-Beltran}}, \bibinfo {author} {\bibfnamefont {A.~E.}\
  \bibnamefont {Fox}}, \bibinfo {author} {\bibfnamefont {S.~P.}\ \bibnamefont
  {Benz}},\ and\ \bibinfo {author} {\bibfnamefont {P.~F.}\ \bibnamefont
  {Hopkins}},\ }\bibfield  {title} {\bibinfo {title} {\textit{Josephson
  Microwave Sources Applied to Quantum Information Systems}},\ }\href
  {https://doi.org/10.1109/TQE.2020.3045682} {\bibfield  {journal} {\bibinfo
  {journal} {IEEE Transactions on Quantum Engineering}\ }\textbf {\bibinfo
  {volume} {1}},\ \bibinfo {pages} {1} (\bibinfo {year} {2020})}\BibitemShut
  {NoStop}%
\bibitem [{\citenamefont {Ball}\ \emph {et~al.}(2016)\citenamefont {Ball},
  \citenamefont {Oliver},\ and\ \citenamefont {Biercuk}}]{ball2016role}%
  \BibitemOpen
  \bibfield  {author} {\bibinfo {author} {\bibfnamefont {H.}~\bibnamefont
  {Ball}}, \bibinfo {author} {\bibfnamefont {W.~D.}\ \bibnamefont {Oliver}},\
  and\ \bibinfo {author} {\bibfnamefont {M.~J.}\ \bibnamefont {Biercuk}},\
  }\bibfield  {title} {\bibinfo {title} {\textit{The role of master clock
  stability in quantum information processing}},\ }\href
  {https://www.nature.com/articles/npjqi201633} {\bibfield  {journal} {\bibinfo
   {journal} {npj Quantum Information}\ }\textbf {\bibinfo {volume} {2}},\
  \bibinfo {pages} {1} (\bibinfo {year} {2016})}\BibitemShut {NoStop}%
\bibitem [{\citenamefont {Rüfenacht}\ \emph {et~al.}(2015)\citenamefont
  {Rüfenacht}, \citenamefont {Howe}, \citenamefont {Fox}, \citenamefont
  {Schwall}, \citenamefont {Dresselhaus}, \citenamefont {Burroughs},
  \citenamefont {Benz},\ and\ \citenamefont {Benz}}]{rufenacht2015cryocooled}%
  \BibitemOpen
  \bibfield  {author} {\bibinfo {author} {\bibfnamefont {A.}~\bibnamefont
  {Rüfenacht}}, \bibinfo {author} {\bibfnamefont {L.~A.}\ \bibnamefont
  {Howe}}, \bibinfo {author} {\bibfnamefont {A.~E.}\ \bibnamefont {Fox}},
  \bibinfo {author} {\bibfnamefont {R.~E.}\ \bibnamefont {Schwall}}, \bibinfo
  {author} {\bibfnamefont {P.~D.}\ \bibnamefont {Dresselhaus}}, \bibinfo
  {author} {\bibfnamefont {C.~J.}\ \bibnamefont {Burroughs}}, \bibinfo {author}
  {\bibfnamefont {S.~P.}\ \bibnamefont {Benz}},\ and\ \bibinfo {author}
  {\bibfnamefont {S.~P.}\ \bibnamefont {Benz}},\ }\bibfield  {title} {\bibinfo
  {title} {\textit{Cryocooled 10 V Programmable Josephson Voltage Standard}},\
  }\href {https://doi.org/10.1109/TIM.2014.2374697} {\bibfield  {journal}
  {\bibinfo  {journal} {IEEE Transactions on Instrumentation and Measurement}\
  }\textbf {\bibinfo {volume} {64}},\ \bibinfo {pages} {1477} (\bibinfo {year}
  {2015})}\BibitemShut {NoStop}%
\bibitem [{\citenamefont {Burroughs}\ \emph {et~al.}(2011)\citenamefont
  {Burroughs}, \citenamefont {Dresselhaus}, \citenamefont {Rufenacht},
  \citenamefont {Olaya}, \citenamefont {Elsbury}, \citenamefont {Tang},\ and\
  \citenamefont {Benz}}]{burroughs2011nist}%
  \BibitemOpen
  \bibfield  {author} {\bibinfo {author} {\bibfnamefont {C.~J.}\ \bibnamefont
  {Burroughs}}, \bibinfo {author} {\bibfnamefont {P.~D.}\ \bibnamefont
  {Dresselhaus}}, \bibinfo {author} {\bibfnamefont {A.}~\bibnamefont
  {Rufenacht}}, \bibinfo {author} {\bibfnamefont {D.}~\bibnamefont {Olaya}},
  \bibinfo {author} {\bibfnamefont {M.~M.}\ \bibnamefont {Elsbury}}, \bibinfo
  {author} {\bibfnamefont {Y.-H.}\ \bibnamefont {Tang}},\ and\ \bibinfo
  {author} {\bibfnamefont {S.~P.}\ \bibnamefont {Benz}},\ }\bibfield  {title}
  {\bibinfo {title} {\textit{NIST 10 V Programmable Josephson Voltage Standard
  System}},\ }\href {https://doi.org/10.1109/TIM.2010.2101191} {\bibfield
  {journal} {\bibinfo  {journal} {IEEE Transactions on Instrumentation and
  Measurement}\ }\textbf {\bibinfo {volume} {60}},\ \bibinfo {pages} {2482}
  (\bibinfo {year} {2011})}\BibitemShut {NoStop}%
\bibitem [{\citenamefont {Brevik}\ \emph {et~al.}(2018)\citenamefont {Brevik},
  \citenamefont {Donnelly}, \citenamefont {Flowers-Jacobs}, \citenamefont
  {Fox}, \citenamefont {Hopkins}, \citenamefont {Dresselhaus},\ and\
  \citenamefont {Benz}}]{brevik2018radio}%
  \BibitemOpen
  \bibfield  {author} {\bibinfo {author} {\bibfnamefont {J.~A.}\ \bibnamefont
  {Brevik}}, \bibinfo {author} {\bibfnamefont {C.~A.}\ \bibnamefont
  {Donnelly}}, \bibinfo {author} {\bibfnamefont {N.~E.}\ \bibnamefont
  {Flowers-Jacobs}}, \bibinfo {author} {\bibfnamefont {A.~E.}\ \bibnamefont
  {Fox}}, \bibinfo {author} {\bibfnamefont {P.~F.}\ \bibnamefont {Hopkins}},
  \bibinfo {author} {\bibfnamefont {P.~D.}\ \bibnamefont {Dresselhaus}},\ and\
  \bibinfo {author} {\bibfnamefont {S.~P.}\ \bibnamefont {Benz}},\ }\bibfield
  {title} {\bibinfo {title} {\textit{Radio-Frequency Waveform Synthesis with
  the Josephson Arbitrary Waveform Synthesizer}},\ }\href
  {https://doi.org/10.1109/CPEM.2018.8501023} {\bibfield  {journal} {\bibinfo
  {journal} {2018 Conference on Precision Electromagnetic Measurements}\ ,\
  \bibinfo {pages} {1}} (\bibinfo {year} {2018})}\BibitemShut {NoStop}%
\bibitem [{\citenamefont {Hopkins}\ \emph {et~al.}(2019)\citenamefont
  {Hopkins}, \citenamefont {Brevik}, \citenamefont {Castellanos-Beltran},
  \citenamefont {Donnelly}, \citenamefont {Flowers-Jacobs}, \citenamefont
  {Fox}, \citenamefont {Olaya}, \citenamefont {Dresselhaus},\ and\
  \citenamefont {Benz}}]{hopkins2019rfwaveform}%
  \BibitemOpen
  \bibfield  {author} {\bibinfo {author} {\bibfnamefont {P.~F.}\ \bibnamefont
  {Hopkins}}, \bibinfo {author} {\bibfnamefont {J.~A.}\ \bibnamefont {Brevik}},
  \bibinfo {author} {\bibfnamefont {M.}~\bibnamefont {Castellanos-Beltran}},
  \bibinfo {author} {\bibfnamefont {C.~A.}\ \bibnamefont {Donnelly}}, \bibinfo
  {author} {\bibfnamefont {N.~E.}\ \bibnamefont {Flowers-Jacobs}}, \bibinfo
  {author} {\bibfnamefont {A.~E.}\ \bibnamefont {Fox}}, \bibinfo {author}
  {\bibfnamefont {D.}~\bibnamefont {Olaya}}, \bibinfo {author} {\bibfnamefont
  {P.~D.}\ \bibnamefont {Dresselhaus}},\ and\ \bibinfo {author} {\bibfnamefont
  {S.~P.}\ \bibnamefont {Benz}},\ }\bibfield  {title} {\bibinfo {title}
  {\textit{RF Waveform Synthesizers With Quantum-Based Voltage Accuracy for
  Communications Metrology}},\ }\href
  {https://doi.org/10.1109/TASC.2019.2898407} {\bibfield  {journal} {\bibinfo
  {journal} {IEEE Transactions on Applied Superconductivity}\ }\textbf
  {\bibinfo {volume} {29}},\ \bibinfo {pages} {1} (\bibinfo {year}
  {2019})}\BibitemShut {NoStop}%
\bibitem [{\citenamefont {R{\"u}fenacht}\ \emph {et~al.}(2018)\citenamefont
  {R{\"u}fenacht}, \citenamefont {Flowers-Jacobs},\ and\ \citenamefont
  {Benz}}]{rufenacht2018impact}%
  \BibitemOpen
  \bibfield  {author} {\bibinfo {author} {\bibfnamefont {A.}~\bibnamefont
  {R{\"u}fenacht}}, \bibinfo {author} {\bibfnamefont {N.~E.}\ \bibnamefont
  {Flowers-Jacobs}},\ and\ \bibinfo {author} {\bibfnamefont {S.~P.}\
  \bibnamefont {Benz}},\ }\bibfield  {title} {\bibinfo {title} {\textit{Impact
  of the latest generation of Josephson voltage standards in ac and dc electric
  metrology}},\ }\href
  {https://iopscience.iop.org/article/10.1088/1681-7575/aad41a} {\bibfield
  {journal} {\bibinfo  {journal} {Metrologia}\ }\textbf {\bibinfo {volume}
  {55}},\ \bibinfo {pages} {S152} (\bibinfo {year} {2018})}\BibitemShut
  {NoStop}%
\bibitem [{\citenamefont {McDermott}\ and\ \citenamefont
  {Vavilov}(2014)}]{mcdermott2014accurate}%
  \BibitemOpen
  \bibfield  {author} {\bibinfo {author} {\bibfnamefont {R.}~\bibnamefont
  {McDermott}}\ and\ \bibinfo {author} {\bibfnamefont {M.~G.}\ \bibnamefont
  {Vavilov}},\ }\bibfield  {title} {\bibinfo {title} {\textit{Accurate Qubit
  Control with Single Flux Quantum Pulses}},\ }\href
  {https://doi.org/10.1103/PhysRevApplied.2.014007} {\bibfield  {journal}
  {\bibinfo  {journal} {Phys. Rev. Applied}\ }\textbf {\bibinfo {volume} {2}},\
  \bibinfo {pages} {014007} (\bibinfo {year} {2014})}\BibitemShut {NoStop}%
\bibitem [{\citenamefont {Liebermann}\ and\ \citenamefont
  {Wilhelm}(2016)}]{liebermann2016optimal}%
  \BibitemOpen
  \bibfield  {author} {\bibinfo {author} {\bibfnamefont {P.~J.}\ \bibnamefont
  {Liebermann}}\ and\ \bibinfo {author} {\bibfnamefont {F.~K.}\ \bibnamefont
  {Wilhelm}},\ }\bibfield  {title} {\bibinfo {title} {\textit{Optimal Qubit
  Control Using Single-Flux Quantum Pulses}},\ }\href
  {https://doi.org/10.1103/PhysRevApplied.6.024022} {\bibfield  {journal}
  {\bibinfo  {journal} {Phys. Rev. Applied}\ }\textbf {\bibinfo {volume} {6}},\
  \bibinfo {pages} {024022} (\bibinfo {year} {2016})}\BibitemShut {NoStop}%
\bibitem [{\citenamefont {McDermott}\ \emph {et~al.}(2018)\citenamefont
  {McDermott}, \citenamefont {Vavilov}, \citenamefont {Plourde}, \citenamefont
  {Wilhelm}, \citenamefont {Liebermann}, \citenamefont {Mukhanov},\ and\
  \citenamefont {Ohki}}]{mcdermott2018quantum}%
  \BibitemOpen
  \bibfield  {author} {\bibinfo {author} {\bibfnamefont {R.}~\bibnamefont
  {McDermott}}, \bibinfo {author} {\bibfnamefont {M.~G.}\ \bibnamefont
  {Vavilov}}, \bibinfo {author} {\bibfnamefont {B.~L.~T.}\ \bibnamefont
  {Plourde}}, \bibinfo {author} {\bibfnamefont {F.~K.}\ \bibnamefont
  {Wilhelm}}, \bibinfo {author} {\bibfnamefont {P.~J.}\ \bibnamefont
  {Liebermann}}, \bibinfo {author} {\bibfnamefont {O.~A.}\ \bibnamefont
  {Mukhanov}},\ and\ \bibinfo {author} {\bibfnamefont {T.~A.}\ \bibnamefont
  {Ohki}},\ }\bibfield  {title} {\bibinfo {title}
  {\textit{Quantum{\textendash}classical interface based on single flux quantum
  digital logic}},\ }\href {https://doi.org/10.1088/2058-9565/aaa3a0}
  {\bibfield  {journal} {\bibinfo  {journal} {Quantum Science and Technology}\
  }\textbf {\bibinfo {volume} {3}},\ \bibinfo {pages} {024004} (\bibinfo {year}
  {2018})}\BibitemShut {NoStop}%
\bibitem [{\citenamefont {Leonard}\ \emph {et~al.}(2019)\citenamefont
  {Leonard}, \citenamefont {Beck}, \citenamefont {Nelson}, \citenamefont
  {Christensen}, \citenamefont {Thorbeck}, \citenamefont {Howington},
  \citenamefont {Opremcak}, \citenamefont {Pechenezhskiy}, \citenamefont
  {Dodge}, \citenamefont {Dupuis}, \citenamefont {Hutchings}, \citenamefont
  {Ku}, \citenamefont {Schlenker}, \citenamefont {Suttle}, \citenamefont
  {Wilen} \emph {et~al.}}]{leonard2019digital}%
  \BibitemOpen
  \bibfield  {author} {\bibinfo {author} {\bibfnamefont {E.}~\bibnamefont
  {Leonard}}, \bibinfo {author} {\bibfnamefont {M.~A.}\ \bibnamefont {Beck}},
  \bibinfo {author} {\bibfnamefont {J.}~\bibnamefont {Nelson}}, \bibinfo
  {author} {\bibfnamefont {B.}~\bibnamefont {Christensen}}, \bibinfo {author}
  {\bibfnamefont {T.}~\bibnamefont {Thorbeck}}, \bibinfo {author}
  {\bibfnamefont {C.}~\bibnamefont {Howington}}, \bibinfo {author}
  {\bibfnamefont {A.}~\bibnamefont {Opremcak}}, \bibinfo {author}
  {\bibfnamefont {I.}~\bibnamefont {Pechenezhskiy}}, \bibinfo {author}
  {\bibfnamefont {K.}~\bibnamefont {Dodge}}, \bibinfo {author} {\bibfnamefont
  {N.}~\bibnamefont {Dupuis}}, \bibinfo {author} {\bibfnamefont
  {M.}~\bibnamefont {Hutchings}}, \bibinfo {author} {\bibfnamefont
  {J.}~\bibnamefont {Ku}}, \bibinfo {author} {\bibfnamefont {F.}~\bibnamefont
  {Schlenker}}, \bibinfo {author} {\bibfnamefont {J.}~\bibnamefont {Suttle}},
  \bibinfo {author} {\bibfnamefont {C.}~\bibnamefont {Wilen}}, \emph {et~al.},\
  }\bibfield  {title} {\bibinfo {title} {\textit{Digital Coherent Control of a
  Superconducting Qubit}},\ }\href
  {https://doi.org/10.1103/PhysRevApplied.11.014009} {\bibfield  {journal}
  {\bibinfo  {journal} {Phys. Rev. Applied}\ }\textbf {\bibinfo {volume}
  {11}},\ \bibinfo {pages} {014009} (\bibinfo {year} {2019})}\BibitemShut
  {NoStop}%
\bibitem [{\citenamefont {Patel}\ \emph {et~al.}(2017)\citenamefont {Patel},
  \citenamefont {Pechenezhskiy}, \citenamefont {Plourde}, \citenamefont
  {Vavilov},\ and\ \citenamefont {McDermott}}]{patel2017phonon}%
  \BibitemOpen
  \bibfield  {author} {\bibinfo {author} {\bibfnamefont {U.}~\bibnamefont
  {Patel}}, \bibinfo {author} {\bibfnamefont {I.~V.}\ \bibnamefont
  {Pechenezhskiy}}, \bibinfo {author} {\bibfnamefont {B.~L.~T.}\ \bibnamefont
  {Plourde}}, \bibinfo {author} {\bibfnamefont {M.~G.}\ \bibnamefont
  {Vavilov}},\ and\ \bibinfo {author} {\bibfnamefont {R.}~\bibnamefont
  {McDermott}},\ }\bibfield  {title} {\bibinfo {title} {\textit{Phonon-mediated
  quasiparticle poisoning of superconducting microwave resonators}},\ }\href
  {https://doi.org/10.1103/PhysRevB.96.220501} {\bibfield  {journal} {\bibinfo
  {journal} {Phys. Rev. B}\ }\textbf {\bibinfo {volume} {96}},\ \bibinfo
  {pages} {220501} (\bibinfo {year} {2017})}\BibitemShut {NoStop}%
\bibitem [{\citenamefont {Martinis}\ \emph {et~al.}(2009)\citenamefont
  {Martinis}, \citenamefont {Ansmann},\ and\ \citenamefont
  {Aumentado}}]{martinis2009energy}%
  \BibitemOpen
  \bibfield  {author} {\bibinfo {author} {\bibfnamefont {J.~M.}\ \bibnamefont
  {Martinis}}, \bibinfo {author} {\bibfnamefont {M.}~\bibnamefont {Ansmann}},\
  and\ \bibinfo {author} {\bibfnamefont {J.}~\bibnamefont {Aumentado}},\
  }\bibfield  {title} {\bibinfo {title} {\textit{Energy Decay in
  Superconducting Josephson-Junction Qubits from Nonequilibrium Quasiparticle
  Excitations}},\ }\href {https://doi.org/10.1103/PhysRevLett.103.097002}
  {\bibfield  {journal} {\bibinfo  {journal} {Phys. Rev. Lett.}\ }\textbf
  {\bibinfo {volume} {103}},\ \bibinfo {pages} {097002} (\bibinfo {year}
  {2009})}\BibitemShut {NoStop}%
\bibitem [{\citenamefont {Ballard}\ \emph {et~al.}(2021)\citenamefont
  {Ballard}, \citenamefont {Iaia}, \citenamefont {McBroom}, \citenamefont
  {Liu}, \citenamefont {Dodge}, \citenamefont {Ku}, \citenamefont {Liu},
  \citenamefont {Opremcak}, \citenamefont {Wilen}, \citenamefont {Leonard}
  \emph {et~al.}}]{ballard2021single}%
  \BibitemOpen
  \bibfield  {author} {\bibinfo {author} {\bibfnamefont {A.}~\bibnamefont
  {Ballard}}, \bibinfo {author} {\bibfnamefont {V.}~\bibnamefont {Iaia}},
  \bibinfo {author} {\bibfnamefont {T.}~\bibnamefont {McBroom}}, \bibinfo
  {author} {\bibfnamefont {Y.}~\bibnamefont {Liu}}, \bibinfo {author}
  {\bibfnamefont {K.}~\bibnamefont {Dodge}}, \bibinfo {author} {\bibfnamefont
  {J.}~\bibnamefont {Ku}}, \bibinfo {author} {\bibfnamefont {C.-H.}\
  \bibnamefont {Liu}}, \bibinfo {author} {\bibfnamefont {A.}~\bibnamefont
  {Opremcak}}, \bibinfo {author} {\bibfnamefont {C.}~\bibnamefont {Wilen}},
  \bibinfo {author} {\bibfnamefont {E.}~\bibnamefont {Leonard}}, \emph
  {et~al.},\ }\bibfield  {title} {\bibinfo {title} {\textit{Single Flux
  Quantum-Based Superconducting Qubit Control and Quasiparticle Mitigation:
  Part I}},\ }\href {https://meetings.aps.org/Meeting/MAR21/Session/P28.4}
  {\bibfield  {journal} {\bibinfo  {journal} {Bulletin of the American Physical
  Society}\ } (\bibinfo {year} {2021})}\BibitemShut {NoStop}%
\bibitem [{\citenamefont {Liu}\ \emph {et~al.}(2021)\citenamefont {Liu},
  \citenamefont {Opremcak}, \citenamefont {Wilen}, \citenamefont {Leonard},
  \citenamefont {Beck}, \citenamefont {Abdullah}, \citenamefont {Ballard},
  \citenamefont {Iaia}, \citenamefont {McBroom}, \citenamefont {Liu} \emph
  {et~al.}}]{liu2021single}%
  \BibitemOpen
  \bibfield  {author} {\bibinfo {author} {\bibfnamefont {C.}~\bibnamefont
  {Liu}}, \bibinfo {author} {\bibfnamefont {A.}~\bibnamefont {Opremcak}},
  \bibinfo {author} {\bibfnamefont {C.}~\bibnamefont {Wilen}}, \bibinfo
  {author} {\bibfnamefont {E.}~\bibnamefont {Leonard}}, \bibinfo {author}
  {\bibfnamefont {M.}~\bibnamefont {Beck}}, \bibinfo {author} {\bibfnamefont
  {S.}~\bibnamefont {Abdullah}}, \bibinfo {author} {\bibfnamefont
  {A.}~\bibnamefont {Ballard}}, \bibinfo {author} {\bibfnamefont
  {V.}~\bibnamefont {Iaia}}, \bibinfo {author} {\bibfnamefont {T.}~\bibnamefont
  {McBroom}}, \bibinfo {author} {\bibfnamefont {Y.}~\bibnamefont {Liu}}, \emph
  {et~al.},\ }\bibfield  {title} {\bibinfo {title} {\textit{Single Flux
  Quantum-Based Superconducting Qubit Control and Quasiparticle Mitigation:
  Part 2}},\ }\href {https://meetings.aps.org/Meeting/MAR21/Session/P28.5}
  {\bibfield  {journal} {\bibinfo  {journal} {Bulletin of the American Physical
  Society}\ } (\bibinfo {year} {2021})}\BibitemShut {NoStop}%
\bibitem [{\citenamefont {Hosseinkhani}\ \emph {et~al.}(2017)\citenamefont
  {Hosseinkhani}, \citenamefont {Riwar}, \citenamefont {Schoelkopf},
  \citenamefont {Glazman},\ and\ \citenamefont
  {Catelani}}]{hosseinkhani2017optimal}%
  \BibitemOpen
  \bibfield  {author} {\bibinfo {author} {\bibfnamefont {A.}~\bibnamefont
  {Hosseinkhani}}, \bibinfo {author} {\bibfnamefont {R.-P.}\ \bibnamefont
  {Riwar}}, \bibinfo {author} {\bibfnamefont {R.~J.}\ \bibnamefont
  {Schoelkopf}}, \bibinfo {author} {\bibfnamefont {L.~I.}\ \bibnamefont
  {Glazman}},\ and\ \bibinfo {author} {\bibfnamefont {G.}~\bibnamefont
  {Catelani}},\ }\bibfield  {title} {\bibinfo {title} {\textit{Optimal
  Configurations for Normal-Metal Traps in Transmon Qubits}},\ }\href
  {https://doi.org/10.1103/PhysRevApplied.8.064028} {\bibfield  {journal}
  {\bibinfo  {journal} {Phys. Rev. Applied}\ }\textbf {\bibinfo {volume} {8}},\
  \bibinfo {pages} {064028} (\bibinfo {year} {2017})}\BibitemShut {NoStop}%
\bibitem [{\citenamefont {Martinis}(2021)}]{martinis2021saving}%
  \BibitemOpen
  \bibfield  {author} {\bibinfo {author} {\bibfnamefont {J.~M.}\ \bibnamefont
  {Martinis}},\ }\bibfield  {title} {\bibinfo {title} {\textit{Saving
  superconducting quantum processors from decay and correlated errors generated
  by gamma and cosmic rays}},\ }\href
  {https://www.nature.com/articles/s41534-021-00431-0} {\bibfield  {journal}
  {\bibinfo  {journal} {npj Quantum Information}\ }\textbf {\bibinfo {volume}
  {7}},\ \bibinfo {pages} {1} (\bibinfo {year} {2021})}\BibitemShut {NoStop}%
\bibitem [{\citenamefont {Smith}\ \emph {et~al.}(2020)\citenamefont {Smith},
  \citenamefont {Mazin}, \citenamefont {Walter}, \citenamefont {Daal},
  \citenamefont {Bailey~III}, \citenamefont {Bockstiegel}, \citenamefont
  {Zobrist}, \citenamefont {Swimmer}, \citenamefont {Steiger},\ and\
  \citenamefont {Fruitwala}}]{smith2020flexible}%
  \BibitemOpen
  \bibfield  {author} {\bibinfo {author} {\bibfnamefont {J.~P.}\ \bibnamefont
  {Smith}}, \bibinfo {author} {\bibfnamefont {B.~A.}\ \bibnamefont {Mazin}},
  \bibinfo {author} {\bibfnamefont {A.~B.}\ \bibnamefont {Walter}}, \bibinfo
  {author} {\bibfnamefont {M.}~\bibnamefont {Daal}}, \bibinfo {author}
  {\bibfnamefont {J.}~\bibnamefont {Bailey~III}}, \bibinfo {author}
  {\bibfnamefont {C.}~\bibnamefont {Bockstiegel}}, \bibinfo {author}
  {\bibfnamefont {N.}~\bibnamefont {Zobrist}}, \bibinfo {author} {\bibfnamefont
  {N.}~\bibnamefont {Swimmer}}, \bibinfo {author} {\bibfnamefont
  {S.}~\bibnamefont {Steiger}},\ and\ \bibinfo {author} {\bibfnamefont
  {N.}~\bibnamefont {Fruitwala}},\ }\bibfield  {title} {\bibinfo {title}
  {\textit{Flexible coaxial ribbon cable for high-density superconducting
  microwave device arrays}},\ }\href
  {https://ieeexplore.ieee.org/stamp/stamp.jsp?tp=&arnumber=9138678&tag=1}
  {\bibfield  {journal} {\bibinfo  {journal} {IEEE Transactions on Applied
  Superconductivity}\ }\textbf {\bibinfo {volume} {31}},\ \bibinfo {pages} {1}
  (\bibinfo {year} {2020})}\BibitemShut {NoStop}%
\bibitem [{\citenamefont {Walter}\ \emph {et~al.}(2018)\citenamefont {Walter},
  \citenamefont {Bockstiegel}, \citenamefont {Mazin},\ and\ \citenamefont
  {Daal}}]{walter2018laminated}%
  \BibitemOpen
  \bibfield  {author} {\bibinfo {author} {\bibfnamefont {A.~B.}\ \bibnamefont
  {Walter}}, \bibinfo {author} {\bibfnamefont {C.}~\bibnamefont {Bockstiegel}},
  \bibinfo {author} {\bibfnamefont {B.~A.}\ \bibnamefont {Mazin}},\ and\
  \bibinfo {author} {\bibfnamefont {M.}~\bibnamefont {Daal}},\ }\bibfield
  {title} {\bibinfo {title} {\textit{Laminated NbTi-on-Kapton Microstrip Cables
  for Flexible Sub-Kelvin RF Electronics}},\ }\href
  {https://doi.org/10.1109/TASC.2017.2773836} {\bibfield  {journal} {\bibinfo
  {journal} {IEEE Transactions on Applied Superconductivity}\ }\textbf
  {\bibinfo {volume} {28}},\ \bibinfo {pages} {1} (\bibinfo {year}
  {2018})}\BibitemShut {NoStop}%
\bibitem [{\citenamefont {Tuckerman}\ \emph {et~al.}(2016)\citenamefont
  {Tuckerman}, \citenamefont {Hamilton}, \citenamefont {Reilly}, \citenamefont
  {Bai}, \citenamefont {Hernandez}, \citenamefont {Hornibrook}, \citenamefont
  {Sellers},\ and\ \citenamefont {Ellis}}]{tuckerman2016flexible}%
  \BibitemOpen
  \bibfield  {author} {\bibinfo {author} {\bibfnamefont {D.~B.}\ \bibnamefont
  {Tuckerman}}, \bibinfo {author} {\bibfnamefont {M.~C.}\ \bibnamefont
  {Hamilton}}, \bibinfo {author} {\bibfnamefont {D.~J.}\ \bibnamefont
  {Reilly}}, \bibinfo {author} {\bibfnamefont {R.}~\bibnamefont {Bai}},
  \bibinfo {author} {\bibfnamefont {G.~A.}\ \bibnamefont {Hernandez}}, \bibinfo
  {author} {\bibfnamefont {J.~M.}\ \bibnamefont {Hornibrook}}, \bibinfo
  {author} {\bibfnamefont {J.~A.}\ \bibnamefont {Sellers}},\ and\ \bibinfo
  {author} {\bibfnamefont {C.~D.}\ \bibnamefont {Ellis}},\ }\bibfield  {title}
  {\bibinfo {title} {\textit{Flexible superconducting Nb transmission lines on
  thin film polyimide for quantum computing applications}},\ }\href
  {https://iopscience.iop.org/article/10.1088/0953-2048/29/8/084007} {\bibfield
   {journal} {\bibinfo  {journal} {Superconductor Science and Technology}\
  }\textbf {\bibinfo {volume} {29}},\ \bibinfo {pages} {084007} (\bibinfo
  {year} {2016})}\BibitemShut {NoStop}%
\bibitem [{\citenamefont {Tinkham}(2004)}]{tinkham2004introduction}%
  \BibitemOpen
  \bibfield  {author} {\bibinfo {author} {\bibfnamefont {M.}~\bibnamefont
  {Tinkham}},\ }\href@noop {} {\emph {\bibinfo {title} {Introduction to
  superconductivity}}}\ (\bibinfo  {publisher} {Courier Corporation},\ \bibinfo
  {year} {2004})\BibitemShut {NoStop}%
\bibitem [{\citenamefont {Lecocq}\ \emph {et~al.}(2021)\citenamefont {Lecocq},
  \citenamefont {Quinlan}, \citenamefont {Cicak}, \citenamefont {Aumentado},
  \citenamefont {Diddams},\ and\ \citenamefont {Teufel}}]{lecocq2021control}%
  \BibitemOpen
  \bibfield  {author} {\bibinfo {author} {\bibfnamefont {F.}~\bibnamefont
  {Lecocq}}, \bibinfo {author} {\bibfnamefont {F.}~\bibnamefont {Quinlan}},
  \bibinfo {author} {\bibfnamefont {K.}~\bibnamefont {Cicak}}, \bibinfo
  {author} {\bibfnamefont {J.}~\bibnamefont {Aumentado}}, \bibinfo {author}
  {\bibfnamefont {S.}~\bibnamefont {Diddams}},\ and\ \bibinfo {author}
  {\bibfnamefont {J.}~\bibnamefont {Teufel}},\ }\bibfield  {title} {\bibinfo
  {title} {\textit{Control and readout of a superconducting qubit using a
  photonic link}},\ }\href {https://www.nature.com/articles/s41586-021-03268-x}
  {\bibfield  {journal} {\bibinfo  {journal} {Nature}\ }\textbf {\bibinfo
  {volume} {591}},\ \bibinfo {pages} {575} (\bibinfo {year}
  {2021})}\BibitemShut {NoStop}%
\bibitem [{\citenamefont {Bianchetti}\ \emph {et~al.}(2009)\citenamefont
  {Bianchetti}, \citenamefont {Filipp}, \citenamefont {Baur}, \citenamefont
  {Fink}, \citenamefont {G\"oppl}, \citenamefont {Leek}, \citenamefont
  {Steffen}, \citenamefont {Blais},\ and\ \citenamefont
  {Wallraff}}]{bianchetti2009dynamics}%
  \BibitemOpen
  \bibfield  {author} {\bibinfo {author} {\bibfnamefont {R.}~\bibnamefont
  {Bianchetti}}, \bibinfo {author} {\bibfnamefont {S.}~\bibnamefont {Filipp}},
  \bibinfo {author} {\bibfnamefont {M.}~\bibnamefont {Baur}}, \bibinfo {author}
  {\bibfnamefont {J.~M.}\ \bibnamefont {Fink}}, \bibinfo {author}
  {\bibfnamefont {M.}~\bibnamefont {G\"oppl}}, \bibinfo {author} {\bibfnamefont
  {P.~J.}\ \bibnamefont {Leek}}, \bibinfo {author} {\bibfnamefont
  {L.}~\bibnamefont {Steffen}}, \bibinfo {author} {\bibfnamefont
  {A.}~\bibnamefont {Blais}},\ and\ \bibinfo {author} {\bibfnamefont
  {A.}~\bibnamefont {Wallraff}},\ }\bibfield  {title} {\bibinfo {title}
  {\textit{Dynamics of dispersive single-qubit readout in circuit quantum
  electrodynamics}},\ }\href {https://doi.org/10.1103/PhysRevA.80.043840}
  {\bibfield  {journal} {\bibinfo  {journal} {Phys. Rev. A}\ }\textbf {\bibinfo
  {volume} {80}},\ \bibinfo {pages} {043840} (\bibinfo {year}
  {2009})}\BibitemShut {NoStop}%
\bibitem [{\citenamefont {Castellanos-Beltran}\ and\ \citenamefont
  {Lehnert}(2007)}]{castellanos2007widely}%
  \BibitemOpen
  \bibfield  {author} {\bibinfo {author} {\bibfnamefont {M.}~\bibnamefont
  {Castellanos-Beltran}}\ and\ \bibinfo {author} {\bibfnamefont
  {K.}~\bibnamefont {Lehnert}},\ }\bibfield  {title} {\bibinfo {title}
  {\textit{Widely tunable parametric amplifier based on a superconducting
  quantum interference device array resonator}},\ }\href
  {https://aip.scitation.org/doi/10.1063/1.2773988} {\bibfield  {journal}
  {\bibinfo  {journal} {Applied Physics Letters}\ }\textbf {\bibinfo {volume}
  {91}},\ \bibinfo {pages} {083509} (\bibinfo {year} {2007})}\BibitemShut
  {NoStop}%
\bibitem [{\citenamefont {Sank}\ \emph {et~al.}(2016)\citenamefont {Sank},
  \citenamefont {Chen}, \citenamefont {Khezri}, \citenamefont {Kelly},
  \citenamefont {Barends}, \citenamefont {Campbell}, \citenamefont {Chen},
  \citenamefont {Chiaro}, \citenamefont {Dunsworth}, \citenamefont {Fowler},
  \citenamefont {Jeffrey}, \citenamefont {Lucero}, \citenamefont {Megrant},
  \citenamefont {Mutus}, \citenamefont {Neeley} \emph
  {et~al.}}]{sank2016measurement}%
  \BibitemOpen
  \bibfield  {author} {\bibinfo {author} {\bibfnamefont {D.}~\bibnamefont
  {Sank}}, \bibinfo {author} {\bibfnamefont {Z.}~\bibnamefont {Chen}}, \bibinfo
  {author} {\bibfnamefont {M.}~\bibnamefont {Khezri}}, \bibinfo {author}
  {\bibfnamefont {J.}~\bibnamefont {Kelly}}, \bibinfo {author} {\bibfnamefont
  {R.}~\bibnamefont {Barends}}, \bibinfo {author} {\bibfnamefont
  {B.}~\bibnamefont {Campbell}}, \bibinfo {author} {\bibfnamefont
  {Y.}~\bibnamefont {Chen}}, \bibinfo {author} {\bibfnamefont {B.}~\bibnamefont
  {Chiaro}}, \bibinfo {author} {\bibfnamefont {A.}~\bibnamefont {Dunsworth}},
  \bibinfo {author} {\bibfnamefont {A.}~\bibnamefont {Fowler}}, \bibinfo
  {author} {\bibfnamefont {E.}~\bibnamefont {Jeffrey}}, \bibinfo {author}
  {\bibfnamefont {E.}~\bibnamefont {Lucero}}, \bibinfo {author} {\bibfnamefont
  {A.}~\bibnamefont {Megrant}}, \bibinfo {author} {\bibfnamefont
  {J.}~\bibnamefont {Mutus}}, \bibinfo {author} {\bibfnamefont
  {M.}~\bibnamefont {Neeley}}, \emph {et~al.},\ }\bibfield  {title} {\bibinfo
  {title} {\textit{Measurement-Induced State Transitions in a Superconducting
  Qubit: Beyond the Rotating Wave Approximation}},\ }\href
  {https://doi.org/10.1103/PhysRevLett.117.190503} {\bibfield  {journal}
  {\bibinfo  {journal} {Phys. Rev. Lett.}\ }\textbf {\bibinfo {volume} {117}},\
  \bibinfo {pages} {190503} (\bibinfo {year} {2016})}\BibitemShut {NoStop}%
\bibitem [{\citenamefont {Benz}\ and\ \citenamefont
  {Hamilton}(1996)}]{benz1996pulse}%
  \BibitemOpen
  \bibfield  {author} {\bibinfo {author} {\bibfnamefont {S.~P.}\ \bibnamefont
  {Benz}}\ and\ \bibinfo {author} {\bibfnamefont {C.~A.}\ \bibnamefont
  {Hamilton}},\ }\bibfield  {title} {\bibinfo {title} {\textit{A pulse-driven
  programmable Josephson voltage standard}},\ }\href
  {https://aip.scitation.org/doi/10.1063/1.115814} {\bibfield  {journal}
  {\bibinfo  {journal} {Applied physics letters}\ }\textbf {\bibinfo {volume}
  {68}},\ \bibinfo {pages} {3171} (\bibinfo {year} {1996})}\BibitemShut
  {NoStop}%
\bibitem [{\citenamefont {Donnelly}\ \emph {et~al.}(2020)\citenamefont
  {Donnelly}, \citenamefont {Flowers-Jacobs}, \citenamefont {Brevik},
  \citenamefont {Fox}, \citenamefont {Dresselhaus}, \citenamefont {Hopkins},\
  and\ \citenamefont {Benz}}]{donnelly2020oneghz}%
  \BibitemOpen
  \bibfield  {author} {\bibinfo {author} {\bibfnamefont {C.~A.}\ \bibnamefont
  {Donnelly}}, \bibinfo {author} {\bibfnamefont {N.~E.}\ \bibnamefont
  {Flowers-Jacobs}}, \bibinfo {author} {\bibfnamefont {J.~A.}\ \bibnamefont
  {Brevik}}, \bibinfo {author} {\bibfnamefont {A.~E.}\ \bibnamefont {Fox}},
  \bibinfo {author} {\bibfnamefont {P.~D.}\ \bibnamefont {Dresselhaus}},
  \bibinfo {author} {\bibfnamefont {P.~F.}\ \bibnamefont {Hopkins}},\ and\
  \bibinfo {author} {\bibfnamefont {S.~P.}\ \bibnamefont {Benz}},\ }\bibfield
  {title} {\bibinfo {title} {\textit{1 GHz Waveform Synthesis With Josephson
  Junction Arrays}},\ }\href {https://doi.org/10.1109/TASC.2019.2932342}
  {\bibfield  {journal} {\bibinfo  {journal} {IEEE Transactions on Applied
  Superconductivity}\ }\textbf {\bibinfo {volume} {30}},\ \bibinfo {pages} {1}
  (\bibinfo {year} {2020})}\BibitemShut {NoStop}%
\bibitem [{\citenamefont {Babenko}\ \emph {et~al.}(2020)\citenamefont
  {Babenko}, \citenamefont {Boaventura}, \citenamefont {Flowers-Jacobs},
  \citenamefont {Brevik}, \citenamefont {Fox}, \citenamefont {Williams},
  \citenamefont {Popović}, \citenamefont {Dresselhaus},\ and\ \citenamefont
  {Benz}}]{babenko2020characterization}%
  \BibitemOpen
  \bibfield  {author} {\bibinfo {author} {\bibfnamefont {A.~A.}\ \bibnamefont
  {Babenko}}, \bibinfo {author} {\bibfnamefont {A.~S.}\ \bibnamefont
  {Boaventura}}, \bibinfo {author} {\bibfnamefont {N.~E.}\ \bibnamefont
  {Flowers-Jacobs}}, \bibinfo {author} {\bibfnamefont {J.~A.}\ \bibnamefont
  {Brevik}}, \bibinfo {author} {\bibfnamefont {A.~E.}\ \bibnamefont {Fox}},
  \bibinfo {author} {\bibfnamefont {D.~F.}\ \bibnamefont {Williams}}, \bibinfo
  {author} {\bibfnamefont {Z.}~\bibnamefont {Popović}}, \bibinfo {author}
  {\bibfnamefont {P.~D.}\ \bibnamefont {Dresselhaus}},\ and\ \bibinfo {author}
  {\bibfnamefont {S.~P.}\ \bibnamefont {Benz}},\ }\bibfield  {title} {\bibinfo
  {title} {\textit{Characterization of a Josephson Junction Comb Generator}},\
  }in\ \href {https://doi.org/10.1109/IMS30576.2020.9223811} {\emph {\bibinfo
  {booktitle} {2020 IEEE/MTT-S International Microwave Symposium (IMS)}}}\
  (\bibinfo {year} {2020})\ pp.\ \bibinfo {pages} {936--939}\BibitemShut
  {NoStop}%
\bibitem [{\citenamefont {Johansson}\ \emph {et~al.}(2012)\citenamefont
  {Johansson}, \citenamefont {Nation},\ and\ \citenamefont
  {Nori}}]{johansson2012qutip}%
  \BibitemOpen
  \bibfield  {author} {\bibinfo {author} {\bibfnamefont {J.~R.}\ \bibnamefont
  {Johansson}}, \bibinfo {author} {\bibfnamefont {P.~D.}\ \bibnamefont
  {Nation}},\ and\ \bibinfo {author} {\bibfnamefont {F.}~\bibnamefont {Nori}},\
  }\bibfield  {title} {\bibinfo {title} {\textit{QuTiP: An open-source Python
  framework for the dynamics of open quantum systems}},\ }\href
  {https://www.sciencedirect.com/science/article/pii/S0010465512000835}
  {\bibfield  {journal} {\bibinfo  {journal} {Computer Physics Communications}\
  }\textbf {\bibinfo {volume} {183}},\ \bibinfo {pages} {1760} (\bibinfo {year}
  {2012})}\BibitemShut {NoStop}%
\bibitem [{Note1()}]{Note1}%
  \BibitemOpen
  \bibinfo {note} {These simulations were repeated using simulated SFQ pulses
  with the JPG $\tau = 98$~ps (Supplementary Material~\ref
  {supp:jpg_simulations}) and show no significant change in
  fidelity.}\BibitemShut {Stop}%
\bibitem [{\citenamefont {Burnett}\ \emph {et~al.}(2019)\citenamefont
  {Burnett}, \citenamefont {Bengtsson}, \citenamefont {Scigliuzzo},
  \citenamefont {Niepce}, \citenamefont {Kudra}, \citenamefont {Delsing},\ and\
  \citenamefont {Bylander}}]{burnett2019decoherence}%
  \BibitemOpen
  \bibfield  {author} {\bibinfo {author} {\bibfnamefont {J.~J.}\ \bibnamefont
  {Burnett}}, \bibinfo {author} {\bibfnamefont {A.}~\bibnamefont {Bengtsson}},
  \bibinfo {author} {\bibfnamefont {M.}~\bibnamefont {Scigliuzzo}}, \bibinfo
  {author} {\bibfnamefont {D.}~\bibnamefont {Niepce}}, \bibinfo {author}
  {\bibfnamefont {M.}~\bibnamefont {Kudra}}, \bibinfo {author} {\bibfnamefont
  {P.}~\bibnamefont {Delsing}},\ and\ \bibinfo {author} {\bibfnamefont
  {J.}~\bibnamefont {Bylander}},\ }\bibfield  {title} {\bibinfo {title}
  {\textit{Decoherence benchmarking of superconducting qubits}},\ }\href
  {https://www.nature.com/articles/s41534-019-0168-5} {\bibfield  {journal}
  {\bibinfo  {journal} {npj Quantum Information}\ }\textbf {\bibinfo {volume}
  {5}},\ \bibinfo {pages} {1} (\bibinfo {year} {2019})}\BibitemShut {NoStop}%
\bibitem [{\citenamefont {Veps{\"a}l{\"a}inen}\ \emph
  {et~al.}(2020)\citenamefont {Veps{\"a}l{\"a}inen}, \citenamefont {Karamlou},
  \citenamefont {Orrell}, \citenamefont {Dogra}, \citenamefont {Loer},
  \citenamefont {Vasconcelos}, \citenamefont {Kim}, \citenamefont {Melville},
  \citenamefont {Niedzielski}, \citenamefont {Yoder} \emph
  {et~al.}}]{vepsalainen2020impact}%
  \BibitemOpen
  \bibfield  {author} {\bibinfo {author} {\bibfnamefont {A.~P.}\ \bibnamefont
  {Veps{\"a}l{\"a}inen}}, \bibinfo {author} {\bibfnamefont {A.~H.}\
  \bibnamefont {Karamlou}}, \bibinfo {author} {\bibfnamefont {J.~L.}\
  \bibnamefont {Orrell}}, \bibinfo {author} {\bibfnamefont {A.~S.}\
  \bibnamefont {Dogra}}, \bibinfo {author} {\bibfnamefont {B.}~\bibnamefont
  {Loer}}, \bibinfo {author} {\bibfnamefont {F.}~\bibnamefont {Vasconcelos}},
  \bibinfo {author} {\bibfnamefont {D.~K.}\ \bibnamefont {Kim}}, \bibinfo
  {author} {\bibfnamefont {A.~J.}\ \bibnamefont {Melville}}, \bibinfo {author}
  {\bibfnamefont {B.~M.}\ \bibnamefont {Niedzielski}}, \bibinfo {author}
  {\bibfnamefont {J.~L.}\ \bibnamefont {Yoder}}, \emph {et~al.},\ }\bibfield
  {title} {\bibinfo {title} {\textit{Impact of ionizing radiation on
  superconducting qubit coherence}},\ }\href
  {https://www.nature.com/articles/s41586-020-2619-8} {\bibfield  {journal}
  {\bibinfo  {journal} {Nature}\ }\textbf {\bibinfo {volume} {584}},\ \bibinfo
  {pages} {551} (\bibinfo {year} {2020})}\BibitemShut {NoStop}%
\bibitem [{\citenamefont {McRae}\ \emph {et~al.}(2020)\citenamefont {McRae},
  \citenamefont {Wang}, \citenamefont {Gao}, \citenamefont {Vissers},
  \citenamefont {Brecht}, \citenamefont {Dunsworth}, \citenamefont {Pappas},\
  and\ \citenamefont {Mutus}}]{mcrae2020materials}%
  \BibitemOpen
  \bibfield  {author} {\bibinfo {author} {\bibfnamefont {C.~R.~H.}\
  \bibnamefont {McRae}}, \bibinfo {author} {\bibfnamefont {H.}~\bibnamefont
  {Wang}}, \bibinfo {author} {\bibfnamefont {J.}~\bibnamefont {Gao}}, \bibinfo
  {author} {\bibfnamefont {M.~R.}\ \bibnamefont {Vissers}}, \bibinfo {author}
  {\bibfnamefont {T.}~\bibnamefont {Brecht}}, \bibinfo {author} {\bibfnamefont
  {A.}~\bibnamefont {Dunsworth}}, \bibinfo {author} {\bibfnamefont {D.~P.}\
  \bibnamefont {Pappas}},\ and\ \bibinfo {author} {\bibfnamefont
  {J.}~\bibnamefont {Mutus}},\ }\bibfield  {title} {\bibinfo {title}
  {\textit{Materials loss measurements using superconducting microwave
  resonators}},\ }\href {https://aip.scitation.org/doi/10.1063/5.0017378}
  {\bibfield  {journal} {\bibinfo  {journal} {Review of Scientific
  Instruments}\ }\textbf {\bibinfo {volume} {91}},\ \bibinfo {pages} {091101}
  (\bibinfo {year} {2020})}\BibitemShut {NoStop}%
\bibitem [{\citenamefont {McEwen}\ \emph {et~al.}(2021)\citenamefont {McEwen},
  \citenamefont {Faoro}, \citenamefont {Arya}, \citenamefont {Dunsworth},
  \citenamefont {Huang}, \citenamefont {Kim}, \citenamefont {Burkett},
  \citenamefont {Fowler}, \citenamefont {Arute}, \citenamefont {Bardin} \emph
  {et~al.}}]{mcewen2021resolving}%
  \BibitemOpen
  \bibfield  {author} {\bibinfo {author} {\bibfnamefont {M.}~\bibnamefont
  {McEwen}}, \bibinfo {author} {\bibfnamefont {L.}~\bibnamefont {Faoro}},
  \bibinfo {author} {\bibfnamefont {K.}~\bibnamefont {Arya}}, \bibinfo {author}
  {\bibfnamefont {A.}~\bibnamefont {Dunsworth}}, \bibinfo {author}
  {\bibfnamefont {T.}~\bibnamefont {Huang}}, \bibinfo {author} {\bibfnamefont
  {S.}~\bibnamefont {Kim}}, \bibinfo {author} {\bibfnamefont {B.}~\bibnamefont
  {Burkett}}, \bibinfo {author} {\bibfnamefont {A.}~\bibnamefont {Fowler}},
  \bibinfo {author} {\bibfnamefont {F.}~\bibnamefont {Arute}}, \bibinfo
  {author} {\bibfnamefont {J.~C.}\ \bibnamefont {Bardin}}, \emph {et~al.},\
  }\bibfield  {title} {\bibinfo {title} {\textit{Resolving catastrophic error
  bursts from cosmic rays in large arrays of superconducting qubits}},\ }\href
  {https://arxiv.org/ftp/arxiv/papers/2104/2104.05219.pdf} {\bibfield
  {journal} {\bibinfo  {journal} {arXiv preprint arXiv:2104.05219}\ } (\bibinfo
  {year} {2021})}\BibitemShut {NoStop}%
\bibitem [{\citenamefont {Wenner}\ \emph {et~al.}(2013)\citenamefont {Wenner},
  \citenamefont {Yin}, \citenamefont {Lucero}, \citenamefont {Barends},
  \citenamefont {Chen}, \citenamefont {Chiaro}, \citenamefont {Kelly},
  \citenamefont {Lenander}, \citenamefont {Mariantoni}, \citenamefont
  {Megrant}, \citenamefont {Neill}, \citenamefont {O'Malley}, \citenamefont
  {Sank}, \citenamefont {Vainsencher}, \citenamefont {Wang} \emph
  {et~al.}}]{wenner2013excitation}%
  \BibitemOpen
  \bibfield  {author} {\bibinfo {author} {\bibfnamefont {J.}~\bibnamefont
  {Wenner}}, \bibinfo {author} {\bibfnamefont {Y.}~\bibnamefont {Yin}},
  \bibinfo {author} {\bibfnamefont {E.}~\bibnamefont {Lucero}}, \bibinfo
  {author} {\bibfnamefont {R.}~\bibnamefont {Barends}}, \bibinfo {author}
  {\bibfnamefont {Y.}~\bibnamefont {Chen}}, \bibinfo {author} {\bibfnamefont
  {B.}~\bibnamefont {Chiaro}}, \bibinfo {author} {\bibfnamefont
  {J.}~\bibnamefont {Kelly}}, \bibinfo {author} {\bibfnamefont
  {M.}~\bibnamefont {Lenander}}, \bibinfo {author} {\bibfnamefont
  {M.}~\bibnamefont {Mariantoni}}, \bibinfo {author} {\bibfnamefont
  {A.}~\bibnamefont {Megrant}}, \bibinfo {author} {\bibfnamefont
  {C.}~\bibnamefont {Neill}}, \bibinfo {author} {\bibfnamefont {P.~J.~J.}\
  \bibnamefont {O'Malley}}, \bibinfo {author} {\bibfnamefont {D.}~\bibnamefont
  {Sank}}, \bibinfo {author} {\bibfnamefont {A.}~\bibnamefont {Vainsencher}},
  \bibinfo {author} {\bibfnamefont {H.}~\bibnamefont {Wang}}, \emph {et~al.},\
  }\bibfield  {title} {\bibinfo {title} {\textit{Excitation of Superconducting
  Qubits from Hot Nonequilibrium Quasiparticles}},\ }\href
  {https://doi.org/10.1103/PhysRevLett.110.150502} {\bibfield  {journal}
  {\bibinfo  {journal} {Phys. Rev. Lett.}\ }\textbf {\bibinfo {volume} {110}},\
  \bibinfo {pages} {150502} (\bibinfo {year} {2013})}\BibitemShut {NoStop}%
\bibitem [{\citenamefont {Jin}\ \emph {et~al.}(2015)\citenamefont {Jin},
  \citenamefont {Kamal}, \citenamefont {Sears}, \citenamefont {Gudmundsen},
  \citenamefont {Hover}, \citenamefont {Miloshi}, \citenamefont {Slattery},
  \citenamefont {Yan}, \citenamefont {Yoder}, \citenamefont {Orlando},
  \citenamefont {Gustavsson},\ and\ \citenamefont {Oliver}}]{jin2015thermal}%
  \BibitemOpen
  \bibfield  {author} {\bibinfo {author} {\bibfnamefont {X.~Y.}\ \bibnamefont
  {Jin}}, \bibinfo {author} {\bibfnamefont {A.}~\bibnamefont {Kamal}}, \bibinfo
  {author} {\bibfnamefont {A.~P.}\ \bibnamefont {Sears}}, \bibinfo {author}
  {\bibfnamefont {T.}~\bibnamefont {Gudmundsen}}, \bibinfo {author}
  {\bibfnamefont {D.}~\bibnamefont {Hover}}, \bibinfo {author} {\bibfnamefont
  {J.}~\bibnamefont {Miloshi}}, \bibinfo {author} {\bibfnamefont
  {R.}~\bibnamefont {Slattery}}, \bibinfo {author} {\bibfnamefont
  {F.}~\bibnamefont {Yan}}, \bibinfo {author} {\bibfnamefont {J.}~\bibnamefont
  {Yoder}}, \bibinfo {author} {\bibfnamefont {T.~P.}\ \bibnamefont {Orlando}},
  \bibinfo {author} {\bibfnamefont {S.}~\bibnamefont {Gustavsson}},\ and\
  \bibinfo {author} {\bibfnamefont {W.~D.}\ \bibnamefont {Oliver}},\ }\bibfield
   {title} {\bibinfo {title} {\textit{Thermal and Residual Excited-State
  Population in a 3D Transmon Qubit}},\ }\href
  {https://doi.org/10.1103/PhysRevLett.114.240501} {\bibfield  {journal}
  {\bibinfo  {journal} {Phys. Rev. Lett.}\ }\textbf {\bibinfo {volume} {114}},\
  \bibinfo {pages} {240501} (\bibinfo {year} {2015})}\BibitemShut {NoStop}%
\bibitem [{\citenamefont {Jeffrey}\ \emph {et~al.}(2014)\citenamefont
  {Jeffrey}, \citenamefont {Sank}, \citenamefont {Mutus}, \citenamefont
  {White}, \citenamefont {Kelly}, \citenamefont {Barends}, \citenamefont
  {Chen}, \citenamefont {Chen}, \citenamefont {Chiaro}, \citenamefont
  {Dunsworth}, \citenamefont {Megrant}, \citenamefont {O'Malley}, \citenamefont
  {Neill}, \citenamefont {Roushan}, \citenamefont {Vainsencher} \emph
  {et~al.}}]{jeffrey2014fast}%
  \BibitemOpen
  \bibfield  {author} {\bibinfo {author} {\bibfnamefont {E.}~\bibnamefont
  {Jeffrey}}, \bibinfo {author} {\bibfnamefont {D.}~\bibnamefont {Sank}},
  \bibinfo {author} {\bibfnamefont {J.~Y.}\ \bibnamefont {Mutus}}, \bibinfo
  {author} {\bibfnamefont {T.~C.}\ \bibnamefont {White}}, \bibinfo {author}
  {\bibfnamefont {J.}~\bibnamefont {Kelly}}, \bibinfo {author} {\bibfnamefont
  {R.}~\bibnamefont {Barends}}, \bibinfo {author} {\bibfnamefont
  {Y.}~\bibnamefont {Chen}}, \bibinfo {author} {\bibfnamefont {Z.}~\bibnamefont
  {Chen}}, \bibinfo {author} {\bibfnamefont {B.}~\bibnamefont {Chiaro}},
  \bibinfo {author} {\bibfnamefont {A.}~\bibnamefont {Dunsworth}}, \bibinfo
  {author} {\bibfnamefont {A.}~\bibnamefont {Megrant}}, \bibinfo {author}
  {\bibfnamefont {P.~J.~J.}\ \bibnamefont {O'Malley}}, \bibinfo {author}
  {\bibfnamefont {C.}~\bibnamefont {Neill}}, \bibinfo {author} {\bibfnamefont
  {P.}~\bibnamefont {Roushan}}, \bibinfo {author} {\bibfnamefont
  {A.}~\bibnamefont {Vainsencher}}, \emph {et~al.},\ }\bibfield  {title}
  {\bibinfo {title} {\textit{Fast Accurate State Measurement with
  Superconducting Qubits}},\ }\href
  {https://doi.org/10.1103/PhysRevLett.112.190504} {\bibfield  {journal}
  {\bibinfo  {journal} {Phys. Rev. Lett.}\ }\textbf {\bibinfo {volume} {112}},\
  \bibinfo {pages} {190504} (\bibinfo {year} {2014})}\BibitemShut {NoStop}%
\bibitem [{\citenamefont {Walter}\ \emph {et~al.}(2017)\citenamefont {Walter},
  \citenamefont {Kurpiers}, \citenamefont {Gasparinetti}, \citenamefont
  {Magnard}, \citenamefont {Poto\ifmmode~\check{c}\else \v{c}\fi{}nik},
  \citenamefont {Salath\'e}, \citenamefont {Pechal}, \citenamefont {Mondal},
  \citenamefont {Oppliger}, \citenamefont {Eichler},\ and\ \citenamefont
  {Wallraff}}]{walter2017rapid}%
  \BibitemOpen
  \bibfield  {author} {\bibinfo {author} {\bibfnamefont {T.}~\bibnamefont
  {Walter}}, \bibinfo {author} {\bibfnamefont {P.}~\bibnamefont {Kurpiers}},
  \bibinfo {author} {\bibfnamefont {S.}~\bibnamefont {Gasparinetti}}, \bibinfo
  {author} {\bibfnamefont {P.}~\bibnamefont {Magnard}}, \bibinfo {author}
  {\bibfnamefont {A.}~\bibnamefont {Poto\ifmmode~\check{c}\else
  \v{c}\fi{}nik}}, \bibinfo {author} {\bibfnamefont {Y.}~\bibnamefont
  {Salath\'e}}, \bibinfo {author} {\bibfnamefont {M.}~\bibnamefont {Pechal}},
  \bibinfo {author} {\bibfnamefont {M.}~\bibnamefont {Mondal}}, \bibinfo
  {author} {\bibfnamefont {M.}~\bibnamefont {Oppliger}}, \bibinfo {author}
  {\bibfnamefont {C.}~\bibnamefont {Eichler}},\ and\ \bibinfo {author}
  {\bibfnamefont {A.}~\bibnamefont {Wallraff}},\ }\bibfield  {title} {\bibinfo
  {title} {\textit{Rapid High-Fidelity Single-Shot Dispersive Readout of
  Superconducting Qubits}},\ }\href
  {https://doi.org/10.1103/PhysRevApplied.7.054020} {\bibfield  {journal}
  {\bibinfo  {journal} {Phys. Rev. Applied}\ }\textbf {\bibinfo {volume} {7}},\
  \bibinfo {pages} {054020} (\bibinfo {year} {2017})}\BibitemShut {NoStop}%
\bibitem [{\citenamefont {Knill}\ \emph {et~al.}(2008)\citenamefont {Knill},
  \citenamefont {Leibfried}, \citenamefont {Reichle}, \citenamefont {Britton},
  \citenamefont {Blakestad}, \citenamefont {Jost}, \citenamefont {Langer},
  \citenamefont {Ozeri}, \citenamefont {Seidelin},\ and\ \citenamefont
  {Wineland}}]{knill2008randomized}%
  \BibitemOpen
  \bibfield  {author} {\bibinfo {author} {\bibfnamefont {E.}~\bibnamefont
  {Knill}}, \bibinfo {author} {\bibfnamefont {D.}~\bibnamefont {Leibfried}},
  \bibinfo {author} {\bibfnamefont {R.}~\bibnamefont {Reichle}}, \bibinfo
  {author} {\bibfnamefont {J.}~\bibnamefont {Britton}}, \bibinfo {author}
  {\bibfnamefont {R.~B.}\ \bibnamefont {Blakestad}}, \bibinfo {author}
  {\bibfnamefont {J.~D.}\ \bibnamefont {Jost}}, \bibinfo {author}
  {\bibfnamefont {C.}~\bibnamefont {Langer}}, \bibinfo {author} {\bibfnamefont
  {R.}~\bibnamefont {Ozeri}}, \bibinfo {author} {\bibfnamefont
  {S.}~\bibnamefont {Seidelin}},\ and\ \bibinfo {author} {\bibfnamefont
  {D.~J.}\ \bibnamefont {Wineland}},\ }\bibfield  {title} {\bibinfo {title}
  {\textit{Randomized benchmarking of quantum gates}},\ }\href
  {https://doi.org/10.1103/PhysRevA.77.012307} {\bibfield  {journal} {\bibinfo
  {journal} {Phys. Rev. A}\ }\textbf {\bibinfo {volume} {77}},\ \bibinfo
  {pages} {012307} (\bibinfo {year} {2008})}\BibitemShut {NoStop}%
\bibitem [{\citenamefont {Magesan}\ \emph {et~al.}(2012)\citenamefont
  {Magesan}, \citenamefont {Gambetta}, \citenamefont {Johnson}, \citenamefont
  {Ryan}, \citenamefont {Chow}, \citenamefont {Merkel}, \citenamefont
  {da~Silva}, \citenamefont {Keefe}, \citenamefont {Rothwell}, \citenamefont
  {Ohki}, \citenamefont {Ketchen},\ and\ \citenamefont
  {Steffen}}]{magesan2012efficient}%
  \BibitemOpen
  \bibfield  {author} {\bibinfo {author} {\bibfnamefont {E.}~\bibnamefont
  {Magesan}}, \bibinfo {author} {\bibfnamefont {J.~M.}\ \bibnamefont
  {Gambetta}}, \bibinfo {author} {\bibfnamefont {B.~R.}\ \bibnamefont
  {Johnson}}, \bibinfo {author} {\bibfnamefont {C.~A.}\ \bibnamefont {Ryan}},
  \bibinfo {author} {\bibfnamefont {J.~M.}\ \bibnamefont {Chow}}, \bibinfo
  {author} {\bibfnamefont {S.~T.}\ \bibnamefont {Merkel}}, \bibinfo {author}
  {\bibfnamefont {M.~P.}\ \bibnamefont {da~Silva}}, \bibinfo {author}
  {\bibfnamefont {G.~A.}\ \bibnamefont {Keefe}}, \bibinfo {author}
  {\bibfnamefont {M.~B.}\ \bibnamefont {Rothwell}}, \bibinfo {author}
  {\bibfnamefont {T.~A.}\ \bibnamefont {Ohki}}, \bibinfo {author}
  {\bibfnamefont {M.~B.}\ \bibnamefont {Ketchen}},\ and\ \bibinfo {author}
  {\bibfnamefont {M.}~\bibnamefont {Steffen}},\ }\bibfield  {title} {\bibinfo
  {title} {\textit{Efficient Measurement of Quantum Gate Error by Interleaved
  Randomized Benchmarking}},\ }\href
  {https://doi.org/10.1103/PhysRevLett.109.080505} {\bibfield  {journal}
  {\bibinfo  {journal} {Phys. Rev. Lett.}\ }\textbf {\bibinfo {volume} {109}},\
  \bibinfo {pages} {080505} (\bibinfo {year} {2012})}\BibitemShut {NoStop}%
\bibitem [{\citenamefont {Magesan}\ \emph {et~al.}(2011)\citenamefont
  {Magesan}, \citenamefont {Gambetta},\ and\ \citenamefont
  {Emerson}}]{magesan2011scalable}%
  \BibitemOpen
  \bibfield  {author} {\bibinfo {author} {\bibfnamefont {E.}~\bibnamefont
  {Magesan}}, \bibinfo {author} {\bibfnamefont {J.~M.}\ \bibnamefont
  {Gambetta}},\ and\ \bibinfo {author} {\bibfnamefont {J.}~\bibnamefont
  {Emerson}},\ }\bibfield  {title} {\bibinfo {title} {\textit{Scalable and
  Robust Randomized Benchmarking of Quantum Processes}},\ }\href
  {https://doi.org/10.1103/PhysRevLett.106.180504} {\bibfield  {journal}
  {\bibinfo  {journal} {Phys. Rev. Lett.}\ }\textbf {\bibinfo {volume} {106}},\
  \bibinfo {pages} {180504} (\bibinfo {year} {2011})}\BibitemShut {NoStop}%
\bibitem [{\citenamefont {Chen}(2018)}]{chen2018metrology}%
  \BibitemOpen
  \bibfield  {author} {\bibinfo {author} {\bibfnamefont {Z.}~\bibnamefont
  {Chen}},\ }\emph {\bibinfo {title} {\textit{Metrology of quantum control and
  measurement in superconducting qubits}}},\ \href
  {https://escholarship.org/uc/item/0g29b4p0} {Ph.D. thesis},\ \bibinfo
  {school} {UC Santa Barbara} (\bibinfo {year} {2018})\BibitemShut {NoStop}%
\bibitem [{\citenamefont {Rol}\ \emph {et~al.}(2017)\citenamefont {Rol},
  \citenamefont {Bultink}, \citenamefont {O'Brien}, \citenamefont {de~Jong},
  \citenamefont {Theis}, \citenamefont {Fu}, \citenamefont {Luthi},
  \citenamefont {Vermeulen}, \citenamefont {de~Sterke}, \citenamefont {Bruno},
  \citenamefont {Deurloo}, \citenamefont {Schouten}, \citenamefont {Wilhelm},\
  and\ \citenamefont {DiCarlo}}]{rol2017restless}%
  \BibitemOpen
  \bibfield  {author} {\bibinfo {author} {\bibfnamefont {M.~A.}\ \bibnamefont
  {Rol}}, \bibinfo {author} {\bibfnamefont {C.~C.}\ \bibnamefont {Bultink}},
  \bibinfo {author} {\bibfnamefont {T.~E.}\ \bibnamefont {O'Brien}}, \bibinfo
  {author} {\bibfnamefont {S.~R.}\ \bibnamefont {de~Jong}}, \bibinfo {author}
  {\bibfnamefont {L.~S.}\ \bibnamefont {Theis}}, \bibinfo {author}
  {\bibfnamefont {X.}~\bibnamefont {Fu}}, \bibinfo {author} {\bibfnamefont
  {F.}~\bibnamefont {Luthi}}, \bibinfo {author} {\bibfnamefont {R.~F.~L.}\
  \bibnamefont {Vermeulen}}, \bibinfo {author} {\bibfnamefont {J.~C.}\
  \bibnamefont {de~Sterke}}, \bibinfo {author} {\bibfnamefont {A.}~\bibnamefont
  {Bruno}}, \bibinfo {author} {\bibfnamefont {D.}~\bibnamefont {Deurloo}},
  \bibinfo {author} {\bibfnamefont {R.~N.}\ \bibnamefont {Schouten}}, \bibinfo
  {author} {\bibfnamefont {F.~K.}\ \bibnamefont {Wilhelm}},\ and\ \bibinfo
  {author} {\bibfnamefont {L.}~\bibnamefont {DiCarlo}},\ }\bibfield  {title}
  {\bibinfo {title} {\textit{Restless Tuneup of High-Fidelity Qubit Gates}},\
  }\href {https://doi.org/10.1103/PhysRevApplied.7.041001} {\bibfield
  {journal} {\bibinfo  {journal} {Phys. Rev. Applied}\ }\textbf {\bibinfo
  {volume} {7}},\ \bibinfo {pages} {041001} (\bibinfo {year}
  {2017})}\BibitemShut {NoStop}%
\bibitem [{\citenamefont {Petit}\ \emph {et~al.}(2020)\citenamefont {Petit},
  \citenamefont {Eenink}, \citenamefont {Russ}, \citenamefont {Lawrie},
  \citenamefont {Hendrickx}, \citenamefont {Philips}, \citenamefont {Clarke},
  \citenamefont {Vandersypen},\ and\ \citenamefont
  {Veldhorst}}]{petit2020universal}%
  \BibitemOpen
  \bibfield  {author} {\bibinfo {author} {\bibfnamefont {L.}~\bibnamefont
  {Petit}}, \bibinfo {author} {\bibfnamefont {H.}~\bibnamefont {Eenink}},
  \bibinfo {author} {\bibfnamefont {M.}~\bibnamefont {Russ}}, \bibinfo {author}
  {\bibfnamefont {W.}~\bibnamefont {Lawrie}}, \bibinfo {author} {\bibfnamefont
  {N.}~\bibnamefont {Hendrickx}}, \bibinfo {author} {\bibfnamefont
  {S.}~\bibnamefont {Philips}}, \bibinfo {author} {\bibfnamefont
  {J.}~\bibnamefont {Clarke}}, \bibinfo {author} {\bibfnamefont
  {L.}~\bibnamefont {Vandersypen}},\ and\ \bibinfo {author} {\bibfnamefont
  {M.}~\bibnamefont {Veldhorst}},\ }\bibfield  {title} {\bibinfo {title}
  {\textit{Universal quantum logic in hot silicon qubits}},\ }\href
  {https://www.nature.com/articles/s41586-020-2170-7} {\bibfield  {journal}
  {\bibinfo  {journal} {Nature}\ }\textbf {\bibinfo {volume} {580}},\ \bibinfo
  {pages} {355} (\bibinfo {year} {2020})}\BibitemShut {NoStop}%
\bibitem [{\citenamefont {McKay}\ \emph {et~al.}(2019)\citenamefont {McKay},
  \citenamefont {Sheldon}, \citenamefont {Smolin}, \citenamefont {Chow},\ and\
  \citenamefont {Gambetta}}]{mckay2019three}%
  \BibitemOpen
  \bibfield  {author} {\bibinfo {author} {\bibfnamefont {D.~C.}\ \bibnamefont
  {McKay}}, \bibinfo {author} {\bibfnamefont {S.}~\bibnamefont {Sheldon}},
  \bibinfo {author} {\bibfnamefont {J.~A.}\ \bibnamefont {Smolin}}, \bibinfo
  {author} {\bibfnamefont {J.~M.}\ \bibnamefont {Chow}},\ and\ \bibinfo
  {author} {\bibfnamefont {J.~M.}\ \bibnamefont {Gambetta}},\ }\bibfield
  {title} {\bibinfo {title} {\textit{Three-Qubit Randomized Benchmarking}},\
  }\href {https://doi.org/10.1103/PhysRevLett.122.200502} {\bibfield  {journal}
  {\bibinfo  {journal} {Phys. Rev. Lett.}\ }\textbf {\bibinfo {volume} {122}},\
  \bibinfo {pages} {200502} (\bibinfo {year} {2019})}\BibitemShut {NoStop}%
\bibitem [{Note2()}]{Note2}%
  \BibitemOpen
  \bibinfo {note} {Note that a small measurement error is present in all the
  JPG RB data, which were taken first and on a separate cooldown than the TSCE
  RB. This error was remedied before the TSCE measurement was performed and is
  responsible for the offset in the JPG curve (Fig.~\ref {fig:jpg_rb}). We
  emphasize this has no influence on the extracted $r$ for either
  setup.}\BibitemShut {Stop}%
\bibitem [{\citenamefont {Tsan}\ \emph {et~al.}(2021)\citenamefont {Tsan},
  \citenamefont {Galitzki}, \citenamefont {Ali}, \citenamefont {Arnold},
  \citenamefont {Coppi}, \citenamefont {Ervin}, \citenamefont {Foote},
  \citenamefont {Keating}, \citenamefont {Lashner}, \citenamefont
  {Orlowski-Scherer}, \citenamefont {Randall}, \citenamefont {Seibert},
  \citenamefont {Spisak}, \citenamefont {Teply}, \citenamefont {Xu} \emph
  {et~al.}}]{tsan2021effects}%
  \BibitemOpen
  \bibfield  {author} {\bibinfo {author} {\bibfnamefont {T.}~\bibnamefont
  {Tsan}}, \bibinfo {author} {\bibfnamefont {N.}~\bibnamefont {Galitzki}},
  \bibinfo {author} {\bibfnamefont {A.~M.}\ \bibnamefont {Ali}}, \bibinfo
  {author} {\bibfnamefont {K.}~\bibnamefont {Arnold}}, \bibinfo {author}
  {\bibfnamefont {G.}~\bibnamefont {Coppi}}, \bibinfo {author} {\bibfnamefont
  {T.}~\bibnamefont {Ervin}}, \bibinfo {author} {\bibfnamefont
  {L.}~\bibnamefont {Foote}}, \bibinfo {author} {\bibfnamefont
  {B.}~\bibnamefont {Keating}}, \bibinfo {author} {\bibfnamefont
  {J.}~\bibnamefont {Lashner}}, \bibinfo {author} {\bibfnamefont
  {J.}~\bibnamefont {Orlowski-Scherer}}, \bibinfo {author} {\bibfnamefont
  {M.~J.}\ \bibnamefont {Randall}}, \bibinfo {author} {\bibfnamefont
  {J.}~\bibnamefont {Seibert}}, \bibinfo {author} {\bibfnamefont
  {J.}~\bibnamefont {Spisak}}, \bibinfo {author} {\bibfnamefont {G.~P.}\
  \bibnamefont {Teply}}, \bibinfo {author} {\bibfnamefont {Z.}~\bibnamefont
  {Xu}}, \emph {et~al.},\ }\bibfield  {title} {\bibinfo {title} {\textit{The
  effects of inclination on a two stage pulse tube cryocooler for use with a
  ground based observatory}},\ }\href
  {https://www.sciencedirect.com/science/article/pii/S0011227521000813}
  {\bibfield  {journal} {\bibinfo  {journal} {Cryogenics}\ }\textbf {\bibinfo
  {volume} {117}},\ \bibinfo {pages} {103323} (\bibinfo {year}
  {2021})}\BibitemShut {NoStop}%
\bibitem [{\citenamefont {Holman}\ \emph {et~al.}(2021)\citenamefont {Holman},
  \citenamefont {Rosenberg}, \citenamefont {Yost}, \citenamefont {Yoder},
  \citenamefont {Das}, \citenamefont {Oliver}, \citenamefont {McDermott},\ and\
  \citenamefont {Eriksson}}]{holman2021three}%
  \BibitemOpen
  \bibfield  {author} {\bibinfo {author} {\bibfnamefont {N.}~\bibnamefont
  {Holman}}, \bibinfo {author} {\bibfnamefont {D.}~\bibnamefont {Rosenberg}},
  \bibinfo {author} {\bibfnamefont {D.}~\bibnamefont {Yost}}, \bibinfo {author}
  {\bibfnamefont {J.}~\bibnamefont {Yoder}}, \bibinfo {author} {\bibfnamefont
  {R.}~\bibnamefont {Das}}, \bibinfo {author} {\bibfnamefont {W.~D.}\
  \bibnamefont {Oliver}}, \bibinfo {author} {\bibfnamefont {R.}~\bibnamefont
  {McDermott}},\ and\ \bibinfo {author} {\bibfnamefont {M.}~\bibnamefont
  {Eriksson}},\ }\bibfield  {title} {\bibinfo {title} {\textit{3D integration
  and measurement of a semiconductor double quantum dot with a high-impedance
  TiN resonator}},\ }\href {https://www.nature.com/articles/s41534-021-00469-0}
  {\bibfield  {journal} {\bibinfo  {journal} {npj Quantum Information}\
  }\textbf {\bibinfo {volume} {7}},\ \bibinfo {pages} {1} (\bibinfo {year}
  {2021})}\BibitemShut {NoStop}%
\bibitem [{\citenamefont {Gold}\ \emph {et~al.}(2021)\citenamefont {Gold},
  \citenamefont {Paquette}, \citenamefont {Stockklauser}, \citenamefont
  {Reagor}, \citenamefont {Alam}, \citenamefont {Bestwick}, \citenamefont
  {Didier}, \citenamefont {Nersisyan}, \citenamefont {Oruc}, \citenamefont
  {Razavi} \emph {et~al.}}]{gold2021entanglement}%
  \BibitemOpen
  \bibfield  {author} {\bibinfo {author} {\bibfnamefont {A.}~\bibnamefont
  {Gold}}, \bibinfo {author} {\bibfnamefont {J.}~\bibnamefont {Paquette}},
  \bibinfo {author} {\bibfnamefont {A.}~\bibnamefont {Stockklauser}}, \bibinfo
  {author} {\bibfnamefont {M.~J.}\ \bibnamefont {Reagor}}, \bibinfo {author}
  {\bibfnamefont {M.~S.}\ \bibnamefont {Alam}}, \bibinfo {author}
  {\bibfnamefont {A.}~\bibnamefont {Bestwick}}, \bibinfo {author}
  {\bibfnamefont {N.}~\bibnamefont {Didier}}, \bibinfo {author} {\bibfnamefont
  {A.}~\bibnamefont {Nersisyan}}, \bibinfo {author} {\bibfnamefont
  {F.}~\bibnamefont {Oruc}}, \bibinfo {author} {\bibfnamefont {A.}~\bibnamefont
  {Razavi}}, \emph {et~al.},\ }\bibfield  {title} {\bibinfo {title}
  {\textit{Entanglement across separate silicon dies in a modular
  superconducting qubit device}},\ }\href
  {https://www.nature.com/articles/s41534-021-00484-1} {\bibfield  {journal}
  {\bibinfo  {journal} {npj Quantum Information}\ }\textbf {\bibinfo {volume}
  {7}},\ \bibinfo {pages} {1} (\bibinfo {year} {2021})}\BibitemShut {NoStop}%
\bibitem [{\citenamefont {Hidaka}(2019)}]{hidakajapanese}%
  \BibitemOpen
  \bibfield  {author} {\bibinfo {author} {\bibfnamefont {M.}~\bibnamefont
  {Hidaka}},\ }\bibfield  {title} {\bibinfo {title} {\textit{Japanese
  Activities for Superconducting Circuits Using Flip-chip Configurations}},\
  }\href
  {https://documents.pub/document/japanese-activities-for-superconducting-circuits-japanese-activities-for-superconducting.html}
  {\bibfield  {journal} {\bibinfo  {journal} {IEEE CSC and ESAS
  Superconductivity News Forum}\ } (\bibinfo {year} {2019})}\BibitemShut
  {NoStop}%
\bibitem [{\citenamefont {Rahamim}\ \emph {et~al.}(2017)\citenamefont
  {Rahamim}, \citenamefont {Behrle}, \citenamefont {Peterer}, \citenamefont
  {Patterson}, \citenamefont {Spring}, \citenamefont {Tsunoda}, \citenamefont
  {Manenti}, \citenamefont {Tancredi},\ and\ \citenamefont
  {Leek}}]{rahamim2017double}%
  \BibitemOpen
  \bibfield  {author} {\bibinfo {author} {\bibfnamefont {J.}~\bibnamefont
  {Rahamim}}, \bibinfo {author} {\bibfnamefont {T.}~\bibnamefont {Behrle}},
  \bibinfo {author} {\bibfnamefont {M.}~\bibnamefont {Peterer}}, \bibinfo
  {author} {\bibfnamefont {A.}~\bibnamefont {Patterson}}, \bibinfo {author}
  {\bibfnamefont {P.}~\bibnamefont {Spring}}, \bibinfo {author} {\bibfnamefont
  {T.}~\bibnamefont {Tsunoda}}, \bibinfo {author} {\bibfnamefont
  {R.}~\bibnamefont {Manenti}}, \bibinfo {author} {\bibfnamefont
  {G.}~\bibnamefont {Tancredi}},\ and\ \bibinfo {author} {\bibfnamefont
  {P.}~\bibnamefont {Leek}},\ }\bibfield  {title} {\bibinfo {title}
  {\textit{Double-sided coaxial circuit QED with out-of-plane wiring}},\ }\href
  {https://arxiv.org/pdf/1703.05828.pdf} {\bibfield  {journal} {\bibinfo
  {journal} {Applied Physics Letters}\ }\textbf {\bibinfo {volume} {110}},\
  \bibinfo {pages} {222602} (\bibinfo {year} {2017})}\BibitemShut {NoStop}%
\bibitem [{\citenamefont {Castellanos-Beltran}\ \emph
  {et~al.}(2021)\citenamefont {Castellanos-Beltran}, \citenamefont {Olaya},
  \citenamefont {Sirois}, \citenamefont {Donnelly}, \citenamefont
  {Dresselhaus}, \citenamefont {Benz},\ and\ \citenamefont
  {Hopkins}}]{castellanos2021single}%
  \BibitemOpen
  \bibfield  {author} {\bibinfo {author} {\bibfnamefont {M.~A.}\ \bibnamefont
  {Castellanos-Beltran}}, \bibinfo {author} {\bibfnamefont {D.~I.}\
  \bibnamefont {Olaya}}, \bibinfo {author} {\bibfnamefont {A.~J.}\ \bibnamefont
  {Sirois}}, \bibinfo {author} {\bibfnamefont {C.~A.}\ \bibnamefont
  {Donnelly}}, \bibinfo {author} {\bibfnamefont {P.~D.}\ \bibnamefont
  {Dresselhaus}}, \bibinfo {author} {\bibfnamefont {S.}~\bibnamefont {Benz}},\
  and\ \bibinfo {author} {\bibfnamefont {P.~F.}\ \bibnamefont {Hopkins}},\
  }\bibfield  {title} {\bibinfo {title} {\textit{Single-Flux-Quantum Multiplier
  Circuits for Synthesizing Gigahertz Waveforms With Quantum-Based Accuracy}},\
  }\href {https://doi.org/10.1109/TASC.2021.3057013} {\bibfield  {journal}
  {\bibinfo  {journal} {IEEE Transactions on Applied Superconductivity}\
  }\textbf {\bibinfo {volume} {31}},\ \bibinfo {pages} {1} (\bibinfo {year}
  {2021})}\BibitemShut {NoStop}%
\bibitem [{\citenamefont {Rylyakov}\ and\ \citenamefont
  {Likharev}(1999)}]{rylyakov1999pulse}%
  \BibitemOpen
  \bibfield  {author} {\bibinfo {author} {\bibfnamefont {A.~V.}\ \bibnamefont
  {Rylyakov}}\ and\ \bibinfo {author} {\bibfnamefont {K.~K.}\ \bibnamefont
  {Likharev}},\ }\bibfield  {title} {\bibinfo {title} {\textit{Pulse jitter and
  timing errors in RSFQ circuits}},\ }\href
  {https://ieeexplore.ieee.org/abstract/document/783794} {\bibfield  {journal}
  {\bibinfo  {journal} {IEEE transactions on applied superconductivity}\
  }\textbf {\bibinfo {volume} {9}},\ \bibinfo {pages} {3539} (\bibinfo {year}
  {1999})}\BibitemShut {NoStop}%
\bibitem [{\citenamefont {Fox}\ \emph {et~al.}(2015)\citenamefont {Fox},
  \citenamefont {Dresselhaus}, \citenamefont {Rüfenacht}, \citenamefont
  {Sanders},\ and\ \citenamefont {Benz}}]{fox2015junction}%
  \BibitemOpen
  \bibfield  {author} {\bibinfo {author} {\bibfnamefont {A.~E.}\ \bibnamefont
  {Fox}}, \bibinfo {author} {\bibfnamefont {P.~D.}\ \bibnamefont
  {Dresselhaus}}, \bibinfo {author} {\bibfnamefont {A.}~\bibnamefont
  {Rüfenacht}}, \bibinfo {author} {\bibfnamefont {A.}~\bibnamefont
  {Sanders}},\ and\ \bibinfo {author} {\bibfnamefont {S.~P.}\ \bibnamefont
  {Benz}},\ }\bibfield  {title} {\bibinfo {title} {\textit{Junction Yield
  Analysis for 10 V Programmable Josephson Voltage Standard Devices}},\ }\href
  {https://doi.org/10.1109/TASC.2014.2377744} {\bibfield  {journal} {\bibinfo
  {journal} {IEEE Transactions on Applied Superconductivity}\ }\textbf
  {\bibinfo {volume} {25}},\ \bibinfo {pages} {1} (\bibinfo {year}
  {2015})}\BibitemShut {NoStop}%
\end{thebibliography}%



\clearpage
\onecolumngrid
\begin{center}
\textbf{\large Supplementary Material: Digital control of a superconducting qubit using a Josephson pulse generator at 3~K}
\end{center}
\setcounter{equation}{0}
\setcounter{figure}{0}
\setcounter{table}{0}
\setcounter{page}{1}
\makeatletter
\renewcommand{\theequation}{S\arabic{equation}}
\renewcommand{\thefigure}{S\arabic{figure}}

\section{\label{supp:full_schematic}Full Experimental Schematic}

\begin{figure*}[ht!]
    \centering
    \includegraphics[width = \textwidth]{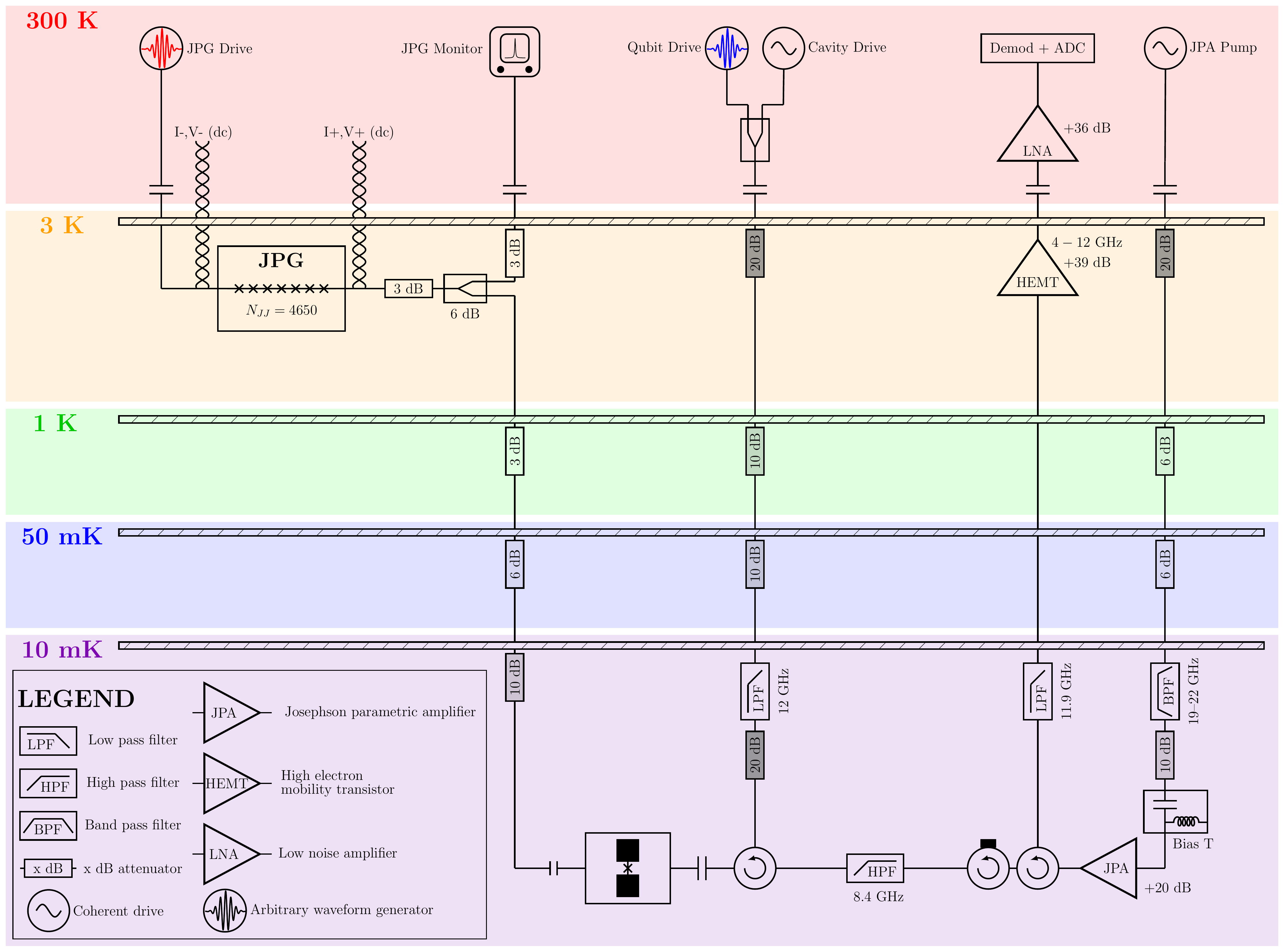}
    \caption{(Color) Detailed schematic of the setup used to evaluate qubit control using a Josephson Pulse Generator (JPG) at 3~K in a commercial DR. The dc connection to the bias T (via twisted pair) is used to provide a dc flux bias to the JPA. Both the JPA dc flux bias and the $I$ and $V$ JPG taps are low-pass filtered at 1.9~MHz and the JPG taps are further filtered with an on-chip inductor network (see Fig~\ref{fig:digital_jpg_qubit_control}(b) in the main body). We insert an attenuator between the JPG splitter output and the 3~K plate to minimize reflections from signals exiting the cryostat and prevent them from affecting the rf JPG signal on-chip.}
    \label{fig:full_schematic}
\end{figure*}

We present the full experimental diagram in Fig.~\ref{fig:full_schematic}. The JPG has two sets of twisted pair leads connected to on-chip inductive taps to permit low-frequency \mbox{$I$-$V$} characterization of the JPG. A splitter is also used on the JPG rf output line for analysis of high frequency characteristics using ambient instrumentation. A 2.85~GHz low pass filter with over 60~dB insertion loss at 5~GHz is used at 300~K to fully prevent any higher harmonic content of the JPG clock signal from driving the qubit.

\section{\label{supp:rabi_chevron_ramsey_scan}Rabi and Ramsey Frequency Sweeps}
\begin{figure}[t]
    \centering
    \includegraphics[width = .98\textwidth]{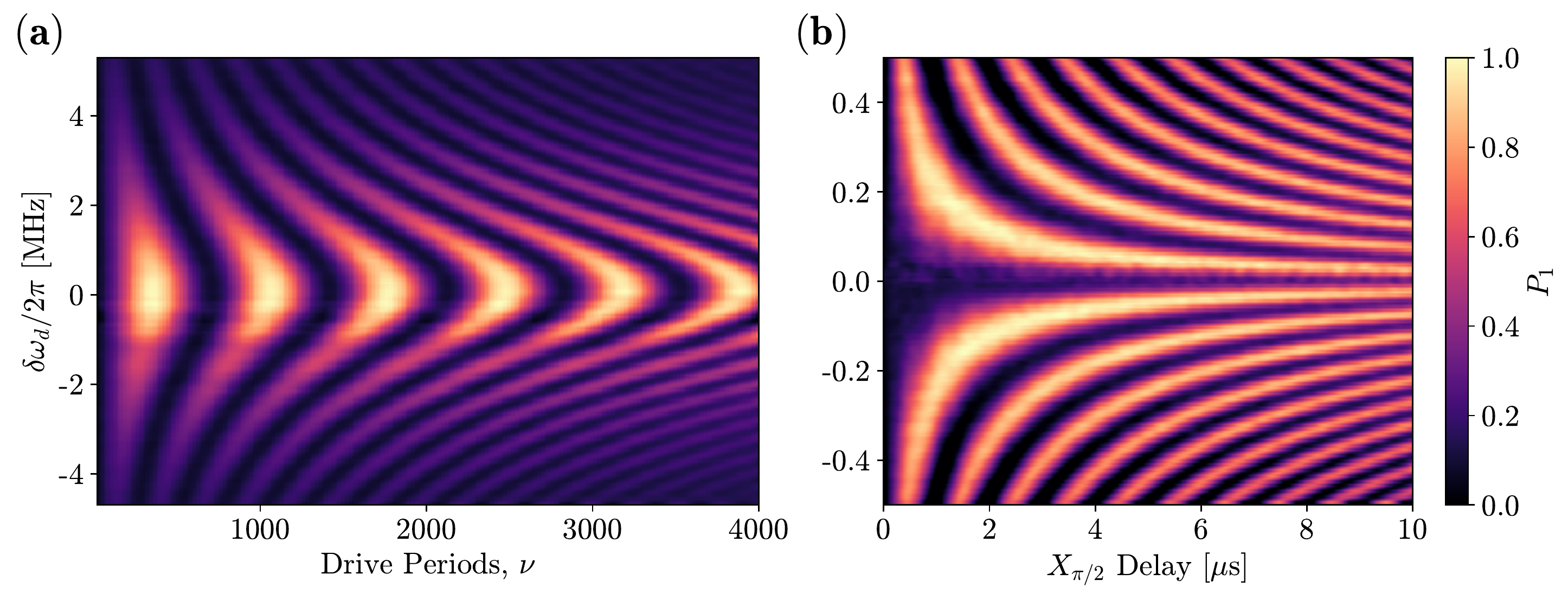}
    \caption{(Color) \textbf{(a)} JPG Rabi chevron at fixed $I_b$. \textbf{(b)} Generalized Ramsey scan. In both measurements the JPG drive frequency $\omega_{d}$ is swept in a small region $\delta \omega_d = \omega_d - \omega_{10} / 2$.}
    \label{fig:rabi_chevron_ramsey_scan}
\end{figure}

In Fig.~\ref{fig:rabi_chevron_ramsey_scan} we present generalized Rabi and Ramsey scans where we sweep the JPG drive frequency $\omega_d + \delta \omega_d = \omega_{10} / 2$. This is simply a demonstration that we may make connection with various routines and protocols with the JPG that are common with TSCE control setups.

\section{\label{supp:ramsey_fringe_colorplots}Ramsey Fringe Compilation}

\begin{figure}[t]
    \centering
    \includegraphics[width = .98\textwidth]{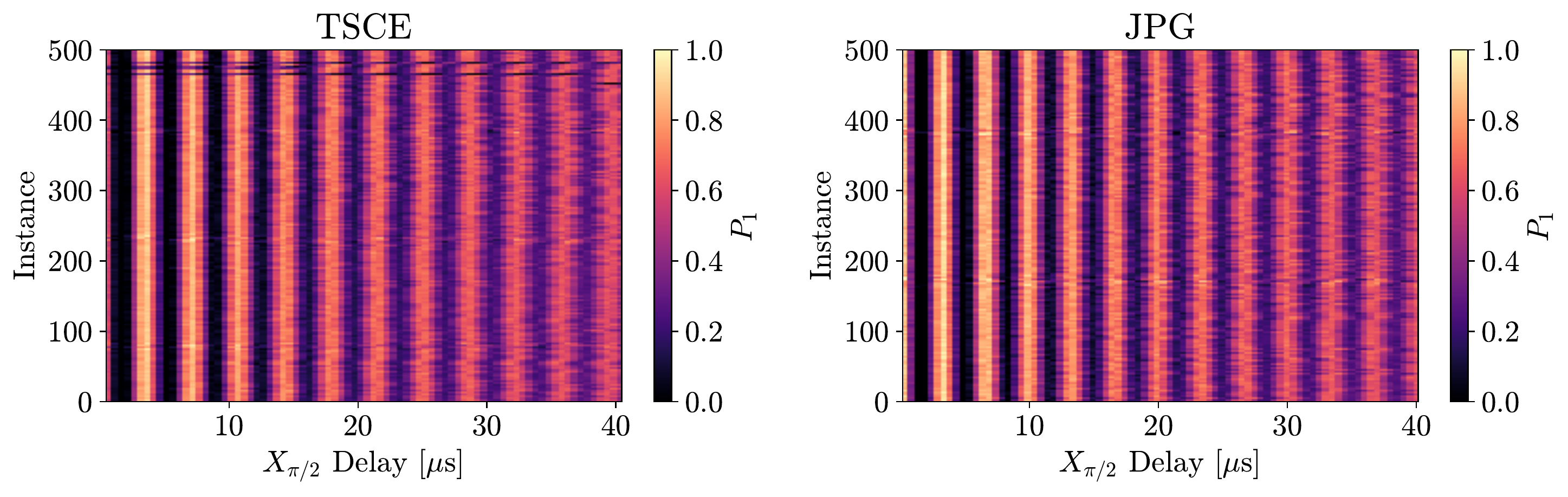}
    \caption{(Color) Compilation colorplot of all Ramsey fringes gathered with both control setups for statistical evaluation of the insensitivity of the qubit coherence time to the control setup used. The three dark, horizontal, stripe-like features in the TSCE compilation are due to instances where the pulse-shaping AWG loses locking during the course of the measurement of a single fringe.}
    \label{fig:ramsey_stats_all_fringes_colorplots}
\end{figure}

In Fig.~\ref{fig:ramsey_stats_all_fringes_colorplots} we compile all Ramsey fringes gathered from the $T_2^*$ measurements in Sec.~\ref{sec:qubit_performance_comparison}. The consistent level of fluctuations in the fringe frequency indicate a similar level of stability in $\omega_{10}$ with both qubit control setups.

\section{\label{supp:jpg_orthogonal_axis_control}Orthogonal Axis Control and Construction of JPG Drive Waveforms}

Here we provide details of JPG drive waveforms and the procedure to establish and verify orthogonal axis control of the qubit. For orthogonal control, first a timing reference is established to define the $\hat{x}$ control axis ($\phi_d^{\hat{x}} = 0$). At subharmonic drive $k \geq 2$ the $\hat{y}$ control axis is realized by phasing the drive signal to allow the qubit to precess by $\phi_q = \pi / 2$ before the control pulse train arrives \cite{leonard2019digital}. Intuitively, there are $k$ qubit periods per drive period and during each qubit period the qubit phase advances relative to the drive by a factor $2 \pi k$. Therefore, to accumulate the required $\phi_q = \pi / 2$ in a single drive period, the $\hat{y}$ drive signal must be phased by $\phi_d^{\hat{y}} = \pi / 2 k$ relative to the $\hat{x}$ drive timing reference.

Drive waveforms are constructed from sine waves with an integer number of samples per sine period with the integer chosen to be compatible with the DAC sample rate of the AWG and also results in synthesis at the proper drive frequency. For all qubit control experiments we generate JPG pulses at the qubit subharmonic $k = 2$ and use sine periods of length 24 samples, leading to an AWG clock frequency of 64.32~GSa/s to synthesize 2.68~GHz clocking sine pulses. For $\omega_d = \omega_{10} / 2$ the expected delay required for $\hat{y}$ rotations is $\phi_d^y = \pi / 4$.

\begin{figure}[t]
    \centering
    \includegraphics[width = .55\textwidth]{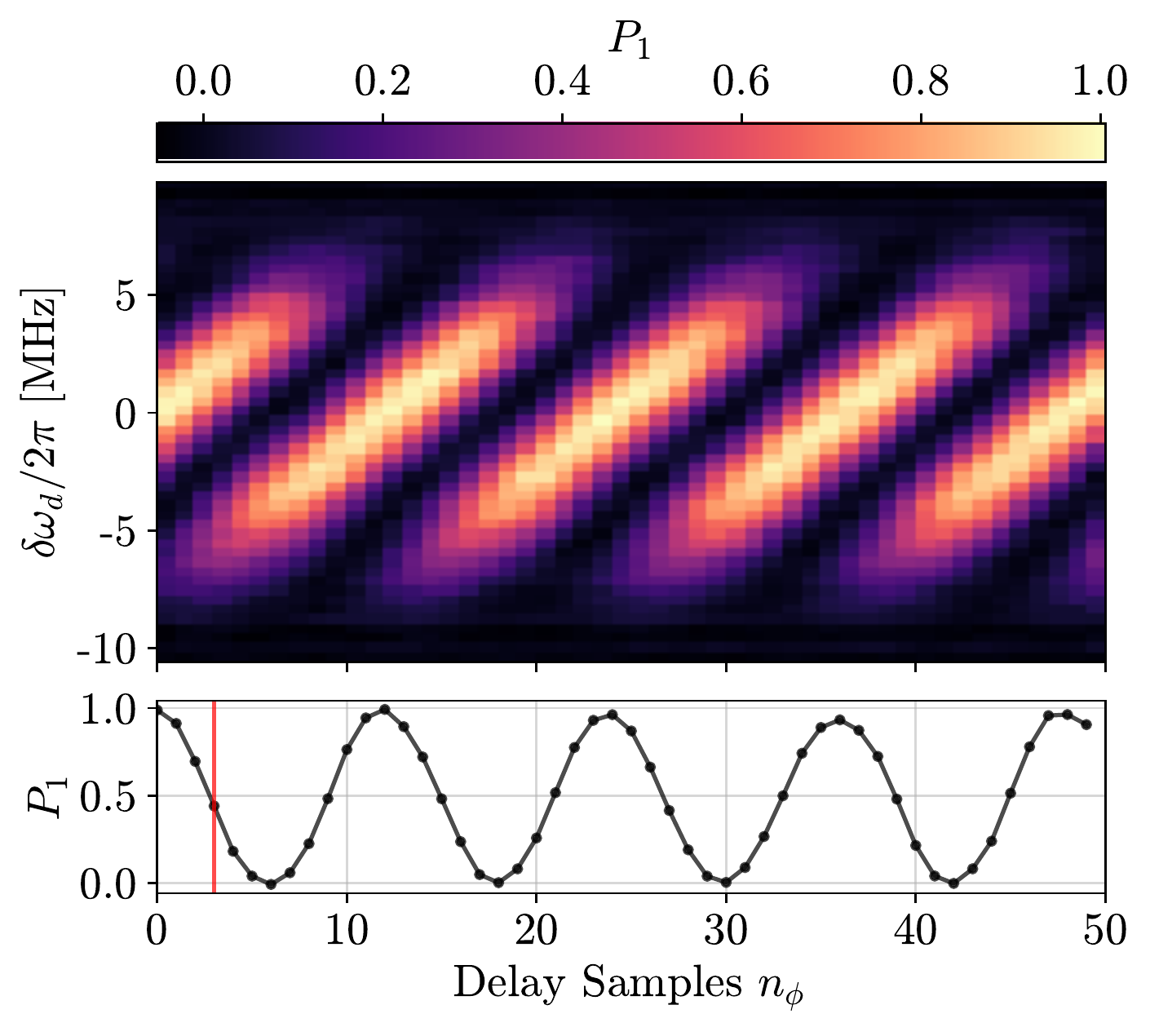}
    \caption{(Color) Calibration of the delay samples $n_\phi$, with respect to the $\hat{x}$ timing reference, which manifests the orthogonal $\hat{y}$ control. In this measurement the clock frequency and number of zero delay samples $n_\phi$ between sine patterns of $\nu_\pi/2$ periods are swept. When $P_1$ progresses from 1 (at $n_\phi = 0$) to 0.5 we have realized an $X_{\pi/2}$+$Y_{\pi/2}$ gate sequence; determining the correct value for $n_\phi$ for $\hat{y}$ control. Using sine patterns with 24 samples per period and with $\omega_d = \omega_{10} / 2$ we expect proper phasing to occur at $n_\phi = 3$ ($\phi_d = \pi / 4$). The black trace (horizontal slice at $\delta \omega_d = 0$) and vertical red line demonstrate correct calibration of $n_\phi$.}
    \label{fig:ramsey_tiger_stripe}
\end{figure}

Phasing between the timing reference (in-phase $\hat{x}$ drive) and the $\hat{y}$ control axis is verified by performing a Ramsey-like experiment. The JPG clock frequency is swept, as well as the delay between two ``$X_{\pi/2}$" pulses. I.e. we begin with zero delay, yielding an $X_\pi$ pulse, and increase the delay by inserting a number $n_\phi$ of zero delay samples until the rotation sequence $X_{\pi/2} + Y_{\pi/2}$ is observed. Thus the signature of the $\hat{y}$ axis calibration experiment will be a measurement of the qubit in the $\ket{1}$ state ($P_1 = 1$) and decreasing to an equal $\ket{0} + \ket{1}$ ($P_1 = 0.5$) superposition when the proper $\hat{y}$ phase ($n_\phi$) is reached. Oscillations then proceed with a period of $4 n_\phi$ delay samples as the required delay in the qubit frame is one-fourth of a period. An example of this calibration routine in shown in Fig.~\ref{fig:ramsey_tiger_stripe}.

For the JPG RB tests we note that there are often subsets of the full pattern resembling $X_\pi+Y_\pi$, or $Y_\pi+X_\pi$ gate combinations. Referring to Fig.~\ref{fig:digital_jpg_qubit_control}(d), we see that to construct a $\hat{y}$ rotation, the phase of the drive must be delayed relative to the initial $\hat{x}$ timing reference by $\phi_d^y = \pi / 2k$. Similarly, to achieve an $\hat{x}$ rotation immediately following a $\hat{y}$ rotation, the drive phase must be advanced by the same amount to re-establish the null timing reference phase. To this end, every $X_\pi$ and $Y_\pi$ JPG waveform is $\nu_\pi + 2$ periods in length. There are $\nu_\pi$ periods of active drive, with the additional two periods of idle divided appropriately at the beginning and end of the waveform to permit delaying and advancing the drive phase as needed. The same procedure is followed for $X_{\pi/2}$ and $Y_{\pi/2}$ rotations. Due to the unidirectional nature of digital qubit control using a JPG, $-\pi/2$ rotations are achieved using $\phi_d^{-x} = 3\pi/2$ and $\phi_d^{-Y} = \pi / 2n + 3 \pi /2$ phases relative to the timing reference (we take $X_\pi = -X_\pi$ and $Y_\pi = -Y_\pi$ as the ideal final states are identical to those from positive gates). Note that, for both positive and negative $\pi/2$ rotations, the drive signals are equal in length as described above.

\section{\label{supp:digital_pulse_fidelity}Digital Qubit Control Fidelity Limits}

Multiple factors arise in a digital qubit control architecture which affect fidelity and are not present in systems using TSCE-based microwave pulses. They are: digitization error, error due to the finite temporal width of the digital pulses, higher state leakage of the pulse train spectrum due to finite qubit anharmonicity, and pulse arrival timing jitter.

\begin{figure}
    \centering
    \includegraphics[width = .55 \textwidth]{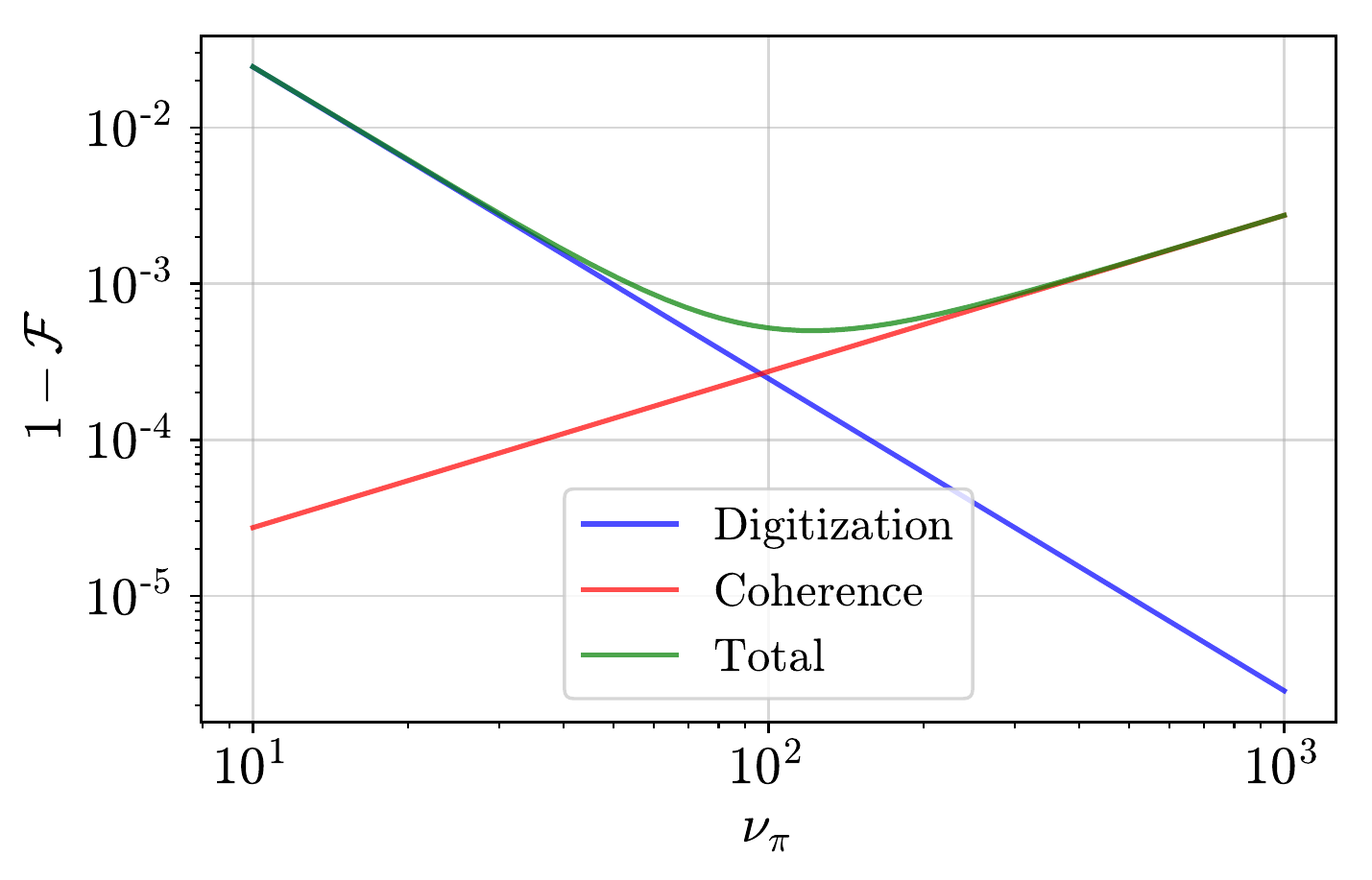}
    \caption{(Color) JPG $X_\pi$ digitization infidelity as a function of $\nu_\pi$, the number of pulses, and thus the gate time, required for an $X_\pi$ rotation. For small $\nu_\pi$ (i.e. fast gates) the digitization error from the finite single pulse rotation angle $\delta \theta$ dominates, whereas for large $\nu_\pi$ the qubit coherence limit becomes dominant. We use representative values for our qubit of $T_1 = T_2^* = 34~\mu$s (i.e. $T_\phi = 68~\mu$s). The qubit frequency is $\omega_{10} / 2 \pi = 5.37$~GHz and drive at $\omega_d = \omega_{10} / 2$.}
    \label{fig:simple_digital_infidelity}
\end{figure}

Digitization error arises from eliminating the continuous variable corresponding to the $X_\pi$ pulse power and manifests as a maximal over- or under-rotation by (at most) $\pm 1/2$ pulse. The digitization error is
\begin{equation}
    1 - \mathcal{F}_{dig} = 1 - \sin^2\left( \pi \frac{\nu_\pi \pm 1/2}{\nu_\pi} \right),
    \label{eq:simple_dig_infidelity}
\end{equation}
which decreases as $\nu_\pi$ increases. With finite qubit coherence time a sweet spot arises which minimizes the infidelity, as is shown in Fig.~\ref{fig:simple_digital_infidelity}. Note that as the qubit lifetime increases the value of $\nu_\pi$ which minimizes the combined digitization and coherence-limited infidelity also increases. For $\nu_\pi = 352$ the digitization infidelity of $3 \times 10^{-5}$ is much smaller than the infidelity due to qubit coherence.

The second affect on the gate fidelity is due to the finite temporal width of the control pulses. As discussed in the main text, Sec.~\ref{sec:gauss_pulse_sims}, we simulate these effects \cite{johansson2012qutip} using a train of Gaussian pulses delivered at $\omega_d = \omega_{10} / 2$ to evolve the qubit state and study the fidelity of $X_\pi$ rotations as a function of the pulse standard deviation $\sigma$. The Hamiltonian for the system is 
\begin{equation}
    \hat{H} = \hat{H}_0 + \hat{H}_d = \frac{1}{2} \hat{\sigma}_z - \Omega_d s_d(t) \hat{\sigma}_x
    \label{eq:sim_hamiltonian}
\end{equation}
where $\Omega_d$ and $s_d(t)$ describe the coupling strength and the pulse train time-dependent amplitude, respectively. Note that due to the pulse drive's time-dependent nature we cannot make a rotating wave approximation. 
Since our simulations use a true two-level system, there is no leakage outside of \{$\ket{0}$, $\ket{1}$\} so $P_1$ can be used to calculate the $X_\pi$ fidelity $\mathcal{F}$. To determine $\mathcal{F}$ we evaluate $P_1$ only at the center of the idle qubit period following a pulse and take the resultant maximum value as $\mathcal{F}$. This also defines the value of $\nu_\pi(\sigma)$, i.e. the number of pulses which maximize the following idle qubit period $P_1$ as a function of $\sigma$. As $\sigma$ is increased, the pulse train power at the second harmonic, and thus the tip angle per pulse, $\delta \theta$, diminishes so more pulses are required to realize a $\pi$ rotation with wider pulses (this is in conjunction with distributed pulse energy delivery as the qubit precesses, discussed below). In turn, $\nu_\pi(\sigma)$ increases so we must recalculate $\nu_\pi(\sigma)$ at each pulse width. For these simulations we include loss with realistic values of $T_1 = T_2^* = 34~\mu$s ($T_\phi = 68~\mu$s) for our qubit. Fig.~\ref{fig:gaussian_pulse_sims} in the main body shows the $X_\pi$ Rabi oscillation curves, $\nu_\pi(\sigma)$, and the corresponding $X_\pi(\sigma)$ fidelity.

If the Gaussian pulse duration ($\sim 4 \sigma$) approaches $2\pi / \omega_{10}$, the pulses no longer deliver the majority of their power when the qubit is aligned with the desired control axes. Rather, the pulses turn on and off slowly relative to the qubit period and their action is distributed over a non-negligible portion of the qubit precession. For a wide pulse near its peak, the pulse delivers the majority of its energy and rotates the qubit state $\ket{\psi}$ along the target control axis. However, the long tails before and after the peak impart lesser rotations while the qubit advances phase at $\omega_{10}$. As the qubit precesses, $\ket{\psi}$ moves through having a component anti-aligned with the target control axis, to being aligned, and then anti-aligned, etc. In the anti-alignment phase the leading and trailing pulse tails induce small negative (back towards $\ket{0}$) rotations on either side of the main positive rotation; corresponding to the pulse maximum and complete alignment of $\ket{\psi}$ and the target control axis. This behavior manifests as a dip in $P_1$ when the pulse begins to arrive, a larger increase in $P_1$ as the pulse peak arrives, and another small reduction in $P_1$ for the trailing pulse tail.

\begin{figure}
    \centering
    \includegraphics[width = .98\textwidth]{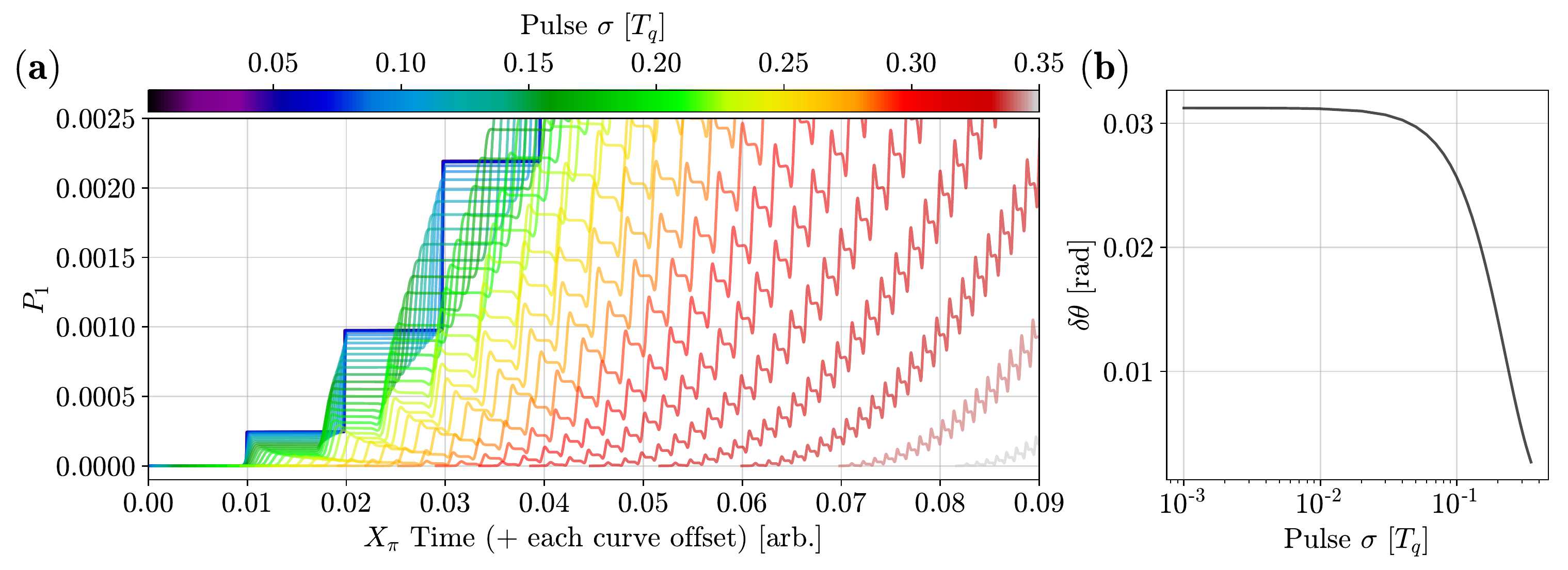}
    \caption{(Color) \textbf{(a)} Depiction of the simulated qubit excited state evolution during delivery of the first few Gaussian pulses and as a function of the pulse $\sigma$. The \mbox{$x$-axis} is scaled to the $X_\pi$ time and each curve is offset relative to their departure from the ideal delta function case of $\nu_\pi = 100$ to better show dynamics. For wide pulses ($4 \sigma \sim 2 \pi / \omega_{10}$) the effects of distributed delivery of the pulse energy during qubit precession can be seen as dips in $P_1$ at the beginning and end of the pulses. \textbf{(b)} Tip angle per pulse $\delta \theta$ (evaluated at the center of the qubit idle period) as a function of $\sigma$. Reduction in $\delta \theta$ is due primarily to pulse energy delivery when the qubit is out of phase of the target control axis, but is also affected by the decreased harmonic tone power for increasing  $\sigma$. Fast pulses do not suffer from these effects and are seen to approach the Dirac delta function limit for $\sigma < 0.02 T_q$.}
    \label{fig:sim_pulse_zoom_offset}
\end{figure}

To demonstrate these effects, Fig.~\ref{fig:sim_pulse_zoom_offset} shows a magnified view of the qubit excited state evolution for the first few pulses and as the pulse $\sigma$ is swept, as well as the reduction in $\delta \theta$ as $\sigma$ increases. We find that for wide pulses ($\sigma \gtrsim 0.1$) the $X_\pi(\sigma)$ infidelity, after subtraction of the coherence limit contribution, roughly corresponds to the rotation lost due to pulse energy delivery while the qubit is anti-aligned with the desired control axis. In the main text we refer to this as the digital-pulse-only infidelity, $1 - \mathcal{F}_{pulse}$. As stated in Fig~\ref{fig:gaussian_pulse_sims}(c), for $\sigma_{\mathrm{JPG}} = 0.19T_q$, we expect $1 - \mathcal{F}_{pulse} = 1 \times 10^{-4}$. This is over an order of magnitude smaller than the coherence-limit contribution as we find $1 - \mathcal{F}_{tot} = 2 \times 10^{-3}$ with finite coherence.

Leakage to the higher transmon level arises from the finite length pulse train used to evolve the qubit state. To estimate the infidelity due to higher state leakage we follow analysis in \cite{mcdermott2014accurate}, which uses $\nu_\pi = 100$ and true Dirac delta function pulses. Since $(1 - \mathcal{F}) \propto \nu_\pi^{-2}$ we rescale by a factor $(352 / 100)^{-2}$. For our qubit, with an anharmonicity $\alpha = 5\%$, this results in higher state leakage infidelity of $7 \times 10^{-4}$. We note that for our case of moderate-width pulses, the ringing in the pulse train spectrum will be further suppressed relative to the Dirac delta function pulse train in \cite{mcdermott2014accurate}.

For our configuration we consider the case of pulse timing jitter when the JPG is driven with a stable external clock \cite{mcdermott2014accurate}. Gaussian-distributed thermal jitter in pulse generation in SFQ circuitry has been measured to have a standard deviation of 3~ps or less at 4.2~K \cite{rylyakov1999pulse}. However,  for an $N_{JJ} = 4650$ array, the effect of thermal jitter at each JJ results only in a slight broadening of the output JPG pulse with little effect on the pulse arrival time.  Timing jitter from the JPG drive signal, however, will cause jitter of the JPG output signal. Measurements of jitter in the JPG drive signal at 300~K indicate jitter of 3~ps (standard deviation), corresponding to a jitter-sourced infidelity estimate of $3 \times 10^{-3}$.

Combining all contributions above we obtain a total expected gate infidelity of $1 - \mathcal{F}_{tot} = 6 \times 10^{-3}$, a factor of three lower than the observed JPG RB error per gate of $2 \times 10^{-2}$. We attribute the unaccounted infidelity to possible systematic and coherent errors in the RB routine and are actively investigating these effects.

\section{\label{supp:jpg_temp_sensitivity}JPG Fabrication details and Reduced $I_c$ Temperature Sensitivity}

\begin{figure}[ht]
    \centering
    \includegraphics[width = .98\textwidth]{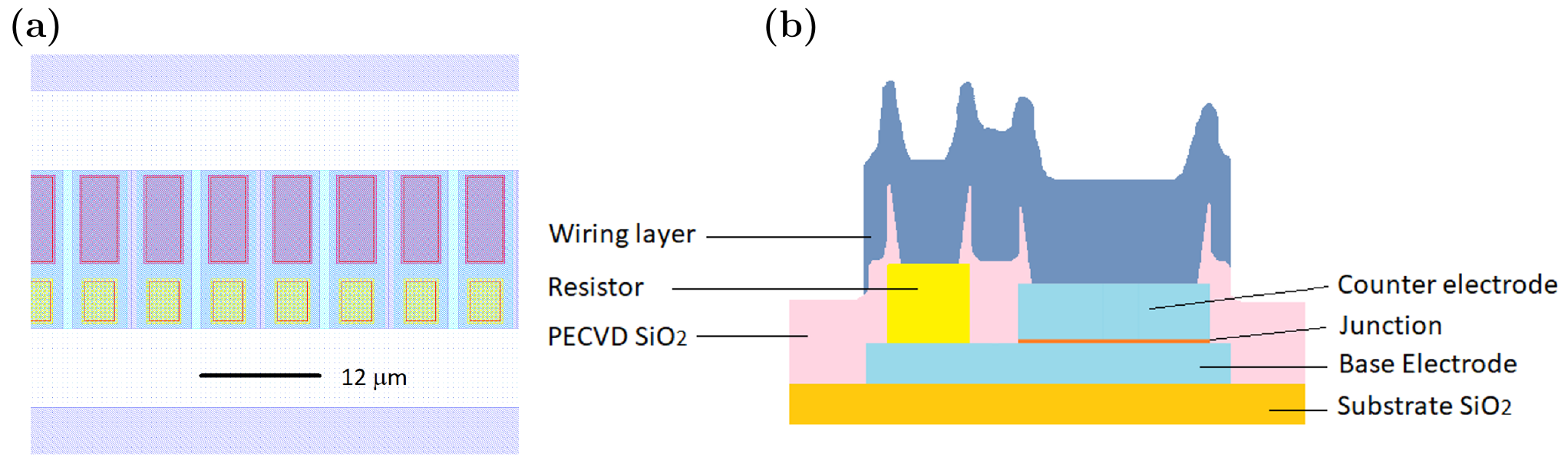}
    \caption{(Color) \textbf{(a)} Rendering of the JJ (red) and shunt resistor $R_n$ (yellow) layout within the center conductor of the coplanar waveguide. The top wiring layer is not shown. \textbf{(b)} Cross-sectional diagram of the JPG layer stack (non-planarized) where the view into the page for (b) corresponds to looking left along the array in (a).}
    \label{fig:aSi_fab}
\end{figure}

\begin{figure}[ht]
    \centering
    \includegraphics[width = .5\textwidth]{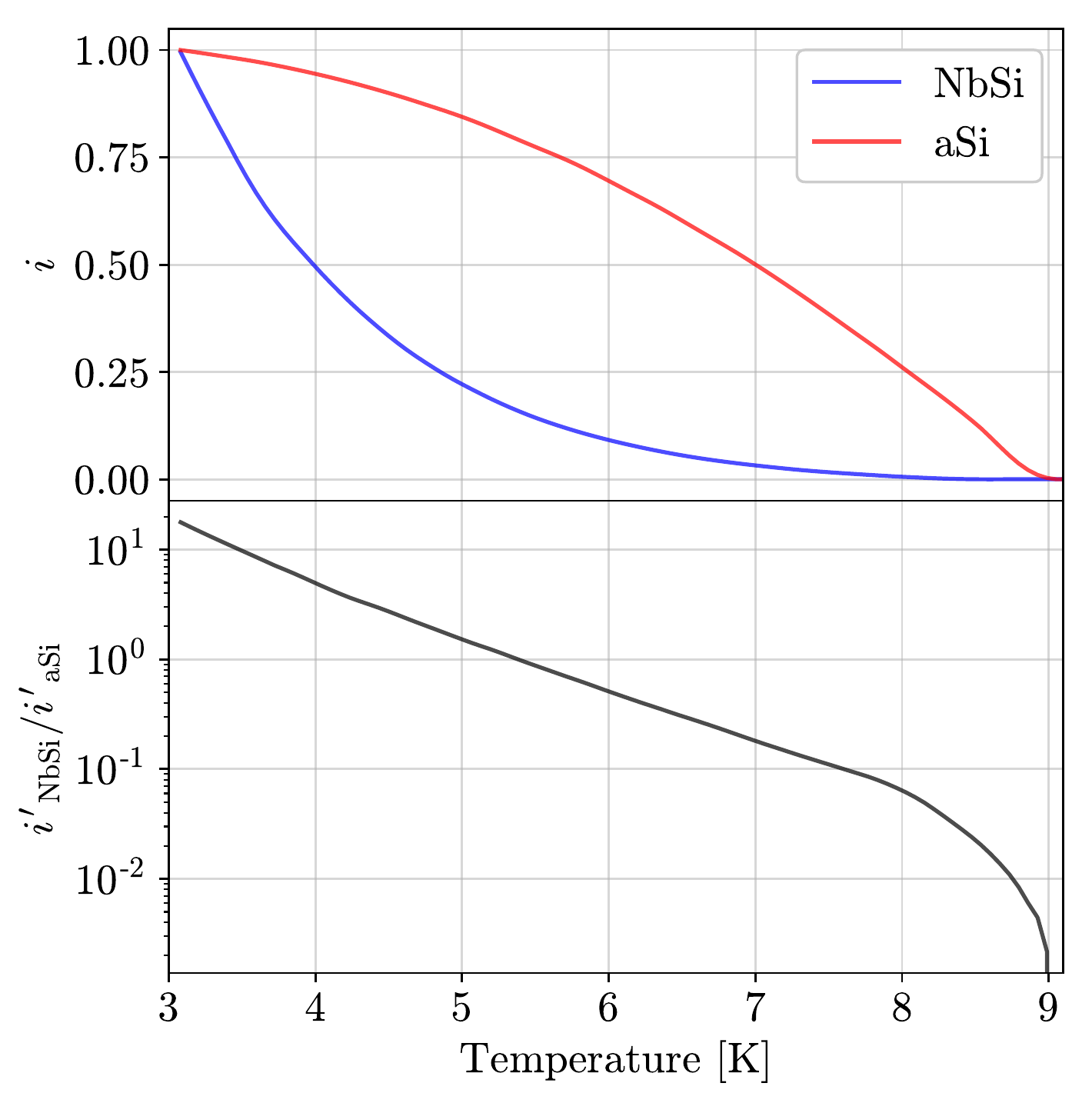}
    \caption{(Color) Reduced critical current $i = I / I_c^{3.1~\mathrm{K}}$ of NbSi and $\alpha$Si barrier JPG devices (top), and the temperature-sensitivity ($i' = di/dT$) ratio of the NbSi and $\alpha$Si devices (bottom). The drastically-reduced 3~K - 4~K temperature sensitivity of the $\alpha$Si JPG provides the necessary stability of the Shapiro step when the JPG is driven with both short patterns for $T_1$, $T_2^*$, and $P_{th}$ measurements, and with long patterns for measuring many Rabi oscillations and performing randomized benchmarking.}
    \label{fig:Ic_vs_T}
\end{figure}

Gating pulses from a JJ array with our scheme of driving with an integer number of pure sine periods means the rf dissipation is toggled rapidly between null and 100\%. Depending on the $I_c$ temperature sensitivity, this can cause an unstable locking range for a JJ array and make calibrations such as those detailed in Fig.~\ref{fig:jpg_bringup} difficult or impossible.

\begin{figure}
    \centering
    \includegraphics[width = .9 \textwidth]{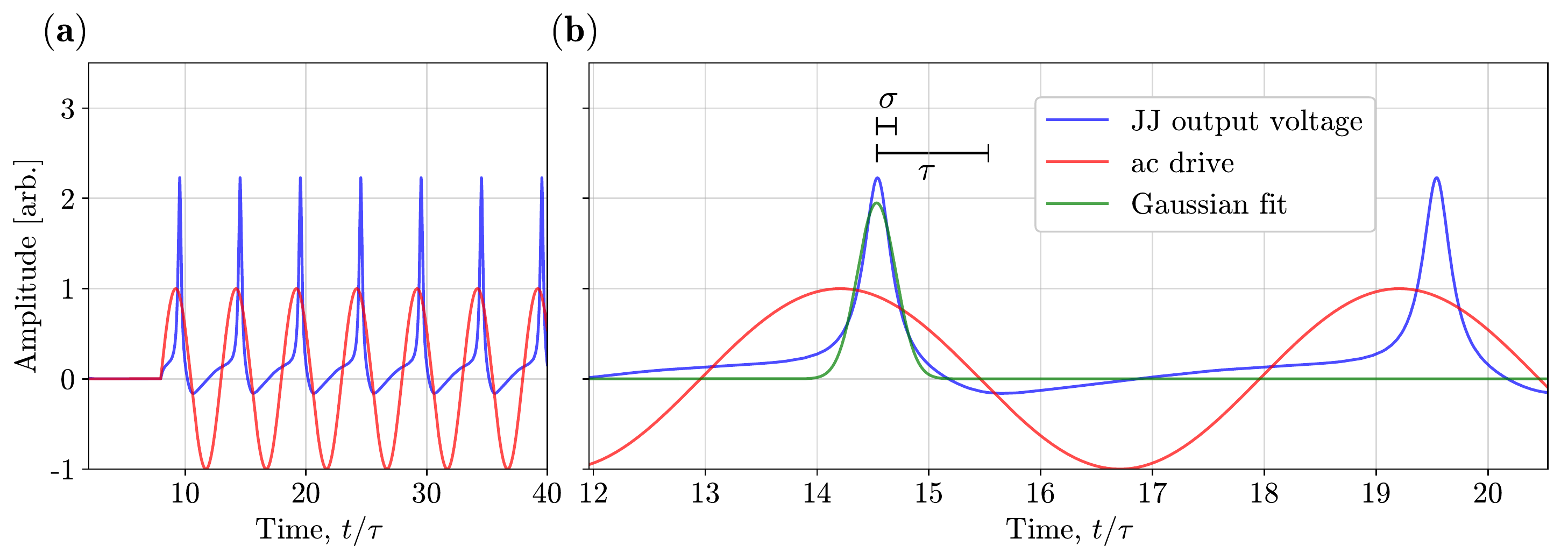}
    \caption{(Color) Simulation of the output of a single-JJ using the RSJ model, driven by an ac drive at frequency $f_d / f_c = 0.2$ and with a dc current bias which places the JJ on the first Shapiro step and results in a periodic train of SFQ pulses. \textbf{(a)} The first seven SFQ pulses and the ac drive signal \textbf{(b)} Magnification of the second and third SFQ pulses with a Gaussian fit to the former. Here we also show pictorially the relative size of the fitted Gaussian $\sigma$ and the JJ characteristic time $\tau$.}
    \label{fig:tau_vs_sigma}
\end{figure}

To reduce the temperature sensitivity of the JPG we have departed from the superconductor-normal metal-superconductor (SNS) JJ technology \cite{fox2015junction} based on triple-stacked self-shunted niobium-doped-silicon barriers (NbSi) embedded in the center conductor of a niobium (Nb) superconducting coplanar waveguide. Instead, we form the JPG array from externally shunted superconductor-insulator-superconductor (SIS) junctions, as can be seen in Fig.~\ref{fig:aSi_fab}(a). The JJs were fabricated using amorphous silicon ($\alpha$Si) for the barriers and palladium-gold (PdAu) alloy for shunting resistors. Due to the small value of the shunting resistors (25~m$\Omega$) required to achieve the target characteristic frequency of $f_c = 10$~GHz, the usual method of fabricating planar resistors could not be used. That method relies on resistor films with sheet resistances of \mbox{1--2 $\Omega / \Box$}. The form factor to achieve 25~m$\Omega$ entails very wide and long arrays to accommodate interconnecting vias to minimize contact resistance. Instead, (vertical) stud resistors were used with a height similar to the junction counter electrode thickness. See a schematic of the cross-section in Fig.~\ref{fig:aSi_fab}(b). This enables significant reduction in the JJ cell size and results in a much more compact array.

The circuits were fabricated on 3-inch silicon wafers with 150 nm of thermal oxide. A trilayer of Nb/$\alpha$Si/Nb was deposited by sputtering. The thickness of the silicon was chosen to obtain a junction critical current density of 12~kA/cm$^2$. PdAu resistors were electron-beam evaporated with a thickness of 240~nm. Silicon oxide insulator was then deposited by electron cyclotron plasma enhanced chemical vapor deposition (EC-PECVD). Lithography followed by dry etching opened vias to the junction counter electrodes, resistors, and base electrodes. The wiring layer was deposited by niobium sputtering and then patterned by dry etching. Resulting devices display $I_c = 3.05$~mA and $R_n = 6.93$~m$\Omega$, giving $f_c = 10.2$~GHz. These devices exploit the much more stable $I_c$ of $\alpha$Si barrier JJs with respect to temperature. Fig~\ref{fig:Ic_vs_T} shows a comparison of the temperature stability of $I_c$ for NbSi and $\alpha$Si devices, demonstrating that in our region of interest of 3--3.3~K the $\alpha$Si barrier device is \mbox{$10$--$20$} times less sensitive.

\section{\label{supp:jpg_simulations}Relation Between JPG $\tau$ and Gaussian-fitted $\sigma$}

Simulations of JPG dynamics can be performed by numerically integrating the resistively shunted junction (RSJ) model with an ac drive term $i_{\mathrm{ac}} \sin(f_d \Theta / f_c)$
\begin{equation}
    \frac{d \phi}{d \Theta^2} = \frac{1}{\beta_c}\left[ i + i_{\mathrm{ac}} \sin\left( \frac{f_d}{f_c} \Theta \right) - \sin\phi - \frac{d \phi}{d \Theta} \right].
    \label{eq:rsj_model}
\end{equation}
Here $\phi$ is the superconducting phase difference across the JJ, $\Theta$ is the dimensionless time variable $\Theta = (2 \pi I_c R_n / \Phi_0) t = 2 \pi t / \tau$, $\beta_c = 2 \pi f_c R_n C$ is the Stewart-McCumber damping parameter with $C$ the intrinsic JJ shunting capacitance, and $i = I_d / I_c$ and $i_{\mathrm{ac}} = I_{\mathrm{ac}} / I_c$ are the normalized dc and ac (amplitude) bias currents, respectively. Note that the ac drive frequency $f_d$ has units of time scaled to the Josephson time constant $\tau = 1 / f_c = \Phi_0 / I_c R_n$ and not to $\Theta$, as these differ by a factor of $2 \pi$.

With the second Josephson equation
\begin{equation}
    V(t) = \frac{\Phi_0}{2 \pi} \frac{d \phi}{dt}
    \label{eq:second_josephson}
\end{equation}
and Eq.~\ref{eq:rsj_model}, we can solve for the time-dependent voltage response of the JJ. We choose $i_{\mathrm{ac}} = 0.6$, and $f_d / f_c = 0.2$, which approximately reflects the relation between the JPG $f_c$ of 10.2~GHz and the typical drive frequency of $\omega_d / 2 \pi = 2.68$~GHz, and ramp $i_\mathrm{dc}$ until pulses are observed. For $\beta_c \ll 1$, the JJ response is overdamped and non-hysteretic so we simulate a heavily overdamped regime of $\beta_c = 0.01$. For our JPG, $\beta_c \lesssim 10^{-4}$. Results are displayed in Fig~\ref{fig:tau_vs_sigma} and we perform a Gaussian fit to one SFQ pulse. Time units are re-scaled from the $\Theta$ time of Eq~\ref{eq:rsj_model} to units of the JJ characteristic time $\tau$. For these parameters we extract $\sigma / \tau = 1.08 / 2 \pi$. This indicates that, given the JPG $\tau$ of 98~ps, the pulses at 3~K (on-chip, prior to any significant dispersion and broadening) have an expected Gaussian sigma of 17~ps. Fits at room temperature of the JPG pulses captured on the oscilloscope (Sec.~\ref{sec:gauss_pulse_sims} of main body) thus display approximately a factor of two in broadening compared to the expected on-chip value. As mentioned in Sec.~\ref{sec:gauss_pulse_sims}, this is the reason we take $\sigma = 35$~ps to be a conservative upper bound on the approximate width of pulses delivered to the qubit, even after traversal of the readout cavity resonance.

\end{document}